\documentclass[fleqn,usenatbib]{mnras}

\usepackage[T1]{fontenc}
\usepackage{ae,aecompl}

\usepackage{graphicx}   
\graphicspath{{AASTEX//}}

\usepackage{amsmath}    
\usepackage{amssymb}    

\usepackage{xfrac}
\usepackage{float}
\usepackage{physics}
\usepackage{amssymb}
\usepackage{savesym}
\savesymbol{tablenum}
\usepackage{siunitx}
\usepackage{rotating}
\usepackage[section]{placeins}
\usepackage[noabbrev]{cleveref}
\usepackage{subfigure}

\newcommand{\HI}{{H}{I} }

\newcommand{\kms}{km s$^{-1}$ }
\newcommand{\msun}{M\textsubscript{\(\odot\) }} 
\newcommand{\dunit}{m \textsuperscript{-3}}
\restoresymbol{SIX}{tablenum}

\newcommand{\tus}[1]{$_{\text{2}}$}
\newcommand{\hh}{H$_2$ }

\usepackage{natbib}
\usepackage{url}
\usepackage{appendix}
\usepackage{hyperref}

\title[Models of Splash Bridges]{Models of Multi-component Splash Bridges in Face-on Galaxy Disc Collisions}

\author[Yeager \& Struck]{
Travis R. Yeager,$^{1}$\thanks{E-mail: yeagertr@iastate.edu (TRY)}
and Curtis Struck,$^{1}$
\\
$^{1}$Department of Physics and Astronomy, Iowa State University, Ames, IA 50014, USA\\
}

\date{Accepted March 26, 2019. Received March 25, 2019; in original form December 21, 2018}

\pubyear{2019}

\begin{document}
\label{firstpage}
\pagerange{\pageref{firstpage}--\pageref{lastpage}}
\maketitle

\begin{abstract}
We use an inelastic particle code with shocks and cooling calculated on a subgrid level to study the gas in direct collisions between galaxy discs.  The interstellar media (ISM) of the discs are modeled with continuous thermal phases.  The models produce many unique structures, collectively called splash bridges.  They range from central bridge discs to swirled sheets, which resemble those observed in interacting galaxies.  These morphologies are sensitive to the rotation, relative mass, disc offsets and the gas structure in the discs.  In the case of the Taffy galaxies - NGC 12914/15, extensive observations have revealed radio continuum emitting gas, \HI gas, hot X-rays from hot diffuse gas and more \hh than exists in the Milky Way coexisting in the bridge.  The origins of the \hh and large asymmetric distribution of ISM are not clear.  We show that for small disc impact parameters, multiple phases of ISM with densities over many orders of magnitude can be removed from their host galaxies into a Taffy-like bridge.  The orientation of the discs initial overlap can have a great effect on the distributions of each phase of ISM.  In some cases, the models also predict the creation of a possible `dark galaxy,' a large flat region of dense ISM far from the stellar disc potential of either galaxy.
\end{abstract}



\section{Introduction: Swing and Splash Bridges}

\indent  Galaxy interactions produce a number of iconic morphologies that are not only diagnostic of the nature of the interaction, but which may also influence the subsequent evolution of the galaxies or their merger remnant \citep{toomre72, struck99gc}.  These morphologies include bridges of stars and gas connecting the visible components (usually the discs) of the two galaxies. There are two types of bridges, which we call swing bridges and splash bridges \citep{struck99gc}. The former are essentially tidal tails whose outer parts have been captured onto the interaction partner, generally in prograde, flyby encounters. These are often continuous and relatively smooth stellar structures, made up of stars like those in the outer disc of the parent galaxy, as well as young clusters in some cases (e.g., \citet{schombert90} and examples in the Arp Atlas).  

\indent  Splash bridges are the result of direct collisions between the gaseous components of two discs, and have a more complex appearance than swing bridges. They do not generally have a smooth old star component, though mixed swing/splash bridges can. They usually contain large masses of HI gas and can have young star clusters \citep{hibbard01}. The generally close proximity of the two parent galaxies suggests that they are very young, but it is difficult to make generalizations about their properties because very few have been studied in detail. This is understandable because, like most interaction induced structures, there are very few nearby examples, so radio telescopes have very limited resolution for many bridges. 

An exception to this rule is the so-called Taffy galaxy system, consisting of two colliding gas-rich discs, UGC 12914 and UGC 12915. This system has been well observed in many different bands over the last 25 years. The nick-name derives from the polarized radio continuum bridge discovered by \citet{condon93}. There is a large amount of \HI gas in the bridge between the two discs, and \citet{gao03} observed CO emission with the BIMA telescope. Both the HI and CO observations show a clear asymmetry near the disc of UGC 12915.  A decade later warm \hh emission was observed in the bridge by \citet{peterson12} with a cooling time scale of only about 5000 years. It was suggested that the source of the emission is turbulence still occurring in the molecular gas.  There is an extraordinary amount of H\tus{2} in the bridge of the Taffy.  Estimates place the mass of molecular material in the bridge at \SI{2e10}{M\textsubscript{\(\odot\)}} \citep{ss01}.  They have also estimated the H\tus{2} masses in the discs at \SI{1.7e10}{M\textsubscript{\(\odot\)}} and \SI{1.9e10}{M\textsubscript{\(\odot\)}}  for UGC12914/15 respectively. Another value by \citet{braine03}, gives a H\tus{2} bridge mass of 2-\SI{9e9}{M\textsubscript{\(\odot\)}}.  This is of order unity with the \HI observed in the bridge and is comparable to the amount of \hh in the host galaxies themselves.  

Soft X-ray emission was observed over the entire bridge with the Chandra telescope as described in \citet{appleton15}.  Several individual X-ray sources were also detected, indicative of recent star formation.  One of these appears to be in an asymmetric region located near UGC 12915.  The bridge was modeled by \citet{vollmer12} who emphasized a prominent X-shaped feature in the models.  Their model explains the warm \hh emission seen by the Spitzer Space telescope by assuming that all the hydrogen gas scattered into the bridge also carries with it a constant proportion of \hh.  

The Taffy galaxies are evidently colliding ring galaxies, and this class of interacting galaxy is a likely place to find more splash bridges. However, there is no evidence for splash bridges in many of the best known colliding ring galaxies. There are a variety of possible reasons for this, beginning with the fact that HI observations are not available for many.  Other reasons include the possibility that: at least one of the galaxies is an early-type with no gas disc, one of the discs is very gas-poor (e.g., possibly Arp 147 and the Lindsay-Shapley ring), or at impact the discs are nearly orthogonal. In the latter case each disc slices through a small part of the other, so a relatively small fraction of the gas is directly impacted. Moreover, in the impacted regions the gas column densities of the two discs are very different, i.e. the vertical column density of one disc versus whole in-plane column density of the other. Thus, the gas from the sliced disc is essentially accreted onto the other galaxy, with little splashed into the bridge (see \citet{struck97simulations}). Arp 271, NGC 7125/26, and ESO 138 - IG029 (the 'Sacred Mushroom') may be examples of slicing impacts based on the fragmentary HI bridges in the HI Rogues Gallery \citep{hibbard01}. Nonetheless, the HI Rogues Gallery does contain a few ring systems with HI bridges, including: Arp 284, Arp 298, the Cartwheel, and VII Zw 466. Most of these probably involved somewhat tilted (partially slicing) impacts, whereas the Taffy impact was evidently nearly face-on for both galaxies. Arp 284, in particular, probably had a somewhat off-centre, partial fly-by encounter \citep{smithstruck97}. Given the extended HI discs of many galaxies, still more off-centre impacts may generate splash bridges, but not a ring galaxy.

\indent \citet{davies08} outline the detection of a dark galaxy candidate, VIRGOHI21, which could be a relic of a splash bridge.  Interestingly, the \HI observations of VIRGOHI21 reveal a \HI galactic disc with a mass of \SI{2e7}{M\textsubscript{\(\odot\)}} embedded in the \SI{2e8}{M\textsubscript{\(\odot\)}} tidal bridge.  Stars are not observed in this object; lower limit for mass to luminosity ratio was found to be \SI{e6}, which the authors note is significantly higher than a typical galactic value of 50 or less.  

Tidal tails and swing bridges often have a rather simple morphology, e.g., long, relatively thin and continuous appearances in optical images. Sometimes there is a pile-up of young star clusters (and presumably gas) along an arc at the outer edge, sometimes the object is wider with two sharp edges. Most of this structure can be readily understood as the result of stars from a region of the initial disc being impulsively launched with a range of tidal velocities in the interaction, and pursuing simple kinematical orbits thereafter. 

Splash bridges result from inelastic collisions between gas elements in the two galaxy discs. Viewed face-on galaxy discs show a wide range of phases and column densities, from HI holes filled with tenuous, hot gas, to diffusely spread HI gas, to an ensemble of dense molecular clouds with a small covering fraction. This structure is superposed on a general decrease of column density with radius, with amplitudes and scalelengths that vary among discs. Colliding gas elements of equal column density will be left near the mid-point of the bridge. Elements of greater or lesser column density than their collision partner will remain near their parent disc or be removed to positions close to the other disc. Given the range of column densities involved, and large fluctuations across the discs, we can see that even face-on, centre-on-centre collisions between two identical non-rotating discs would generate complex bridge structure and turbulence. If we further consider rotations of different magnitudes and possibly different senses in the two discs, then we add a global swirl as well as increased turbulence to the bridge. 

In addition, increasing the offset between the two disc centres at impact yields a more equal mix of regions that impact prograde versus retrograde (or vice versa) until the discs start to separate. If the discs are not face-on, but have a relative tilt, then the cloud-cloud impacts occur over a prolonged interval. Altogether these factors can lead to a great deal of complexity, and turbulent driving, within splash bridges. To date, there has been little study of this complexity. 

Splash bridges are rare and short-lived on galaxy timescales, so examples generally must be found at greater distances than the Taffy, where they are harder to study. Yet, their importance may have been underestimated. Reasons to believe this include the fact that, like galaxy mergers, cosmological structure simulations show that splash bridges were much more common at early times (and more gas rich). Other reasons include the possibility that star formation in splash or mixed splash/swing bridges contributes significantly to stellar halo populations, or that, like tidal tails, delayed fallback from splash bridges fuels and triggers disc star formation.

Given these motivations, the question is how best to study these structures? \citet{vollmer12} have used a SPH particle code to study the Taffy Galaxies, and succeeded in reproducing a number of features of that system.  However, conventional particle codes have some disadvantages in simulating splash bridges. First of all, it takes very large spatial and particle resolution to accurately model the small scale structure, including shocks, whose strength varies rapidly with position. Secondly, the discussion above suggests that there will be a great deal of kinematic mixing and swirling in the bridge, which is easiest to study if there is some direct information on proximate gas elements. This can be hard to discern in images of identical particles (or particle density), though it can be reconstructed. 

Given these difficulties, in the present paper we have opted to focus on the complex global kinematics, and simplify the hydrodynamics by assuming highly inelastic collisions between gas elements in the colliding discs. In addition, we compute the subgrid heating and cooling effects of the shocks driven into each gas element, estimate the shock propagation time through the gas elements, and model the subsequent rarefactions. Algorithmic details are given in the following section. There are many limitations to this simple approach, and it can only be used to model bridge evolution for a relatively short time before the neglected effects are expected to become important. However, it has the advantage of providing good resolution with modest computer resources, and is scalable for better resolution. It is also able to reveal the sensitivity of morphological and thermo-chemical bridge structures to parameter changes expected from the effects described above. These structural sensitivities are described in the results section below. 

\section{The Model}
\label{themodel}

\subsection{Numerical Methods}
\label{methods}

\indent We have created an inelastic particle code to study the distributions of multi-component ISM in galaxy collisions. The two initial galaxy discs are modeled as Sc galaxies with the size and mass parameters like those of the Taffy system.  The code has sufficient flexibility to recreate a large range of face-on disc on disc interactions.  We start with two galaxy discs and collide them with a preset overlap.  The density or temperature of the larger disc, modeled after UGC 12914, will be referred to as G1 and the gas from the other slightly smaller UGC 12915 will be referred to as G2.  Fixed potentials centered on each galaxies evolving center of mass are used to represent the dark haloes and stellar discs of each galaxy.  These potentials and the method of numerical integration are described below.  A Cartesian grid is initially laid across each galaxy disc.  The material in each grid cell is evolved as a test particle.  In grid cells where the galactic discs collide, the cells containing gas from each galaxy are merged, see below and figure \cref{fig:clouds}.  The code also utilizes subgrid physics to approximate shock waves and cooling of individual and merged gas elements.  A contact discontinuity is assumed to form between the gas of each galaxy.  Two shocks are produced at this discontinuity and travel back through each cloud. (We use the terms particle, cloud and grid element interchangeably, though `cloud' is often preferred in discussions of subgrid processes.)   The gas behind the shock is both compressed and heated, the subgrid treatment of this is described below.  Evolution of the galaxy discs is begun at the time of collision between the galaxies and particle collisions are treated as completely inelastic. \\

Each gravitational potential is centred on its galaxy's centre of mass, this centre is treated like an effective point mass and moved accordingly subject to external potentials.  The halo potential takes the form of  \citet{hernquist1990},
\begin{equation} \phi_{halo} = \frac{-G M_{halo}}{(r + l_{halo})} \end{equation} 
where $l_{halo}$ is the scalelength for the halo and r is the three dimensional radius.  The disc potentials use the scalelength $l_{disc}$ and scaleheight  $h_{disc}$ to calculate a best fit sum of three Miyamoto-Nagai (MN) potentials, each with new scaling constants and of the form:

\begin{equation} \phi_{disc} = \frac{-G M_{n}}{\sqrt{(R^2 + (l_{n}+\sqrt{z^2+h_{n}^2})^2}} \end{equation} 

Here r is a two dimensional radius and the parameters, $M_{n}$, $l_{n}$ and $h_{n}$ are the calculated for three separate MN disc potentials following the method in \citet{smith15}. They have found a convenient set of solutions that produce an exponential disc potential within a few percent mass deviation, and have faster run speeds than multi-exponentials.

\FloatBarrier
\subsection{Initial Conditions}
\label{initialconditions}
The total galaxy masses are taken as \SI{4.4e11}{M\textsubscript{\(\odot\)}} and \SI{2.4e11}{M\textsubscript{\(\odot\)}} for UGC 12914/15 respectively (i.e., Taffy system determinations from \citealt{condon93}).  The disc and halo masses are assumed to be 23.5$\%$ and 76.5$\%$, respectively, of the total galaxy's mass.  In our attempt to compare/reproduce results of Vollmer et al. (2012), we adopt the same values for scale lengths and masses.  For the \HI disc scale lengths ($l_{disc}$) for UGC 12914/15 are 6 kpc and 3.8 kpc with scale heights of 1.6 and 1 kpc.  The halo scale lengths ($l_{halo}$) are 10 kpc and 6.3 kpc, respectively.  The specific parameters of each MN disc potential is given in \cref{tab:MNconsts}.

\begin{center}
\begin{table}
\caption{Constants calculated for MN disc potentials.  Subscript notation (n,m) where n denotes the disc component and m the galaxy.}
\label{tab:MNconsts}
\begin{tabular}{| c | c | c | c |}
\hline
Constant & Value (m) & Constant & Value (kg) \\
\hline
 $l_{11}$ & $6.8341x10^{19}$ & $M_{11}$ & $4.1762x10^{40}$ \\ 
 $l_{21}$ & $4.3337x10^{20}$ & $M_{21}$ & $-1.7082x10^{42}$ \\
 $l_{31}$ & $3.8206x10^{20}$ & $M_{31}$ & $2.0079x10^{42}$ \\
 $l_{12}$ & $4.3282x10^{19}$ & $M_{12}$ & $1.8375x10^{40}$ \\
 $l_{22}$ & $2.7447x10^{20}$ & $M_{22}$ & $-7.5162x10^{41}$ \\
 $l_{32}$ & $2.4197x10^{20}$ & $M_{32}$ & $8.8348x10^{41}$ \\ [1ex] 
\hline
\end{tabular}
\end{table}
\end{center}

\indent The initial gas densities and temperatures conditions for the larger galaxy can be seen in \cref{fig:g1initialconditions}, the second galaxy is set up in a similar way.  The \HI gas disc is given a $\sfrac{1}{R}$ profile with the density peak dependent on total mass, but of order 10\textsuperscript{7} \dunit. The total mass of \HI is \SI{1.72e10}{M\textsubscript{\(\odot\)}} and \SI{1.13e10}{M\textsubscript{\(\odot\)}} for G1 and G2 respectively.  Added to the basic \HI disc are four other components which are scaled with respect to the initial \HI disc.  Spiral arms are also added into the \HI disc.  The medium in the spiral arms is a randomly mixed Cold Neutral Medium of 1-10 times the density of the local \HI and H\tus{2} that is 10-100 times the local density.  Throughout the entire disc we have also created several "holes" of diffuse warm and hot ionized HI.  Here the warm gas is taken as \SI{e-2} and hot \SI{e-4} times the local \HI density.  An average initial effective temperature is assigned to the gas depending on what component it is.  This temperature is currently only relevant in calculating the local sound speed of the medium before it is shocked, which affects the Mach number of the incoming waves.  These initial temperatures are chosen to be 50, 100, 6000, 85000, $\SI{e6} K$ in the various phases.  These values are typical of gas in these phases observed in the Milky Way.

\indent Initial orbital velocities for each particle include circular orbit velocities around their galaxy centre, and the relative motion of the galaxies. The circular velocity in the plane of the galaxy is:
\begin{equation} v_{circ}=\sqrt{R \ \va{\nabla} \phi(R,z)}
\end{equation} 

Each galaxy is also assigned a rotation sense. The maximum orbital velocities are 325 and -275 \kms for UGC 14/15 respectively.

\subsection{Cloud Collisions}
At the outset a single x-y Cartesian grid crossing the simulation volume is projected onto both the galactic discs. Cell centers with the same (x,y) values in the two disc projections directly collide, and are assumed to stick (completely inelastically) at their contact face. Cells at the outer edge of each disc are assumed to have one curved face, rather than being perfectly rectangular. For all the models in this paper we have used a 100 pc grid cell length.  

\begin{figure}
\caption{The set up of each cloud collision.}
\label{fig:clouds}
  \centerline{\resizebox{0.4\textwidth}{!}{\includegraphics{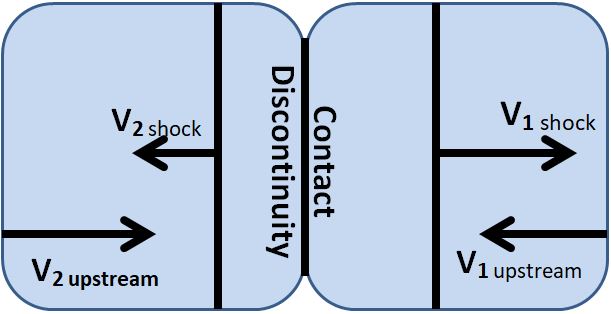}}}
\end{figure}

\begin{figure}
\caption{Initial conditions set pre-collision for the model of UGC12914.  The numbers in ISM type refer to the different phase components. 1) H\tus{2}, 2) cold \HI, 3) \HI, 4) warm HII, 5) Hot diffuse HII}
\label{fig:g1initialconditions}
  \centerline{\resizebox{.5\textwidth}{!}{\includegraphics{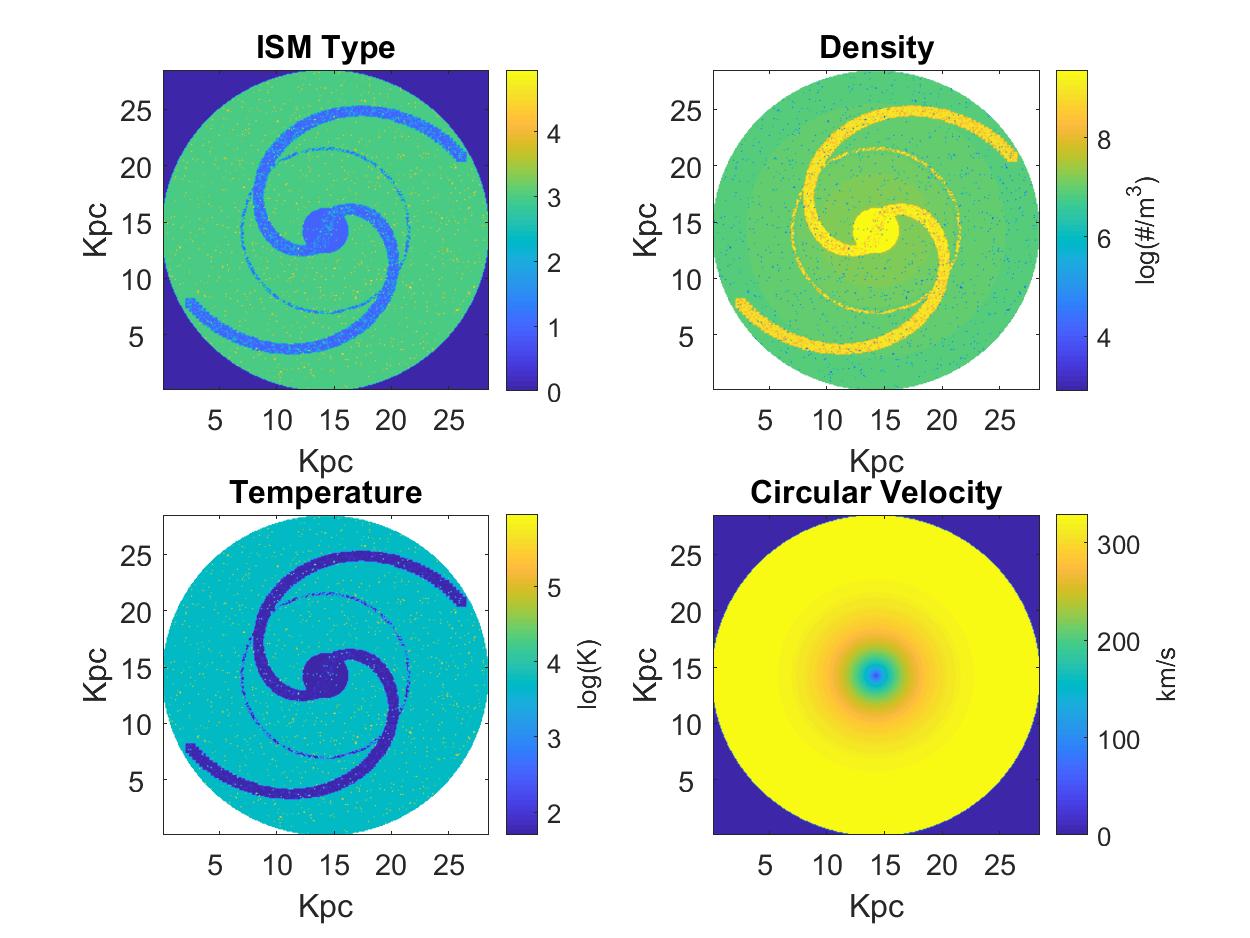}}}
\end{figure}

\Cref{fig:clouds} diagrams the treatment of shocks in each cloud collision.  A contact discontinuity is created at the moment of collision.  On each side of the discontinuity we assume the gas remains in distinct phases, without mixing.  Shock waves are produced at this contact discontinuity and travel through each cloud.  After colliding and sticking the gas elements on both sides of the contact discontinuity are assumed to move together at their centre of mass velocity.  This velocity is determined by conserving momentum across the discontinuity.  The centre of mass velocity is used as the initial condition for integrating the equations of motion of the merged gas elements, henceforth referred to as particles or clouds.  We decouple the 2nd order Newtonian equations of motion and use the Matlab ode23 integration package for solving.
\begin{equation} \dv{q}{t}= v_{q}
\end{equation} 
\begin{equation} \dv{v_{q}}{t}= - \dv{q} \phi(x,y,z),
\end{equation}

\noindent where $q$ represents a coordinate direction, and $v_q$ the corresponding velocity, and $\phi$ is the total gravitational potential.  

Gas particles in both galaxies feel the forces from all potentials.  Gas particles are evolved with a single timestep across the grid.  This allows us to minimize integration errors.  

\subsection{Shocks and Heating}
Shocks are produced at the contact discontinuity between colliding clouds and travel into each cloud.  

Assuming that the gas is steady state, inviscid, non-accelerating, and ideal we have, 

\begin{equation} \epsilon = \frac{3 k_{B} T}{2 m},
\end{equation} 
\begin{equation} P = \frac{2}{3} \rho \epsilon = n k_{B} T,
\end{equation} 

\noindent where $\rho$ is the local gas density, $n$ the number density, $P$ the pressure, $\epsilon$ the specific energy, $k_B$ is the Boltzmann constant, and $m$ is the mean particle mass.

Then the Rankine-Hugoniot jump conditions are (with heating and cooling $\Gamma = \Lambda = 0$ across shocks),
\begin{equation} \rho_{up} v_{up} = \rho_{d} v_{d},
\end{equation} 
\begin{equation} \rho_{up} v_{up}^2 + P_{up}= \rho_{d} v_{d}^2 + P_{d},
\end{equation} 
\begin{equation} \frac{1}{2} \rho_{up} v_{up} + \epsilon_{up} + \frac{P_{up}}{\rho_{1}}=\frac{1}{2} \rho_{d} v_{d} + \epsilon_{d} + \frac{P_{d}}{\rho_{d}},
\end{equation}

\noindent where $v$ terms are the flow velocities (e.g., \citet{drainephysics}). The use of up and d as subscripts refer to upstream and downstream gas flow in front and behind the shocks.  

In regions where strong shocks are not produced the heating is minimal.  We only consider the effects of strong shocks, since we are only interested in gas that takes a significant amount of time to cool. For strong shocks the following are good approximations, 

\begin{equation} c = \sqrt{\gamma \frac{k_{B} T}{\mu m_{h}}}
\label{eq:soundc}.
\end{equation}

\begin{equation} a = \frac{v_{up}}{c_{up}} = \sqrt{\frac{\rho_{up} v_{up}^2}{\gamma P_{up}}} \ (Mach \ number)
\end{equation}
\begin{equation} \frac{\rho_{d}}{\rho_{up}} = \frac{v_{up}}{v_{d}}
\end{equation}
\begin{equation} P_{d} = \frac{3 \rho_{up} v_{up}^2}{4} 
\end{equation}
\begin{equation} T_{d} = \frac{3 m v_{up}^2}{16 k_{B}}
\end{equation}
\noindent where $a$ is the Mach number, $c$ is the speed of sound and $\gamma$ is the ratio of specific heats taken as 5/3 for all of the shocks.

\indent The jump conditions are all applied in the moving frame of the shock.  The velocity of the gas behind the shock is constrained to be at rest with respect to the cloud system's centre of mass, see \cref{fig:clouds}. From these conditions we are able to calculate the velocity of each shock with respect to the centre of mass.  We have post-shock temperatures approaching \SI{e7} K in the most extreme cases.  

\subsection{Gas Cooling} \label{gascooling}
The shock heated gas is cooled through adiabatic expansion, hydrogen and helium line cooling, Bremsstrahlung and fine structure cooling.  Where these processes occur is determined by the local density and temperature of the shocked gas.  

We use the results of \citet{wang15} to calculate the rate for optically thin cooling of hydrogen, helium and Bremsstrahlung. \citet{wang15} provide a best fit analytic cooling function that deviates only by a maximum of 5$\%$ from reality. The cooling function is (see \cref{tab:cfconsts}),

\begin{equation} \Lambda (T) = \frac{{x_1} T^{x_2} + ({x_3} T)^{x_4} ({x_5} T^{x_6} + x_7 T^{x_7})}{1 + ({x_3} T)^{x_4}} + {x_9} T^{x_{10}}
\label{eq:coolingfunction}
\end{equation}

This function is valid at temperatures of \SI{2e4} to \SI{e10} K and densities of less than \SI{e18} m\textsuperscript{-3}, which is appropriate for our models.  For each grid element in the code we use a loop that calculates how much energy is lost in a time step in the range \SI{e-5} - 1 year at the current cooling rate $\Lambda$.  How large the time step is depends on the amount of energy that will be lost, this is done to ensure that the peaks the peaks of the cooling function are resolved in the range of \SI{2e4} - \SI{2e5} K and prevent the temperature dropping beyond \SI{2e4} K from this function.  The pressure of the cloud is held constant as it cools by line or Bremsstrahlung, so that as the temperature drops the density is increased proportionally. 

\indent Since the cloud is not shocked all at once and we do not resolve the gas beyond a single cloud, this cooling is only allowed to begin after half the crossing time of the shock has elapsed.  In reality one side of the cloud will be in a different state than the other, but will only be separated by a typical shock propagation time of \SI{5e4} years.

\begin{table}
\caption{Cooling function constants for \cref{eq:coolingfunction}.}
\label{tab:cfconsts}
\begin{tabular}{| c | c | c | c |}
\hline
Constant & Value & Constant & Value \\
\hline
$x_1$ & \SI{4.86567e-13} & $x_2$ & -2.21974 \\
$x_3$ & \SI{1.35332e-5} & $x_4$ & 9.64775 \\
$x_5$ & \SI{1.11401e-9} & $x_6$ & -2.66528 \\
$x_7$ & \SI{6.91908e-21} & $x_8$ & -0.571255 \\
$x_9$ & \SI{2.45596e-27} & $x_{10}$ & 0.49521 \\
\hline
\end{tabular}
\end{table}

\indent Adiabatic expansion begins cooling the cloud only after a full crossing time has passed.  To calculate the energy lost by adiabatic expansion the cloud is allowed to expand at the current speed of sound, \cref{eq:soundc}.  We calculate the work done by the cloud in a free expansion as,

\begin{equation} E_{adb} = 3 n k_{B} T (((R+c\Delta t)^{3} R^{-3})^{1-\gamma} - 1)
\label{eq:workbyadb}
\end{equation}

The average particle mass $m = \mu m_{h}$ is calculated for a completely ionized gas of 25$\%$ Helium and 75$\%$ hydrogen and the ratio of specific heats ($\gamma$) is taken as $\frac{5}{3}$.  The change in radius $\Delta R$ of the cloud is incremented at the beginning of each time step by c$\Delta t$.  We then calculate change in internal energy of the cloud using \cref{eq:workbyadb}.  The cloud number density is recalculated at the end of each step from the new cloud radius.

Adiabatic expansion ceases if the pressure of the cloud reaches a number density times temperature less than $\SI{e8} m^{-3}$ K.  This basement is based on a pressure ($nT$) of $\SI{1e2} m^{-3}$ and \SI{1e6} K, typical of the observed intergalactic medium, \citet{Nicastro2018}. This limit could be crossed by a cold diffuse gas as well as a hot cloud, but expansion would also be relatively unimportant in that former case.

\indent C II emission is estimated by a constant cooling rate of $\SI{e-39} J m^{3} s^{-1}$ and is only allowed for gas under a critical density of $\SI{2e9} m^{-3}$.  The cooling rate is estimated from figure 30.1 and the critical density from table 17.1 of \citet{drainephysics}.  Pressure is also kept constant during this process, meaning as the temperature decreases from C II emission the cloud density proportionally increases.

The thermal energy is calculated from the current gas temperature, and after each time step the energy lost from each type of cooling is subtracted from the current energy and a new temperature is determined.  The total energy lost is computed by,

\begin{equation} \Lambda_{total} = \Lambda(T)_{H,He,Bremmstralung}+\Lambda_{C II}
\end{equation}

\begin{equation} E = n^2 \Lambda_{total} \Delta t -E_{adb}
\label{eq:energylost}
\end{equation}
The effect of UV photo-heating from young stars in the galaxies is approximated with a constant heating function of $\SI{3e-40} J m^{3} s^{-1}$, \citet{drainephysics}.  The heating is important in clouds that reach the critical density for CII cooling suppression and are below the pressure basement set for adiabatic expansion.  This is where the slow rate of UV heating can overcome cooling and maintain the cloud temperature at several thousand Kelvin, where hydrogen line cooling is effective.
This then repeats until a minimum temperature of \SI{150} K is reached. We do not attempt to calculate low temperature, and particularly molecular, cooling in any detail, but see \cref{cloudpops} below.

\begin{figure}
\caption{Temperatures over time for eight different particles from a 7 kpc counter sense collision with varying initial post shock temperatures.}
\label{fig:particletemperatures}
  \centerline{\resizebox{.5\textwidth}{!}{\includegraphics{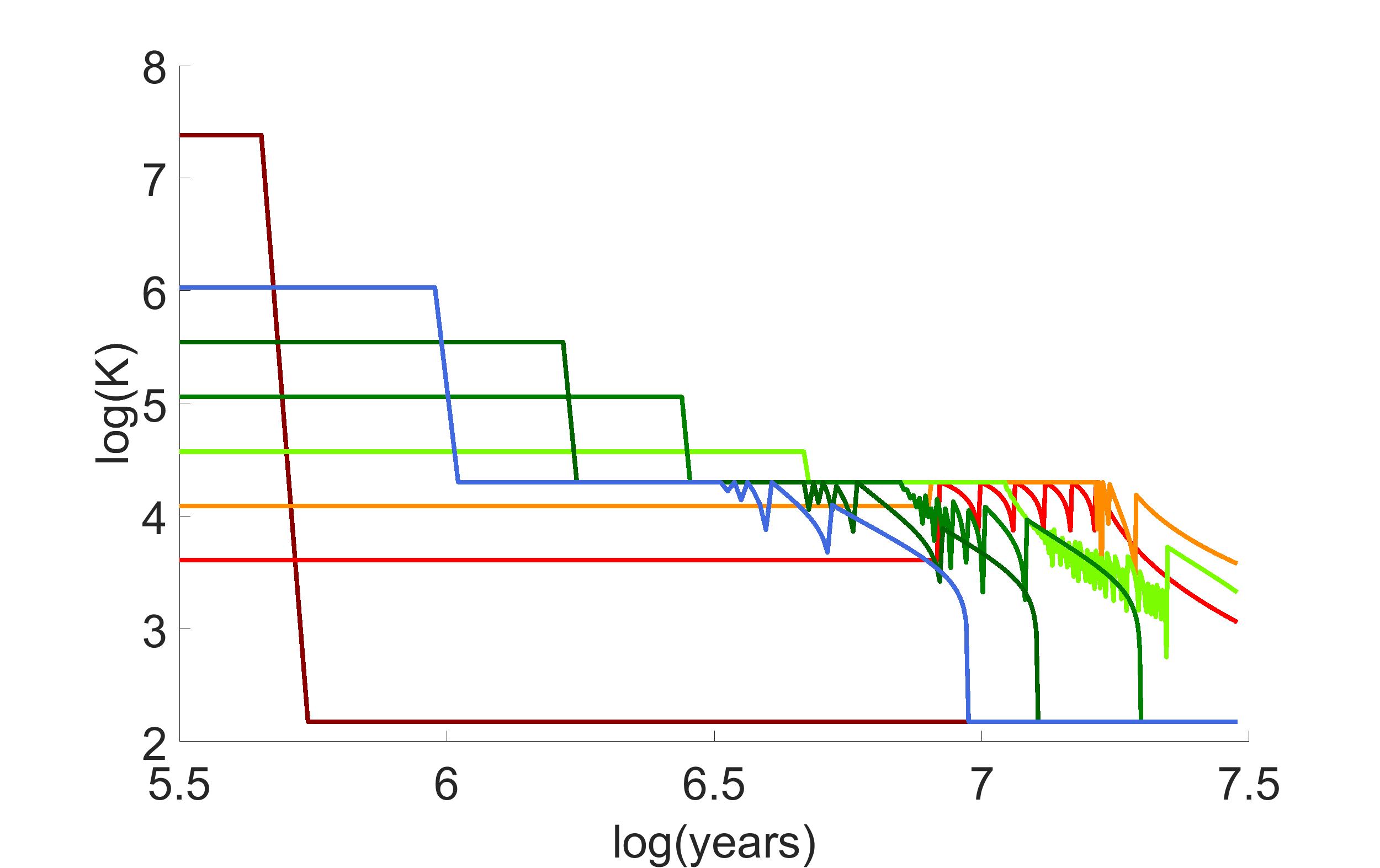}}}
\end{figure}

\section{Results}
\indent  In applying the modeling code to the problem of understanding the complexity of splash bridges described in the Introduction we proceed in stages. The first is the simple case of two discs consisting entirely of diffuse HI, identical surface density profiles, and colliding in first a retrograde and then prograde sense, with only a 500 pc offset between centres at impact. In the next subsection this case is rerun with a greater offset, which begins to show the swirls generated by inelastic collisions of gas elements with different column densities and angular momenta. The next two subsections study similar runs, but with 5-phase interstellar medium, and thus, more realistic thermal effects. In the fifth subsection we look specifically at a model with many of the features of the Taffy system. 

\indent  Before proceeding to specific model results, we consider a few general, qualitative characteristics related to the shock dynamics and cooling. Overall, it is somewhat surprising that high speed disc-disc collisions can produce a full range of phases in the post-collision gas within bridge evolution timescales of order $30$ Myr. In the following we make some estimates to illustrate this. For example, low density (e.g., n < \SI{e4} \ m\textsuperscript{-3}) warm/hot gas will be shock heated (e.g., by a 500\ km\ s\textsuperscript{-1} shock) to temperatures where the cooling time equals or exceeds $30$ Myr. 

\indent  On the other hand, rare collisions between molecular clouds in both discs will generate very high post-shock densities, with rapid cooling. This is a very complicated circumstance with large scale shocks degrading into turbulence in the clumpy clouds, and difficult questions of dust grain survival, and molecule reformation (see Sec. 3.6). Nonetheless, the likely outcome is the prompt reformation of a massive cold cloud \citep{braine03}. 

\indent  The collision of two cool/warm layers of medium density \HI ( $10^7$ < n < $10^8$ m\textsuperscript{-3}) will yield (via recombination) cooling times in the $10^5$ - $3 \times 10^6$ yr range. In the models below this timescale is resolved and cooling fronts are seen to evolve and spread continuously through the discs, often in non-symmetric ways.  Post-shock temperatures are distributed over several orders of magnitude.  These temperature variations highlight the fact that the resulting shocks are dependent on both the relative velocities of each cloud pair (the vector sum of impact and galactic rotation velocities) and on their relative densities.

\indent  When the permitted line cooling cuts off at \SI{2e4} K, fine-structure line cooling can take the gas down to much lower temperatures. However, in strongly shock compressed layers of gas with initial densities of about n > $10^9$\ m\textsuperscript{-3} collisional de-excitation of metastable fine-structure levels may slow this process, again up to 10 Myr timescales.  Gas elements at initial densities of order \SI{e7} m\textsuperscript{-3} reach densities of order \SI{e9} from line cooling isobaric contraction, preventing the dense gas from cooling through CII emission. At temperatures $>2000 K$ molecular cooling (not included in the models) is unlikely.

\indent  Lower density material impacting molecular clouds will drive weak, relatively diffuse shocks into the molecular clouds. The clouds will generally cool rapidly, but with enhanced shock-produced particle fluxes, like cosmic rays (not included in the models).

\indent  Once shocks propagate through a gas layer, they are followed by an expansion wave.  Adiabatic cooling can cool these dense clouds from a few thousand Kelvin to a few hundred relatively quickly. The timescale to push, for example, a $500\ km\ s^{-1}$ shock through a $50\ pc$ gas layer is about $10^5\ yr$. This timescale is low enough for the expansion of initially low density atomic gas to begin before permitted line cooling is complete. Limited expansion (because the compression was not large) and slow continued line cooling would leave this gas in a warm, low-density state, like before the impact. Gas layers that were initially denser may be caught in the fine-structure cooling phase at the onset of expansion. Since the expansion begins at a lower temperature than the previous case, the sound speed will be lower. The expansion timescale would be of order $10\ pc/5\ km\ s^{-1} = 2$ Myr. This gas will end up in a cold, diffuse state, which is primarily atomic.

\indent  The main goals of the present simulations are to `do the bookkeeping' on the thermal outcomes just qualitatively described on a Monte Carlo initial disc ensemble of gas element collisions, while also following the kinematic motions. 

\indent  Note that in the following figures we have binned the post-shock density and temperatures for clarity, though this can give an impression of artificial discreteness.  Using a five component medium produces a maximum of fifteen distinct types of collision, each with their own continuous range of post-shock temperatures and densities.  As a reminder, each particle in these figures has two discrete phases of ISM, separated by their contact discontinuity.  G1 and G2 will be used to refer to gas specifically from each galaxy. Since gas from each galaxy is mixed together, gas of many different phases overlaps in the same regions we find it best to plot separately low and high density and temperature gas.  Density is separated at $\SI{1E7} m^{-3}$ and temperature at $\SI{1E6} K$.  References to high density will be to all gas elements from either G1 or G2 that are above the $\SI{e7} m^{-3}$.  Low density is everything below this limit.  Temperatures are divided the same way with a the dividing value is of $\SI{e6} K$.

\FloatBarrier

\subsection{Single Component ISM Models of HI discs with 500 pc offset}
\label{single500pc}

\begin{figure}
  \caption{Face-on views immediately after the impact, showing post shock \textbf{a)} densities and \textbf{b)} temperature.%
     }%
  \label{fig:0x500face0myrfigures}
     \begin{center}
        \subfigure[]{%
           \label{fig:0x500face0myrlthddensity}
           \includegraphics[width=0.4\textwidth]{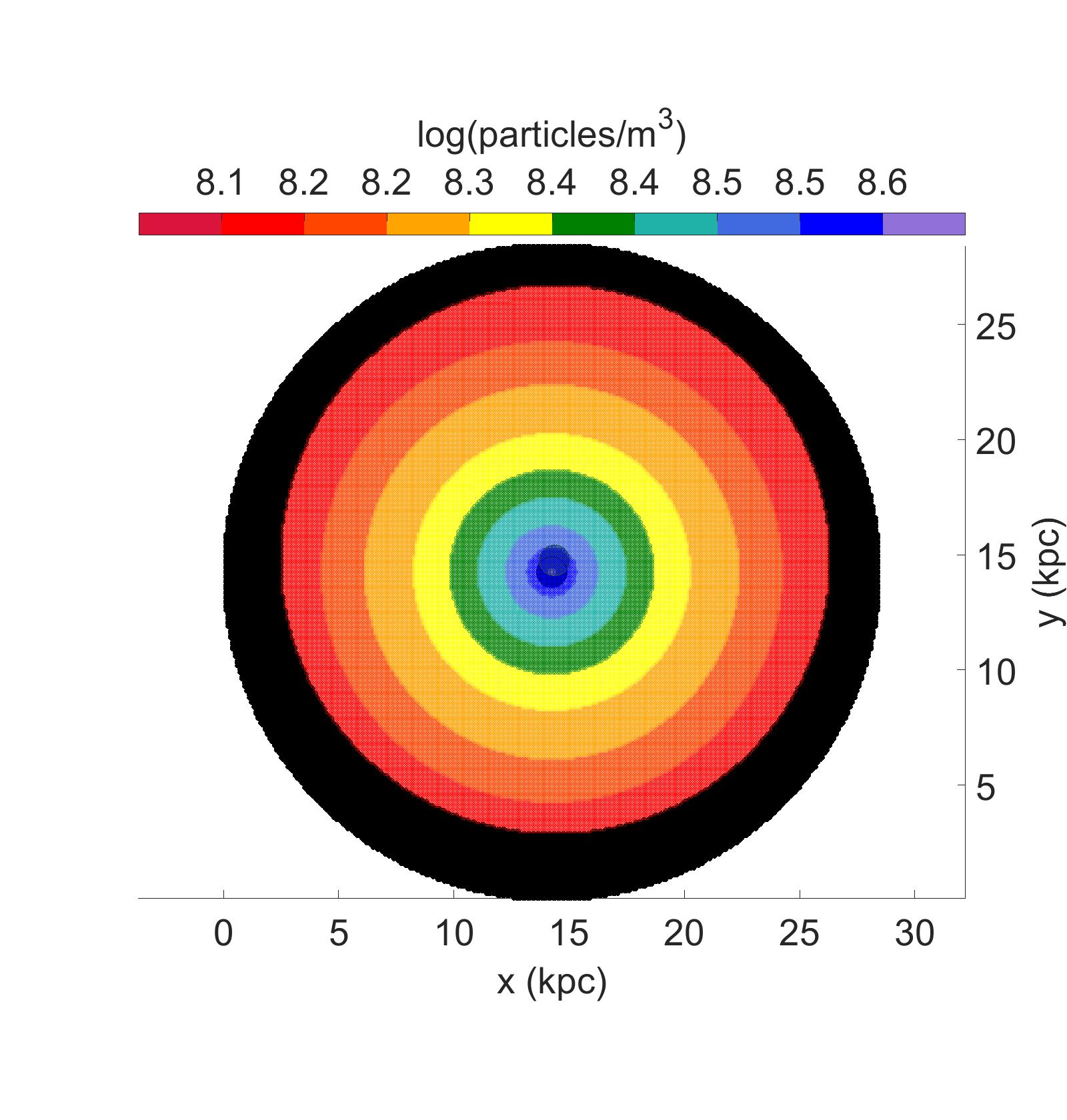}
        }\\ 
        
        \subfigure[]{%
           \label{fig:0x500face0myrlthdtemperature}
           \includegraphics[width=0.4\textwidth]{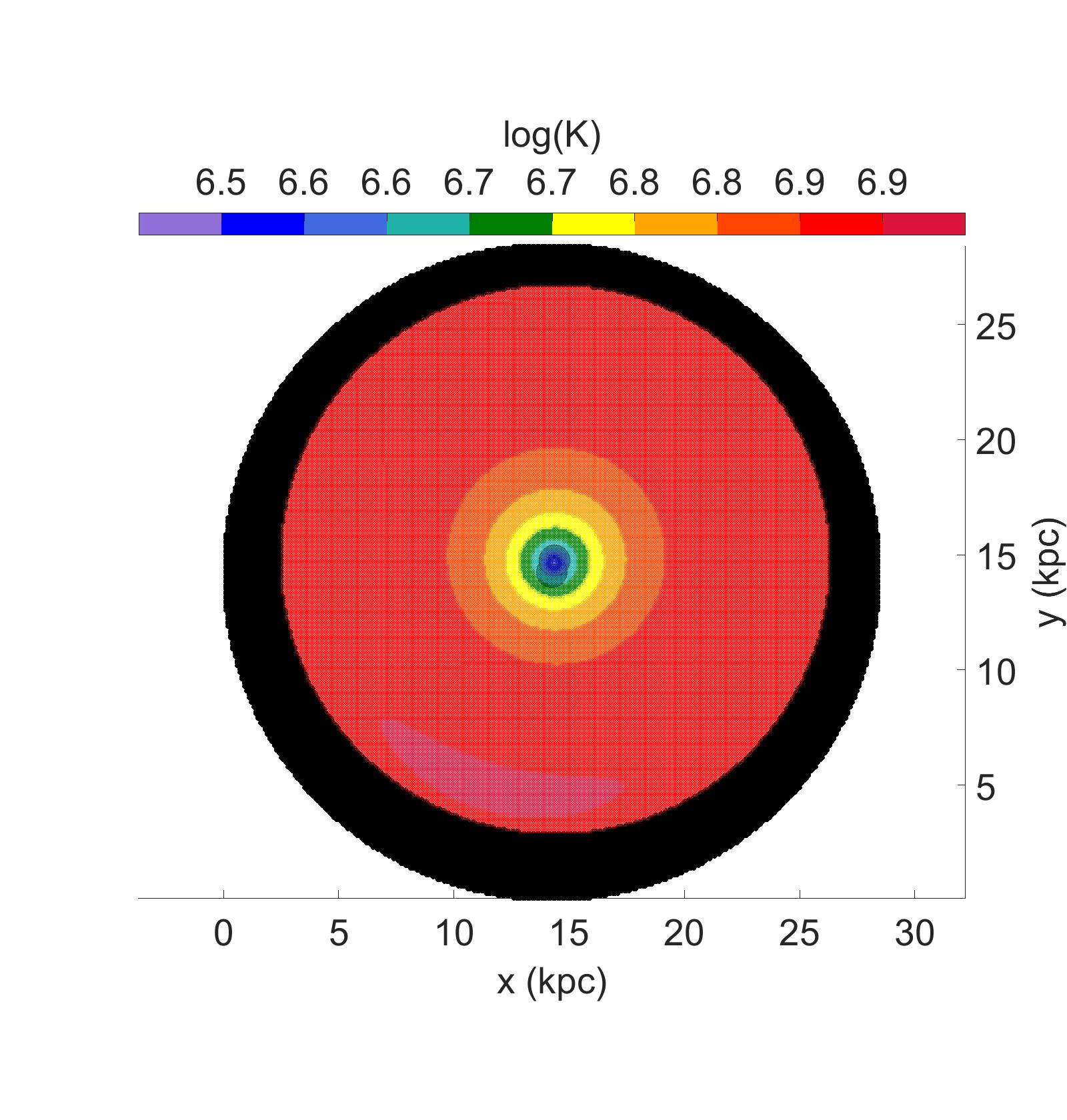}
        }\\ 
    \end{center}
 \end{figure}

\begin{figure}
  \caption{\textbf{a)} Edge-on view of CBD 15 Myr after collision.  \textbf{b)} Edge-on view 30 Myr after collision.%
     }%
  \label{fig:0x500face30myrfigures} 
     \begin{center}
        \subfigure[]{%
           \label{fig:0x500edge5myrlthd}
           \includegraphics[width=0.4\textwidth]{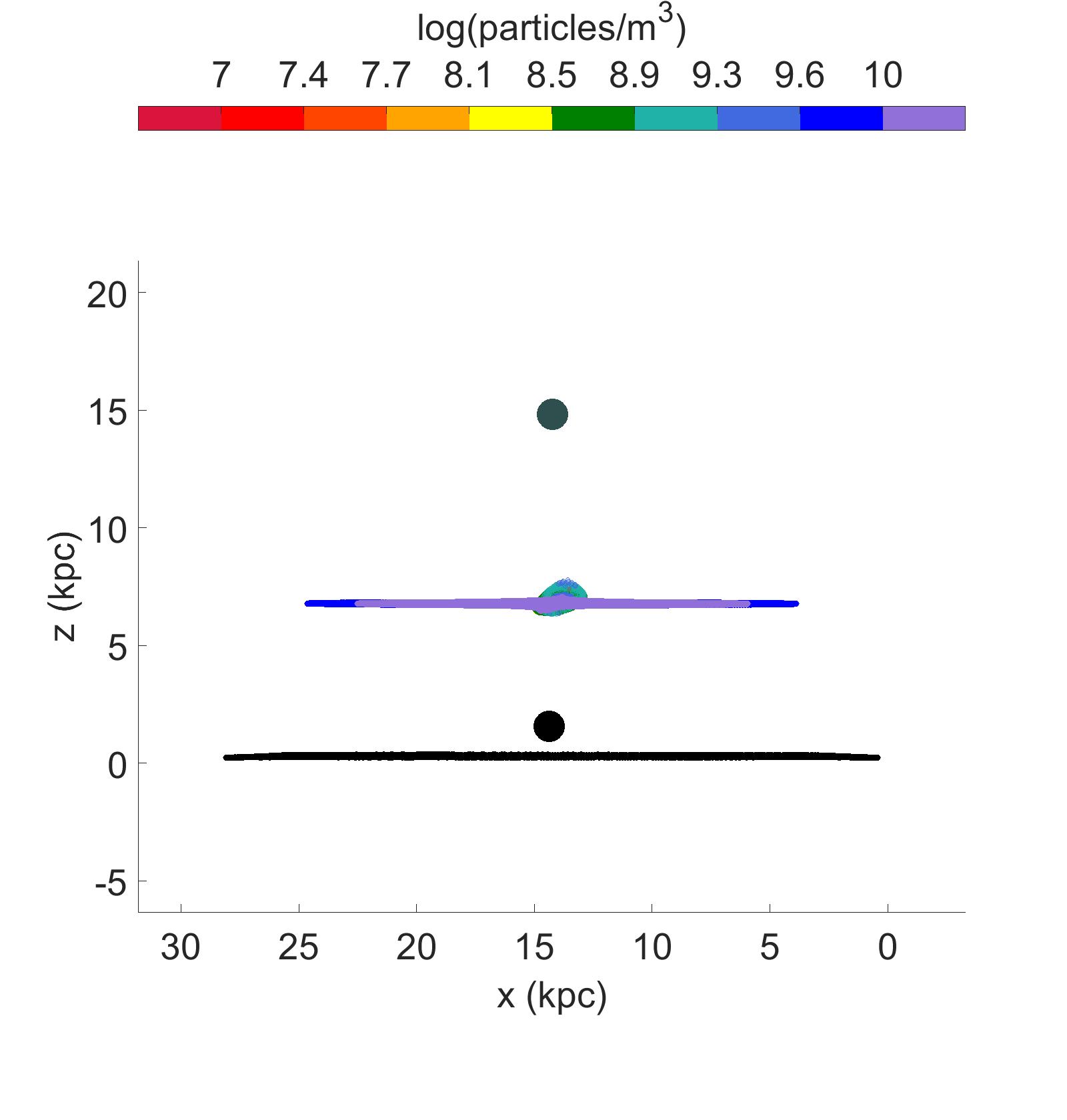}
        }\\ 
        
        \subfigure[]{%
           \label{fig:0x500edge30myrlthd}
           \includegraphics[width=0.4\textwidth]{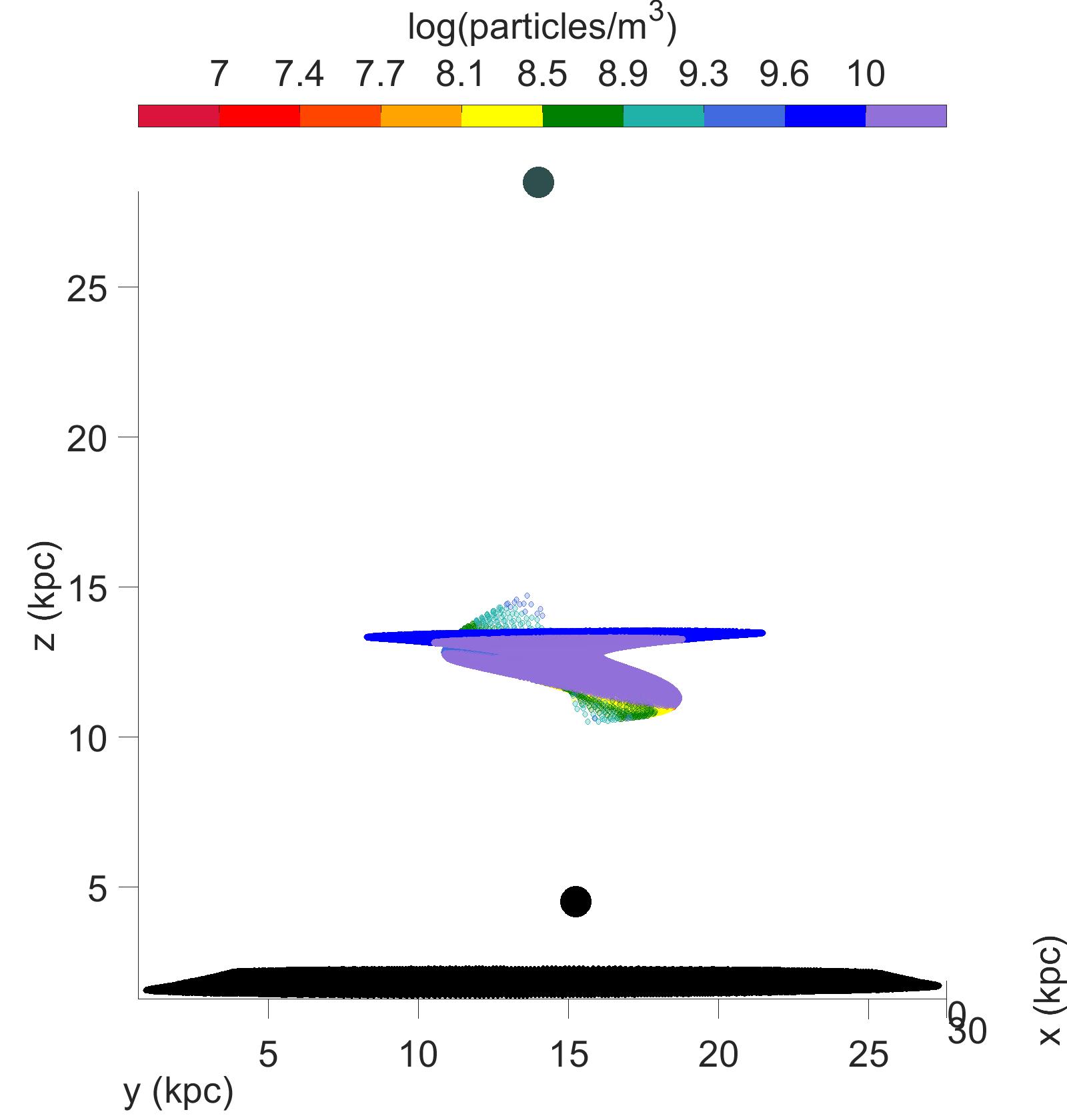}
        }%
    \end{center}
   
   \end{figure}

\begin{figure}
\caption{Same sense rotating discs \textbf{a)} edge-on view of CBD 15 Myr after collision.  \textbf{b)} Edge-on view 30 Myr after collision. 
\label{fig:eqsense0x500edge30myrlthd}}
\begin{center}
\subfigure[]{%
\includegraphics[width=0.4\textwidth]{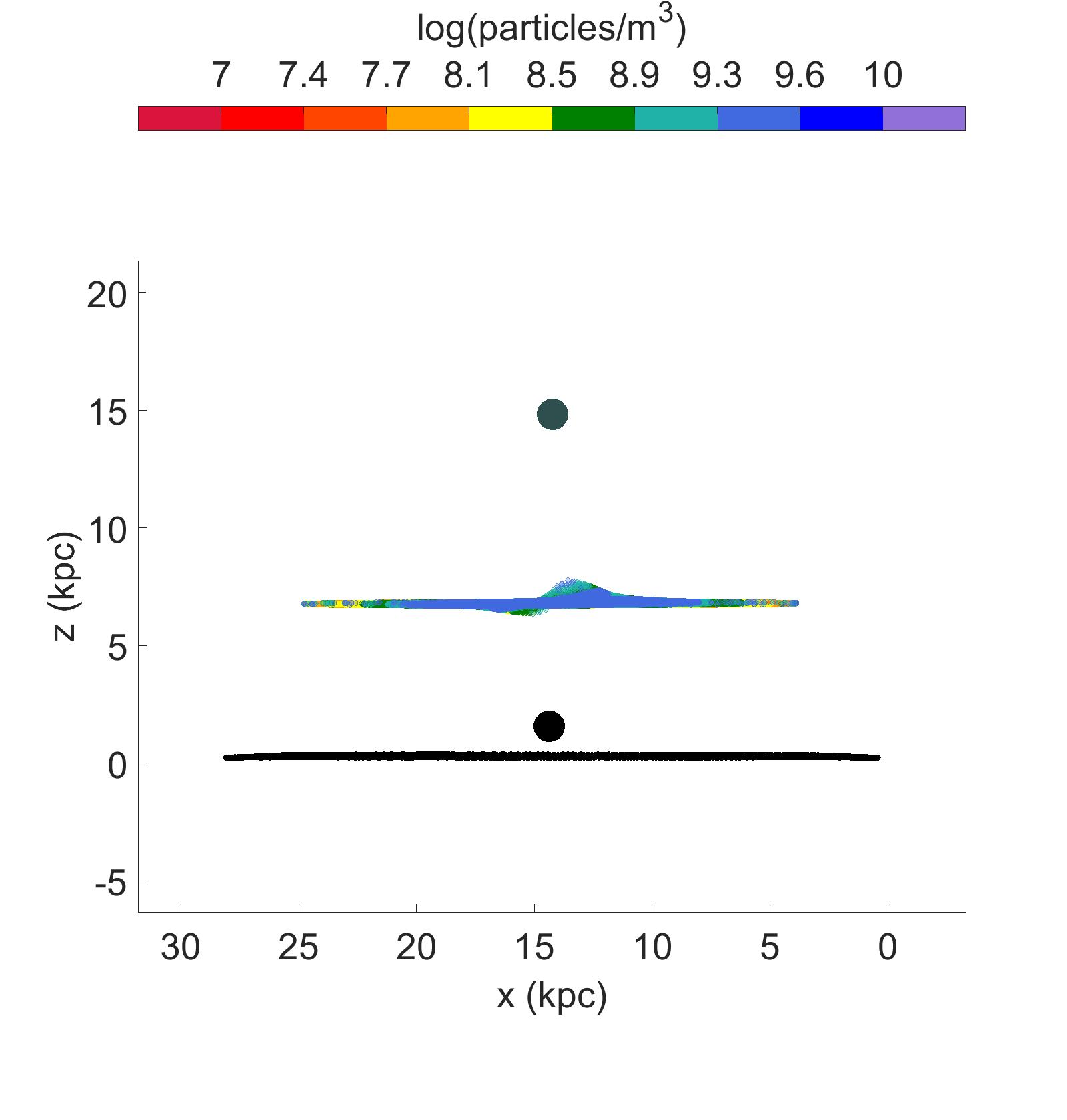}
}\\
\subfigure[]{%
\includegraphics[width=0.4\textwidth]{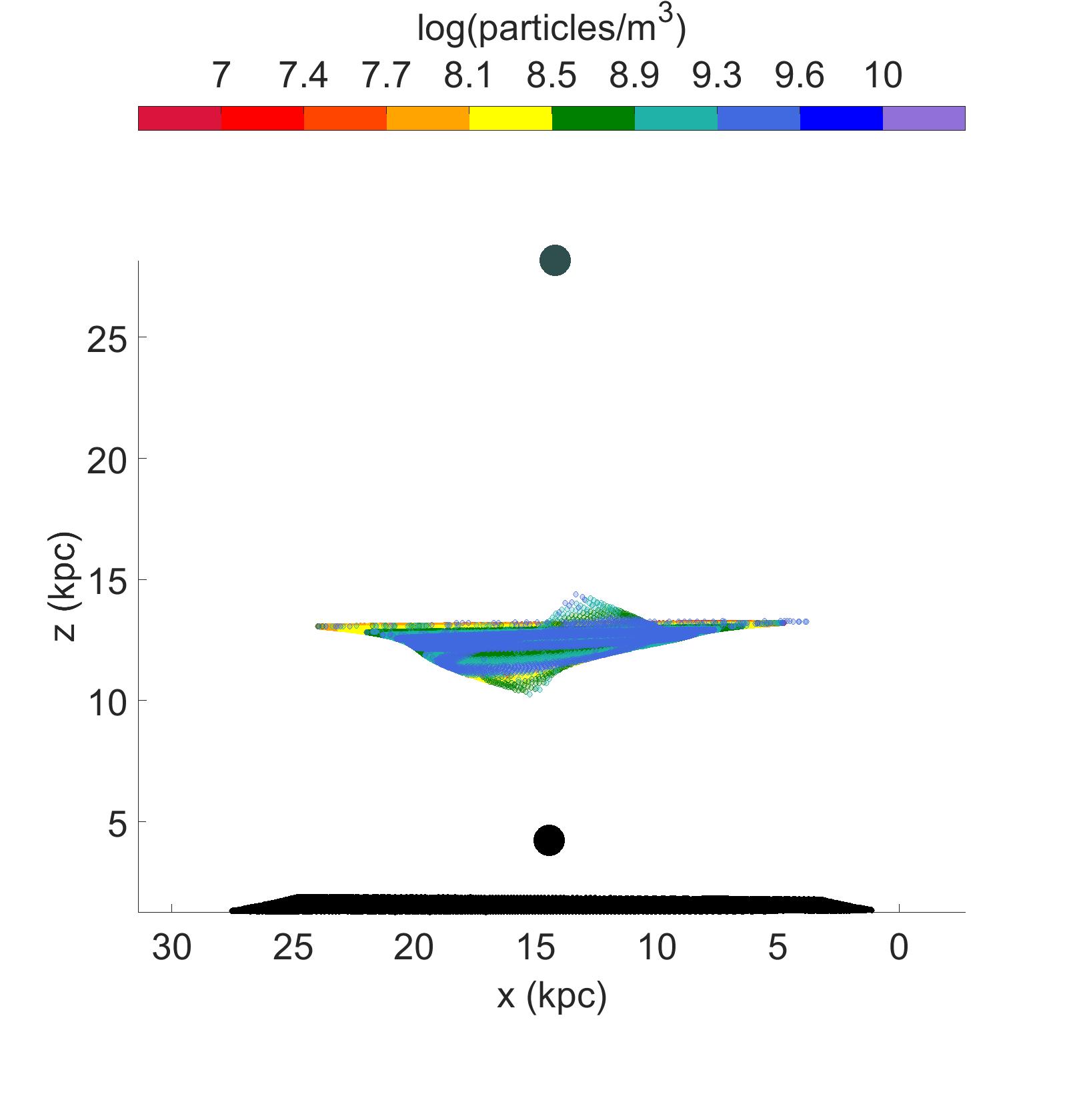}
}\\

\end{center}
\end{figure}

\indent The first case is a model of a collision of pure \HI discs with $\frac{1}{R}$ profiles.  The \HI gas mass removed from both galaxies is approximately \SI{2.45e10}{M\textsubscript{\(\odot\)}} which accounts for 86\% of the total gas mass.  \Cref{fig:0x500face0myrfigures,fig:0x500face30myrfigures,fig:eqsense0x500edge30myrlthd} show a collision with a small 500 pc impact offset toward the left of the page.  Particles that are not involved in the direct collision from either galaxy are represented as black in the figures.  The outer ring of particles belong to G1 and were not involved in the direct collision.  The warping seen in this ring is due only to gravitational effects.  \textit{The gaseous disc of G2 is entirely stripped away from its stellar disc potential and remains in what we refer to as a Central Bridge Disk (CBD).}  In this nearly direct centre-on-centre single component ISM collision we do not see the development of a Taffy-like bridge structure.

\indent \Cref{fig:0x500face0myrfigures} a) shows the post-shock density immediately after the collision and \cref{fig:0x500face0myrfigures} b) is the corresponding post-shock temperatures.  The shocks velocities in the clouds range from a few \kms to nearly 1100 \kms for the 500 pc offset.  The lowest velocity shocks are in dense molecular clouds that are impacted by low density \HI gas, where as the highest velocity shocks are in rare cases where dense molecular clouds collide with other molecular clouds.  The majority of shocks are near 600 \kms which is caused by \HI on \HI collisions.  Gas is heated to temperatures up to \SI{e7}K, which with the given post-shock densities leads to cooling times of 1-3 Myr.

\indent The gas involved in the direct collision is left  between the galaxies.  The larger black dot markers seen in every figure represents the current location of the COM of each galaxy, which is where the fixed gravitational potentials are acting from.  The central bridge disc (CBD), depicted in \cref{fig:0x500face30myrfigures,fig:eqsense0x500edge30myrlthd} is unique to low offset collisions.  In runs with a greater offset a bridge of material is left behind stretching between the two galaxies the entire time as they drift apart.  Low offsets however create a CBD that remains near the systems centre of mass which remains as a disc for several million years.  After about 5 Myr distortion to the flat disc becomes noticeable as material is drawn back toward each galaxy.  For counter rotating collisions the CBD begins a radial contraction immediately.  This occurs because of a loss of rotational momentum due to the merger of counter rotating disc elements.  The rotation velocity difference affects how much the disc will contract after the collision.  In the runs the rotation velocities differ by approximately 50 \kms , resulting in some left over angular momentum in the central bridge gas disc.  \Cref{fig:eqsense0x500edge30myrlthd} is the same system but with galaxies rotating in the same sense.  The \HI disc retains much of its original diameter since the galaxies only transfer a small fraction of their angular momentum.  The same sense rotation case is eventually distorted but its outer disc remains extended and extremely flat.

\indent The densest part of each galaxy is the centre.  This means the central region of each galaxy is able to push through the slightly less dense outer disc of the other, leading to an unequal distribution of post collision momentum in the gas.  \Cref{fig:0x500face30myrfigures} shows the position of gas particles at 15 Myr and later at 30 Myr.  This highlights the slight distortion of the flat CBD which at higher offsets will turn into a taffy-like bridge.  \Cref{fig:eqsense0x500edge30myrlthd} is a snapshot at 30 Myr where the colliding discs were rotating in the same sense.  Compare the diameter of the CBD in \cref{fig:eqsense0x500edge30myrlthd} to the diameter in \cref{fig:0x500face30myrfigures} to see the effect of the different angular momenta.

\subsection{HI Bridges with Various Offsets}
\label{hibridges}
\begin{figure*}
\caption{Counter sense rotating discs at 30 Myr with offsets: 5 kpc, 10 kpc and 15 kpc for each row respectively.  The two figures are not orthogonal views.  The highest of the two densities in each cloud pair is shown.  Low density is not well represented.}%
\label{fig:HIonlyBridgesSubFigures}     
     \begin{center}

        \subfigure[]{%
          \label{fig:onlyHI5kpc30myrlthd}
          \includegraphics[width=0.4\textwidth]{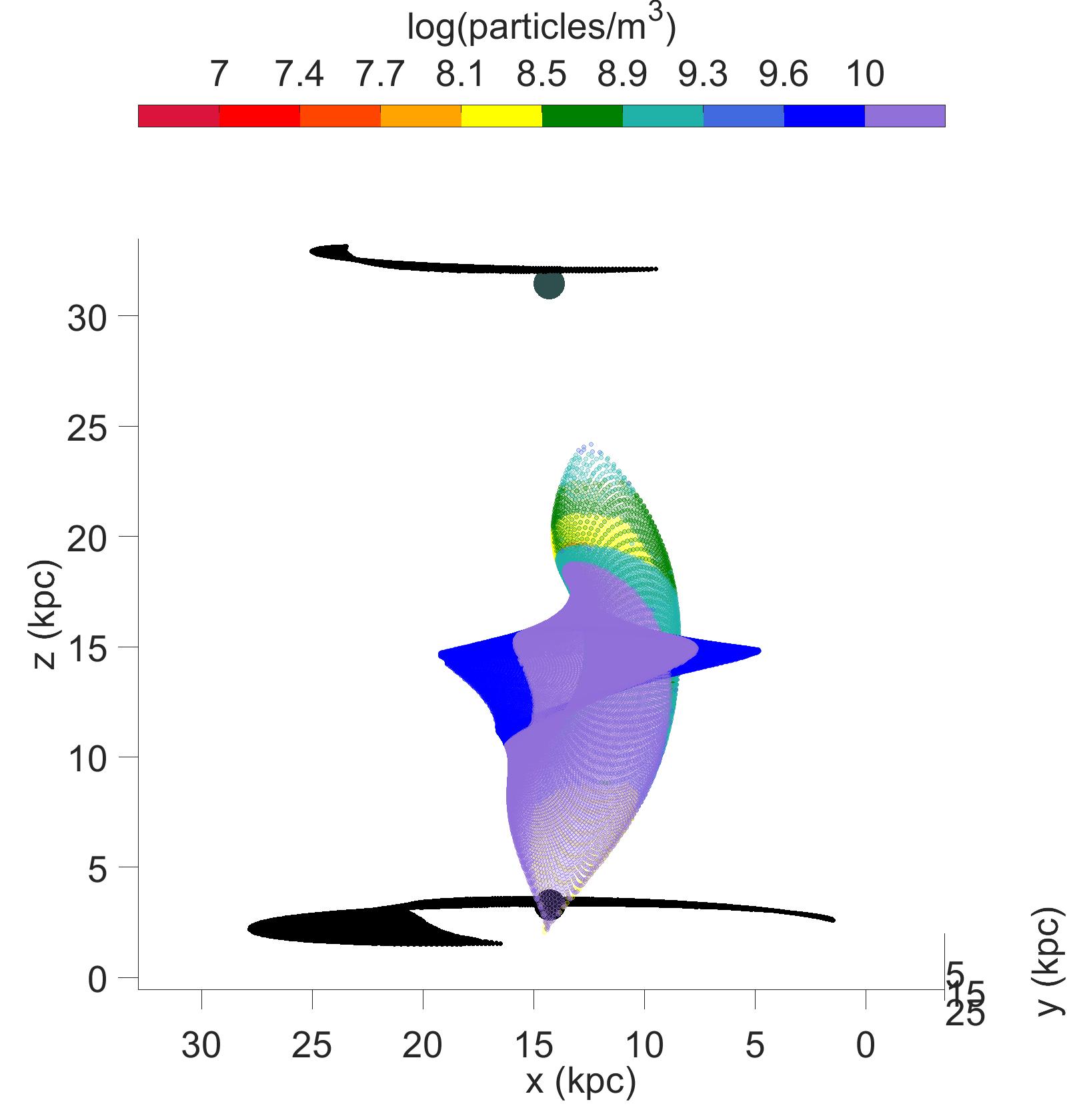}
        }
                \subfigure[]{%
          \label{fig:onlyHI5kpc30myrlthd2}
          \includegraphics[width=0.4\textwidth]{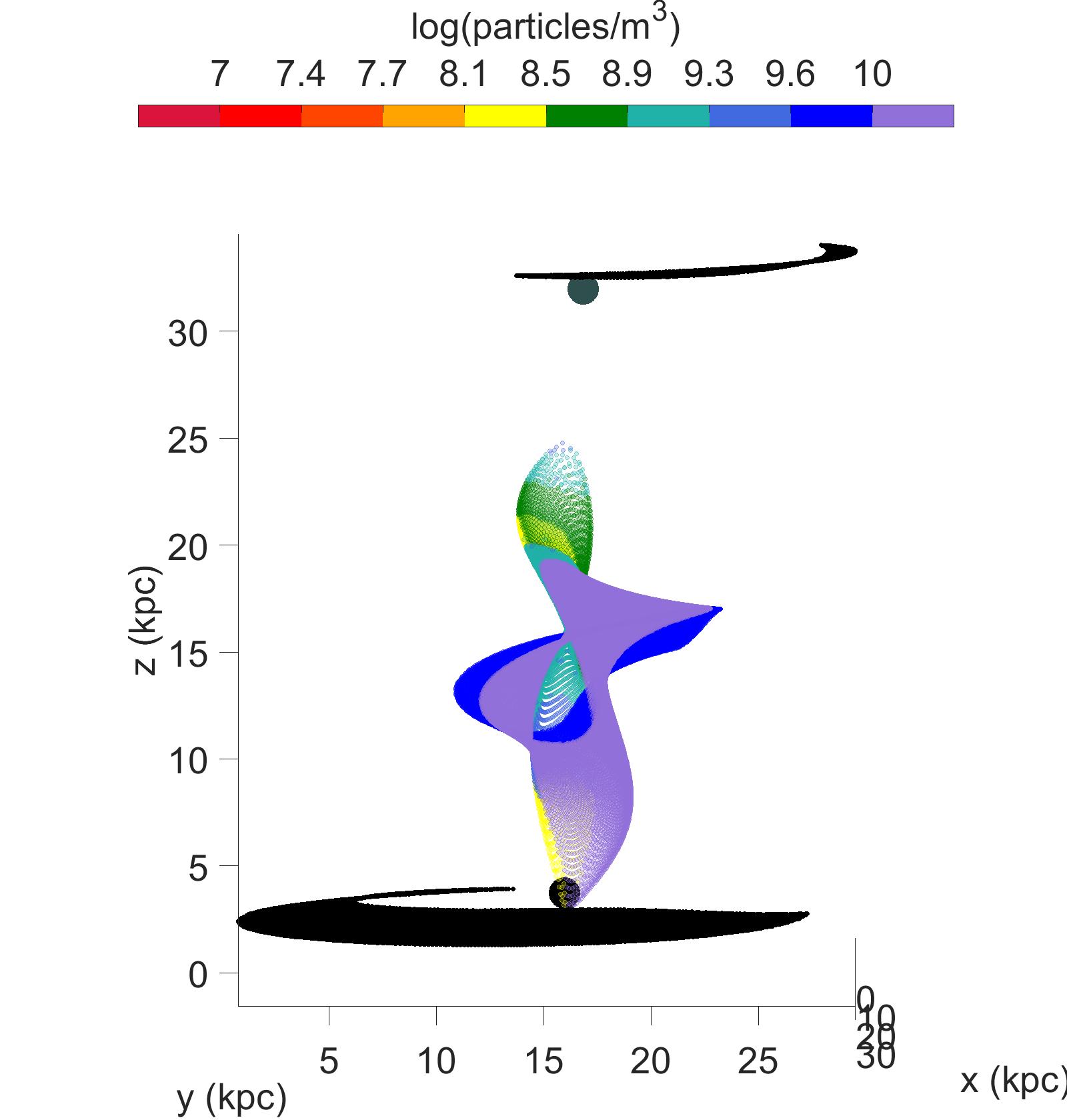}
        }\\ 
                \subfigure[]{%
          \label{fig:onlyHI10kpc30myrlthd}
          \includegraphics[width=0.4\textwidth]{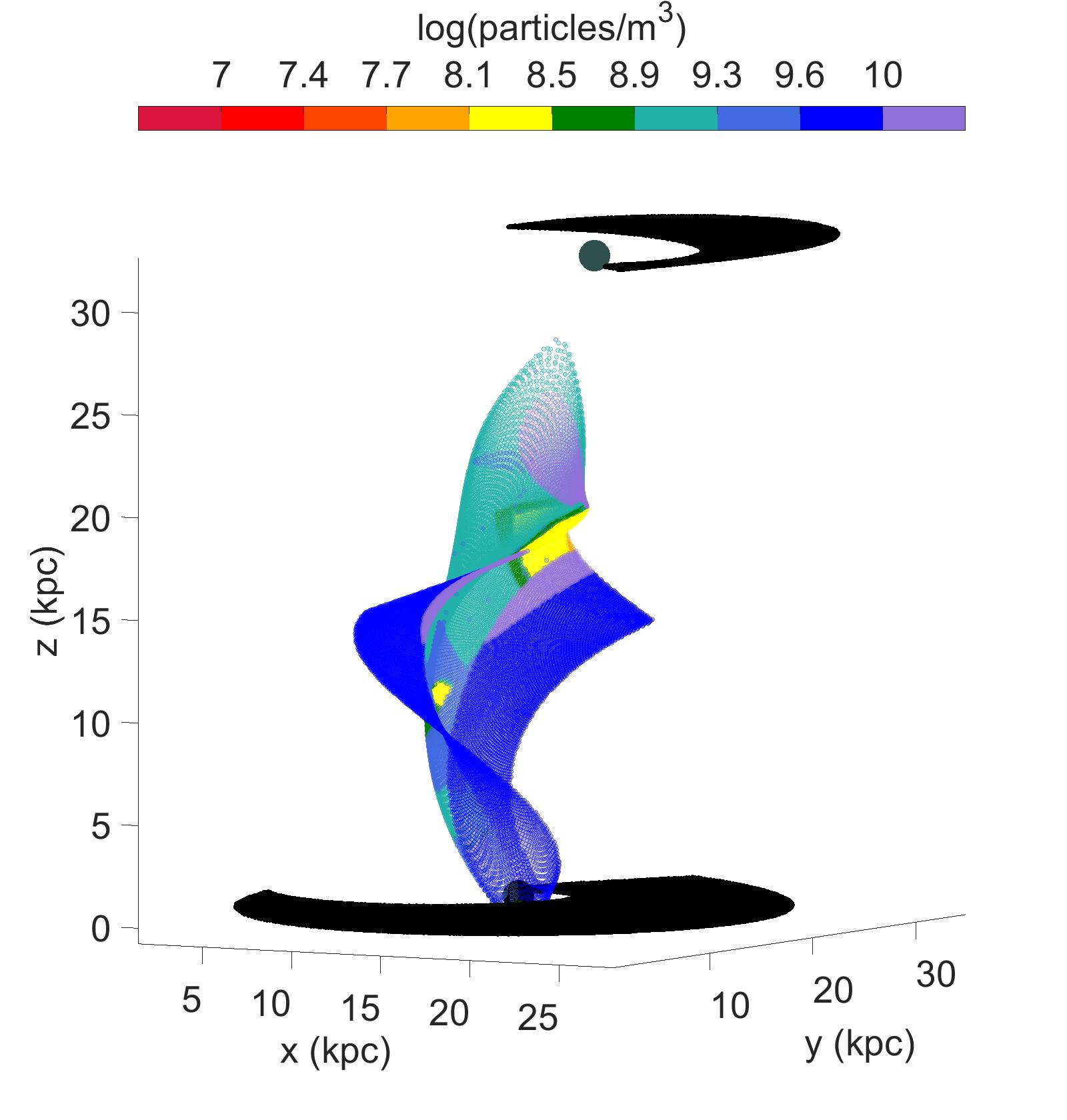}
        }
                \subfigure[]{%
          \label{fig:onlyHI10kpc30myrlthd2}
          \includegraphics[width=0.4\textwidth]{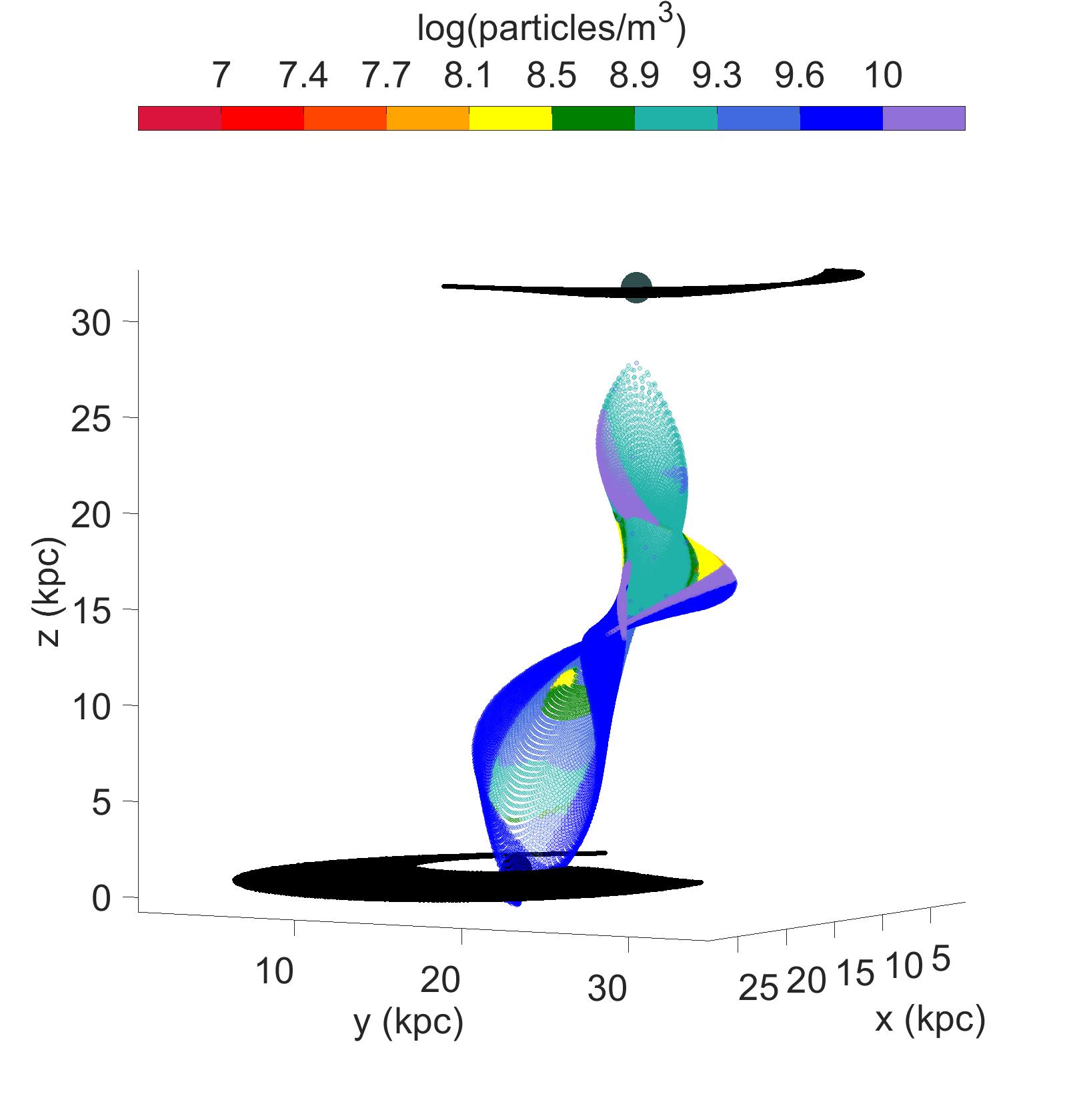}
        }\\ 
                \subfigure[]{%
          \label{fig:onlyHI15kpc30myrlthd}
          \includegraphics[width=0.4\textwidth]{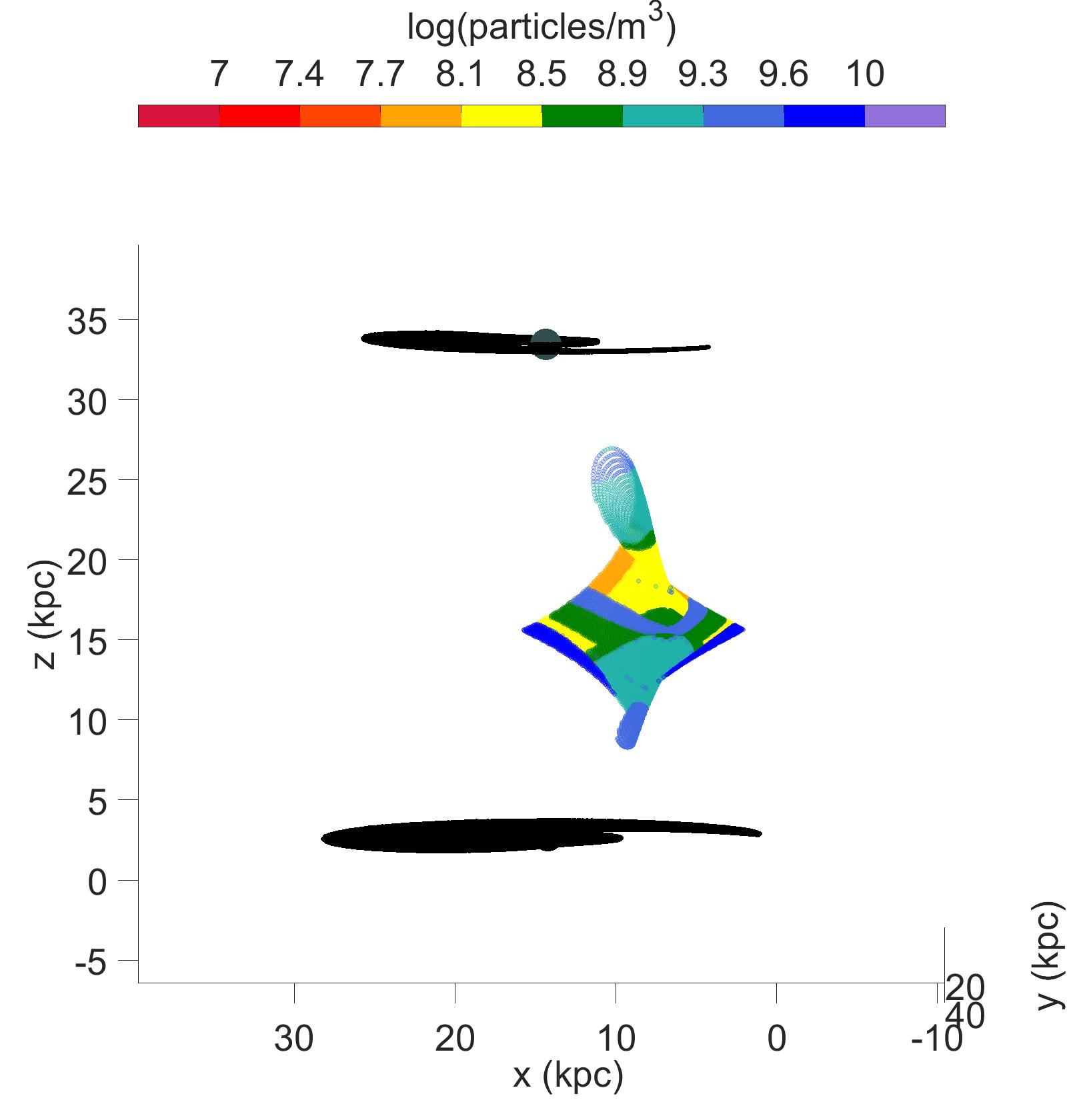}
        }
        \subfigure[]{%
            \label{fig:onlyHI15kpc30myrlthd2}
            \includegraphics[width=0.4\textwidth]{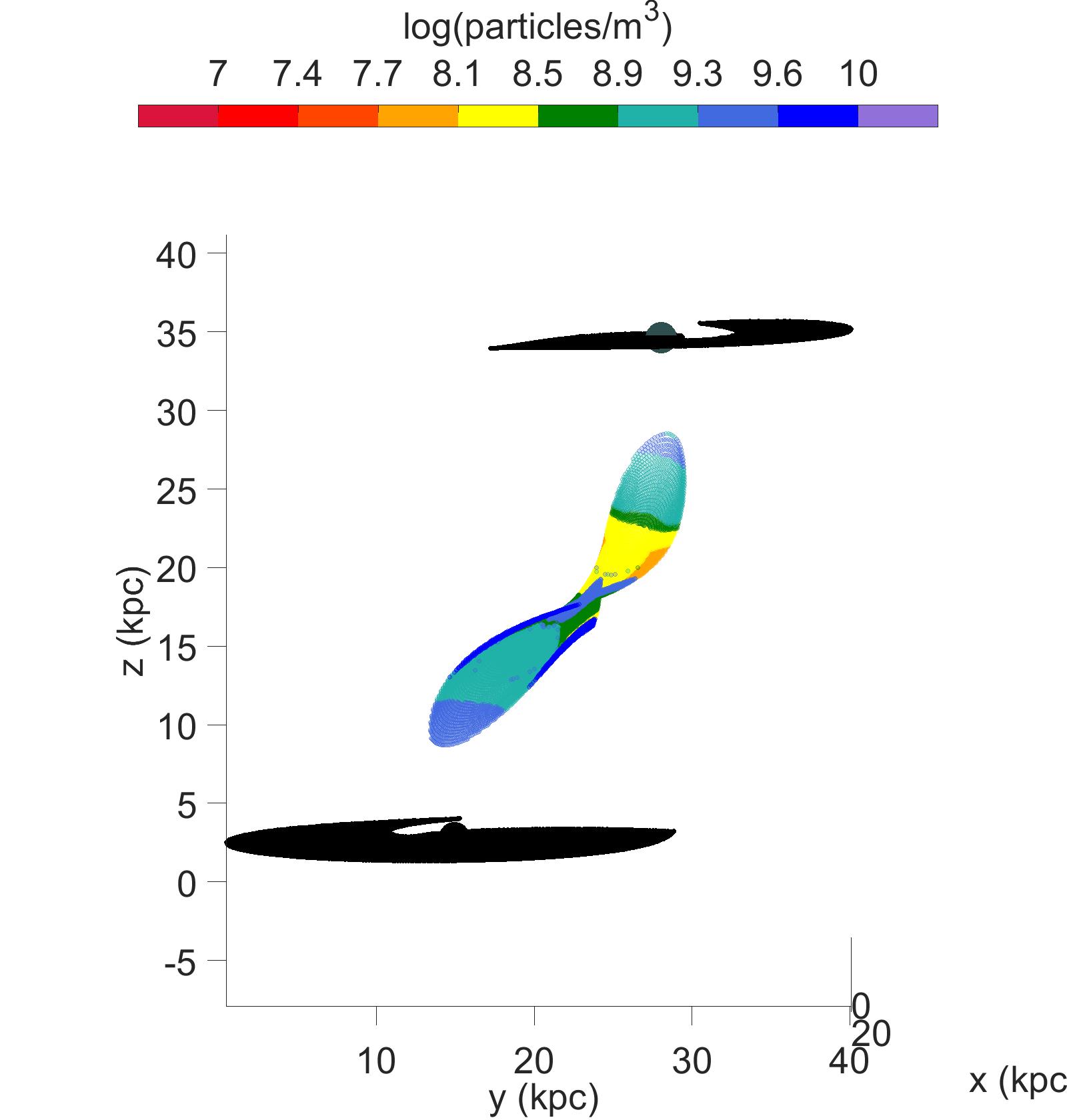}
        }%
    \end{center}

\end{figure*}

 \Cref{fig:HIonlyBridgesSubFigures} shows how the single component \HI gas slowly evolves from a flat disc to a more and more twisted helicoid as the offset between the (counter-sense) gas discs increases.  As the offset grows from 5 kpc to 15 kpc, density and therefore momentum differences between clouds grow, allowing for the colliding gas elements to stretch out in a taffy-like way.  The \HI is relatively flat and smooth and appears to twist as it evolves.  These single component \HI structures are also created in all of the five component runs.  In each \HI bridge the cloud densities saturate at \SI{2e9} \dunit.  This is the cut off for CII fine structure cooling, which prevents clouds from contracting further by cooling.  A cloud seen above this density will have reached that density from contraction due to permitted line cooling only.  Clouds that reach this point may only rely on adiabatic expansion to cool until they return below the CII critical density.  Depending on the clouds adiabatic expansion rate, it will likely oscillate around a density \SI{2e9} \dunit as CII cooling is reappears, and leads to repeated contraction.  Only when the rate of cooling by adiabatic expansion with UV heating exceeds CII cooling will the density of the cloud drop below \SI{2e9} \dunit. The main \HI component began with initial preshock densities of \SI{8e6}-\SI{7e7} \dunit, which were shock heated to temperatures from \SI{2e6} to \SI{7e6}K, illustrated in \cref{fig:0x500face0myrfigures}.  This component of neutral \HI ISM happens to be at a high enough initial density that permitted line cooling can still cool and contract with in a collision time scale.  Gas at initial densities of order \SI{1e6} \dunit no longer cools through line emission rapidly enough before adiabatic expansion begins and significantly slows or prevents the clouds from contracting.  The inability of a gas cloud to isobarically contract drastically increases its total cooling time.
 
\begin{figure}
\caption{Same sense rotating discs with 5 kpc offset at impact showing only the neutral \HI - age: 30 Myr.}
\label{fig:eqseonlyHI5kpc30myrlthd}
\includegraphics[width=0.4\textwidth]{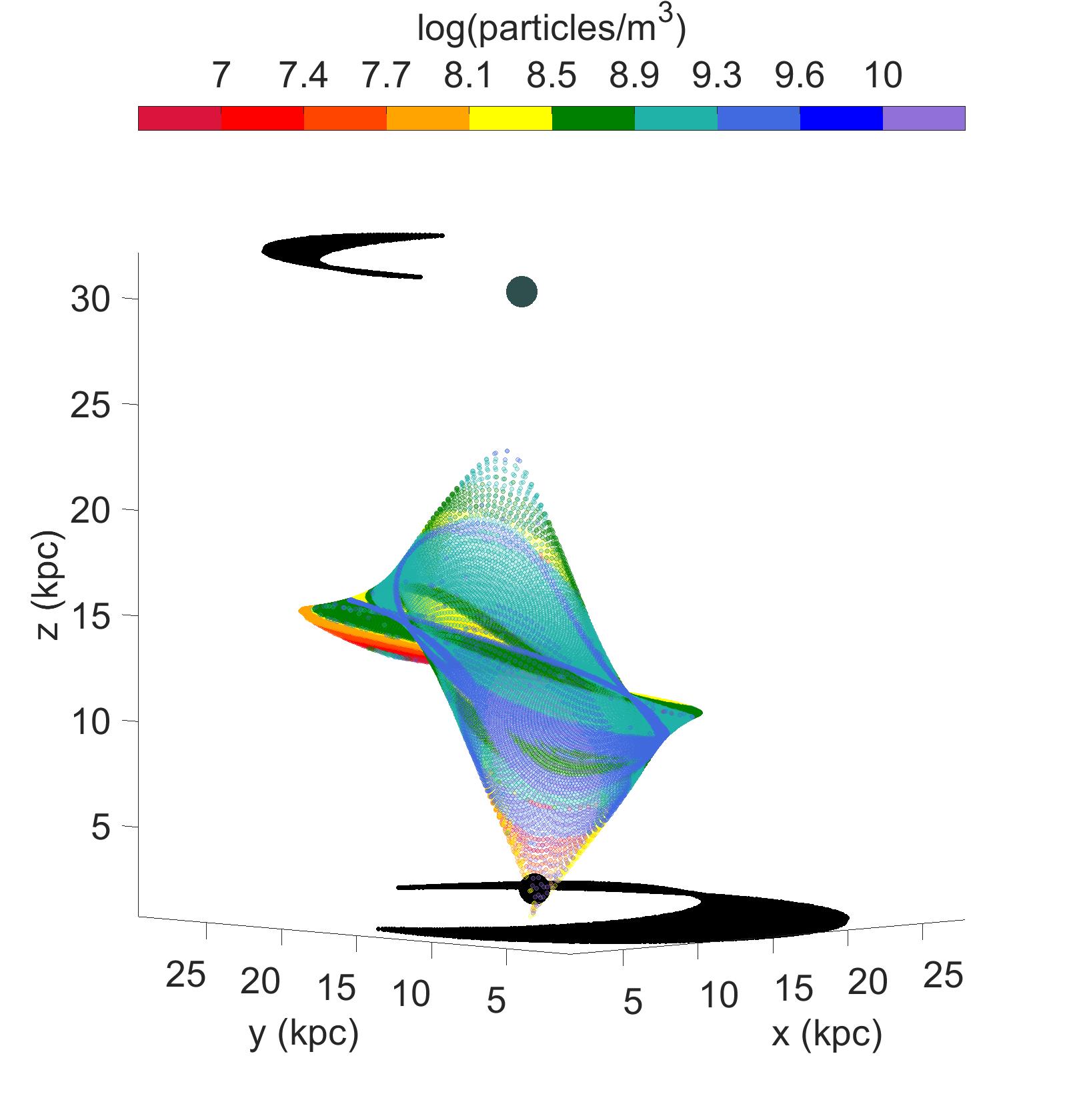}
\end{figure}

 \indent \Cref{fig:eqseonlyHI5kpc30myrlthd} is an example of the single component \HI structure produced from the dense centres of each disc pushing through the lower density outer disc of the other.  This seems to be about the limit at which the CBD is nearly lost and a transition to a twisted bridge structure begins, as seen in the following figures.

\subsection{Five component ISM models with 500 pc offset}
\indent In this section we describe a run with all five discrete types of ISM included.  The discs rotate counter to one another, unless stated otherwise in the figures.  There is an offset of 500 pc in the negative x-direction.  As mentioned in section \cref{themodel}, the initial density of each component of ISM is scaled above or below that of the local \HI $\sfrac{1}{R}$ disc. This means we expect to see a greater spread of momentum between colliding gas elements along with a much more complicated distributions of collided ISM imposed within the \HI structures shown in section \cref{single500pc,hibridges}.

\indent For given offset and rotation directions a change in the initial conditions of the \HI discs in the single component model does not change the morphology of the bridge significantly.  In general single component models of the same offset will end in a taffy-like twisted bridge.  This is not true of all five component runs of the same offset.  Varying the density and velocities can result in two completely different bridges, with only their \HI bridge component remaining similar.  On top of this, with multiple phases of the ISM the orientation of the discs begins to play a much more important role in how the resulting bridge will appear.  For example, how the spiral arms are aligned between the two galaxies can have a strong effect on the abundance of each phase of ISM observed in the bridge.

\begin{figure}

\caption{%
        Face-on views showing post shock temperatures.  \textbf{a)} 0 Myr \textbf{b)} 15 Myr and \textbf{c)} 30 Myr showing CII emission.  See  \cref{gascooling} for more details.
     }%
     \label{fig:0x500subfigurestemphtld}
     \begin{center}
        \subfigure[]{%
           
           \includegraphics[width=0.4\textwidth]{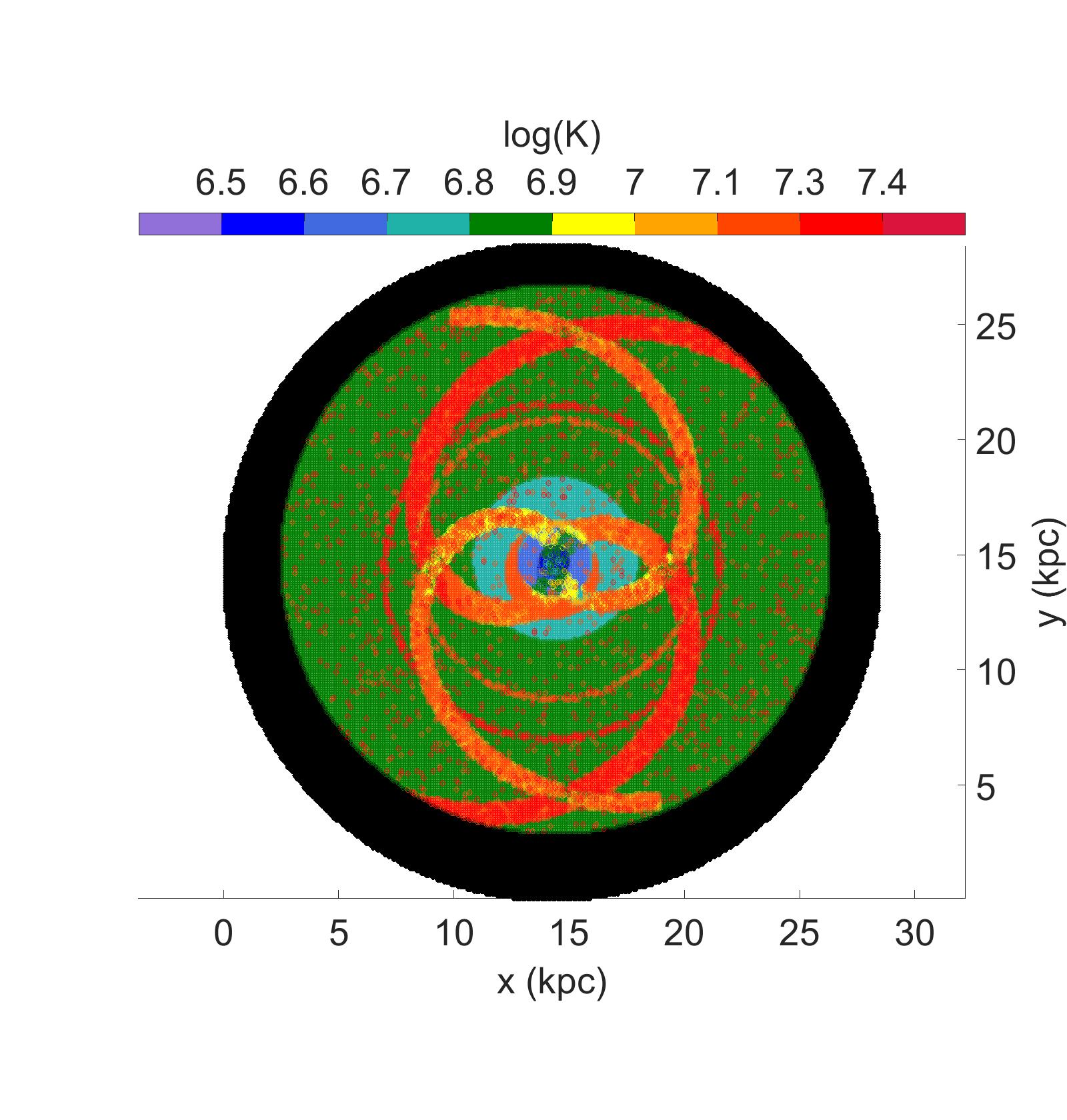}
           \label{fig:0x500temps0myrfacetemphtld}
        }\\ 
        \subfigure[]{%
           \label{fig:0x500face30myrfacetemphtld}
           \includegraphics[width=0.4\textwidth]{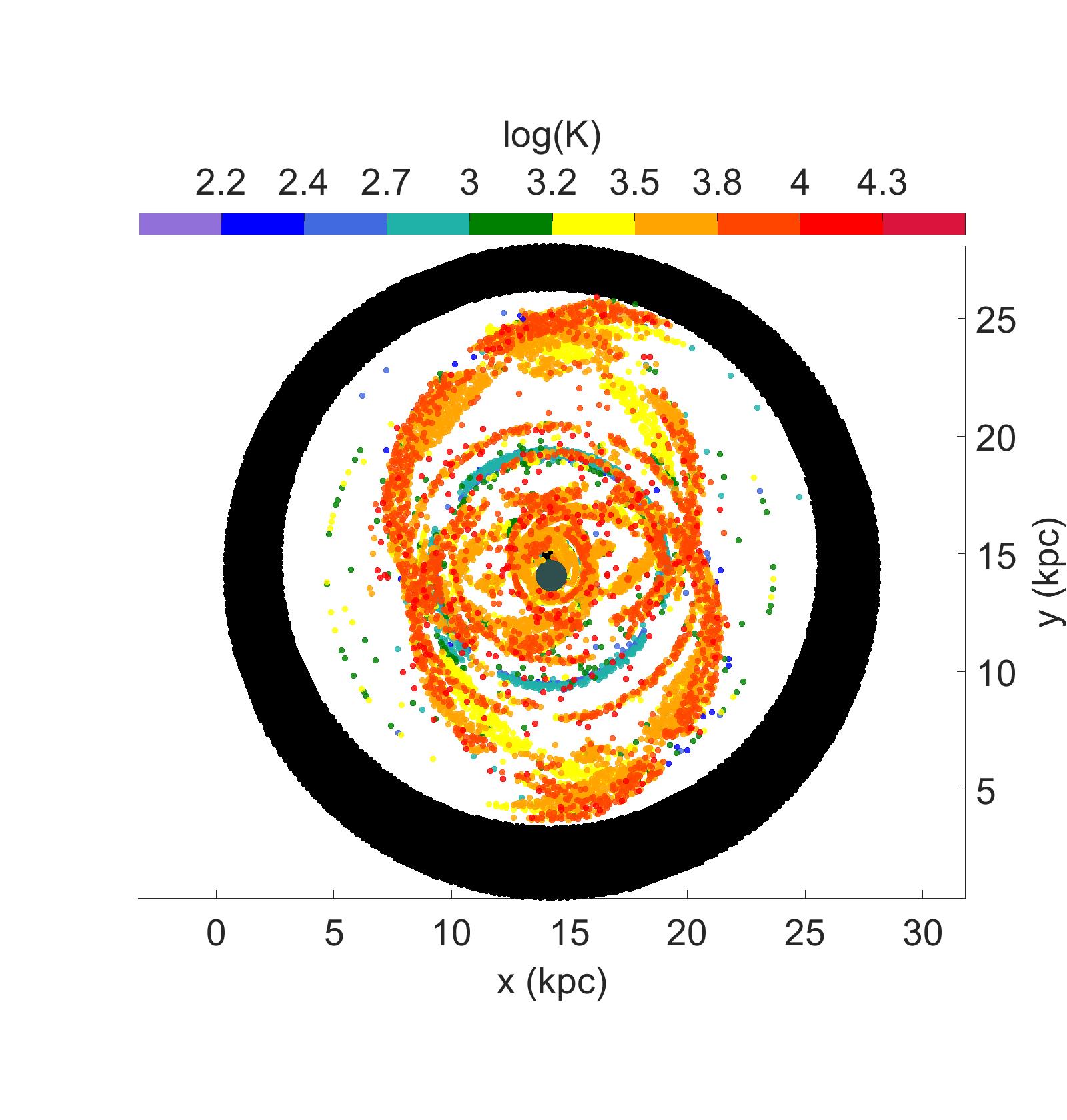}
        }\\ 
        \subfigure[]{%
           \label{fig:0x500edge30myredgetemphtld2}
           \includegraphics[width=0.4\textwidth]{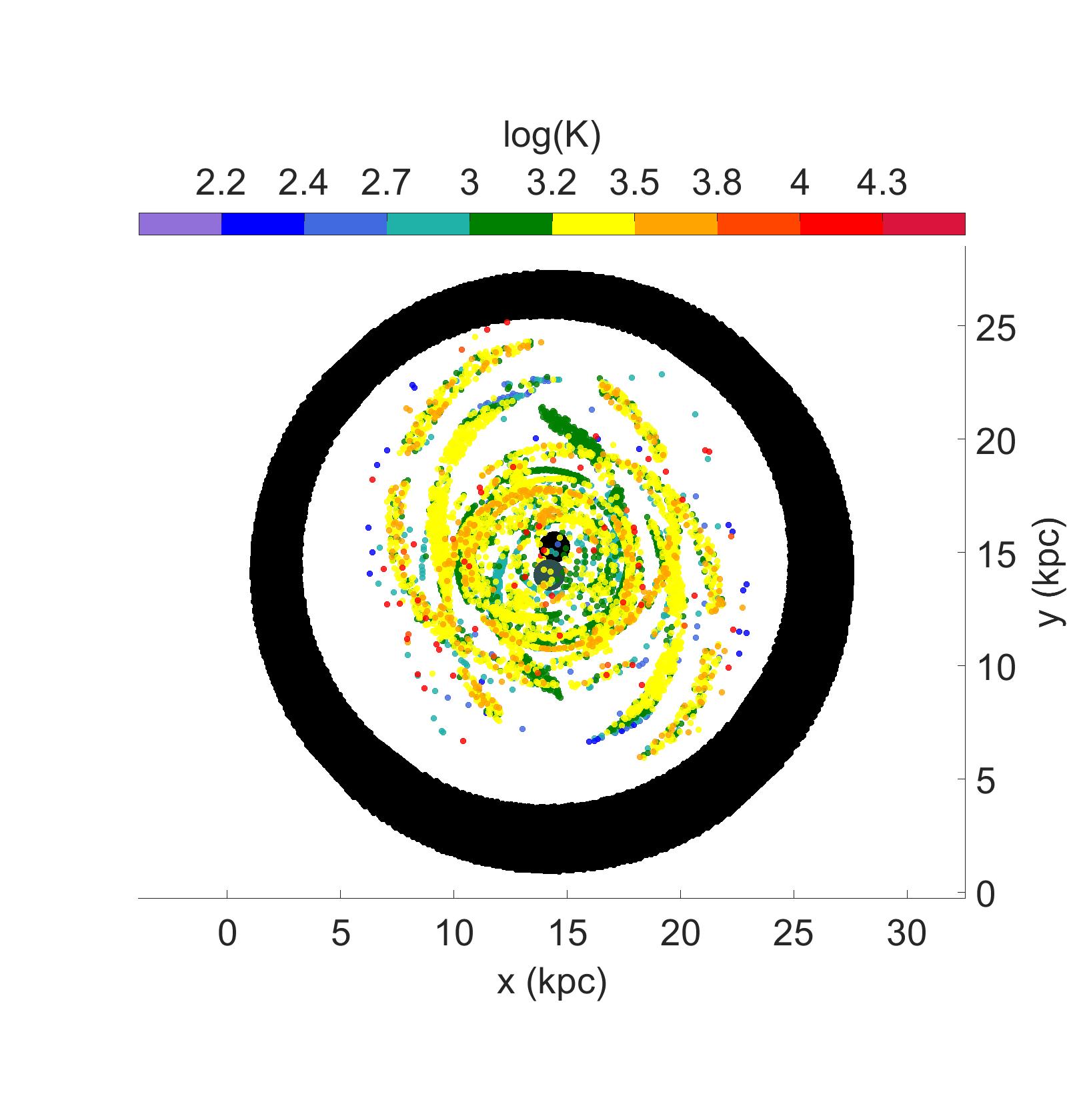}
        }%
    \end{center}

\end{figure}

\begin{figure}

\caption{%
        Only densities $>\SI{e7} m^{-3}$ are shown in each face-on view at time: \textbf{a)} 0 Myr  \textbf{b)} 15 Myr  \textbf{c)} 30 Myr.  
     }%
     \label{fig:0x500subfiguresdensitylthd}
     \begin{center}
        \subfigure[]{%
           \label{fig:0x500density0myrfacelthd}
           \includegraphics[width=0.4\textwidth]{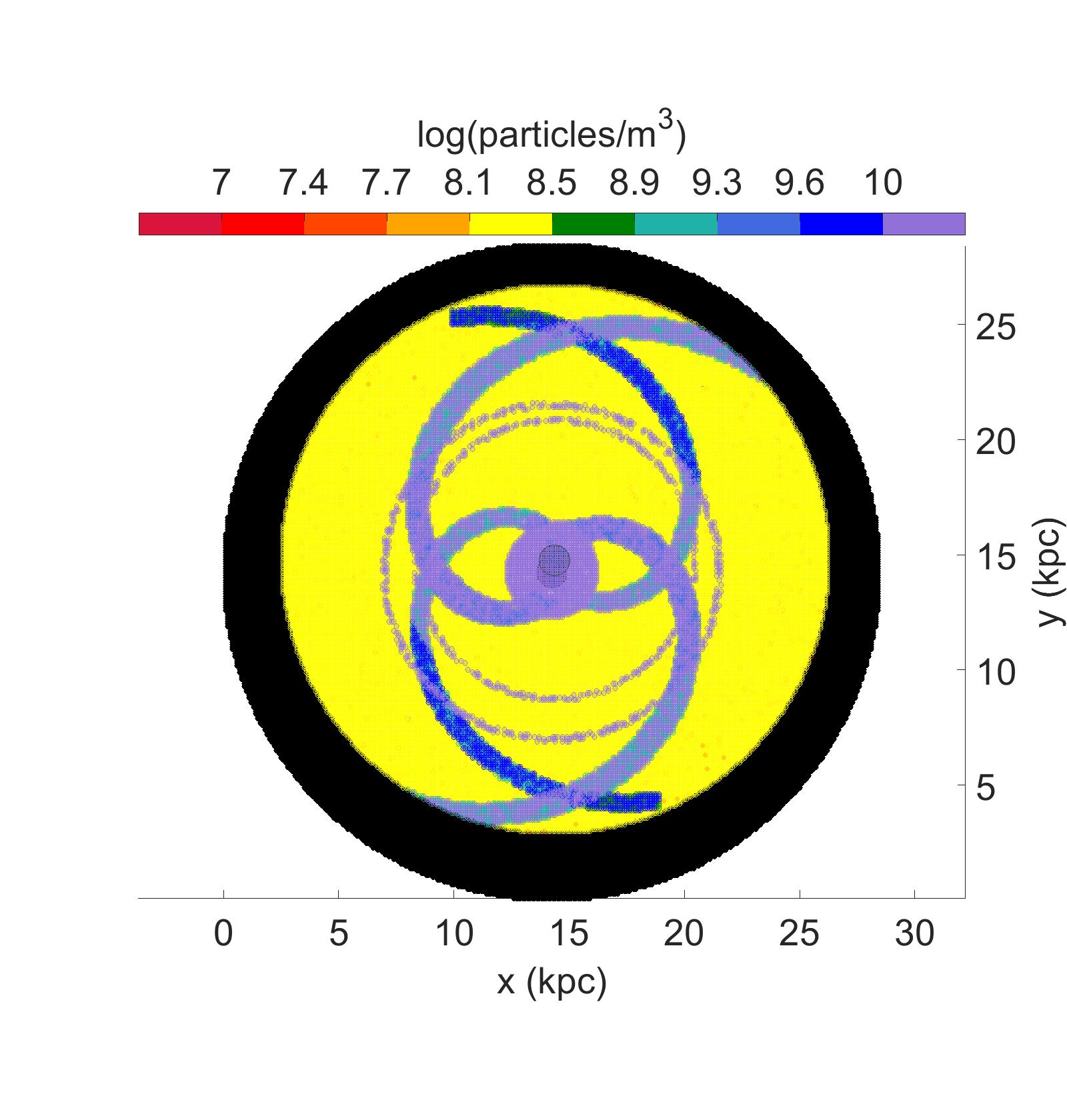}
        }\\ 
        \subfigure[]{%
           \label{fig:0x500density15myrfacelthd}
           \includegraphics[width=0.4\textwidth]{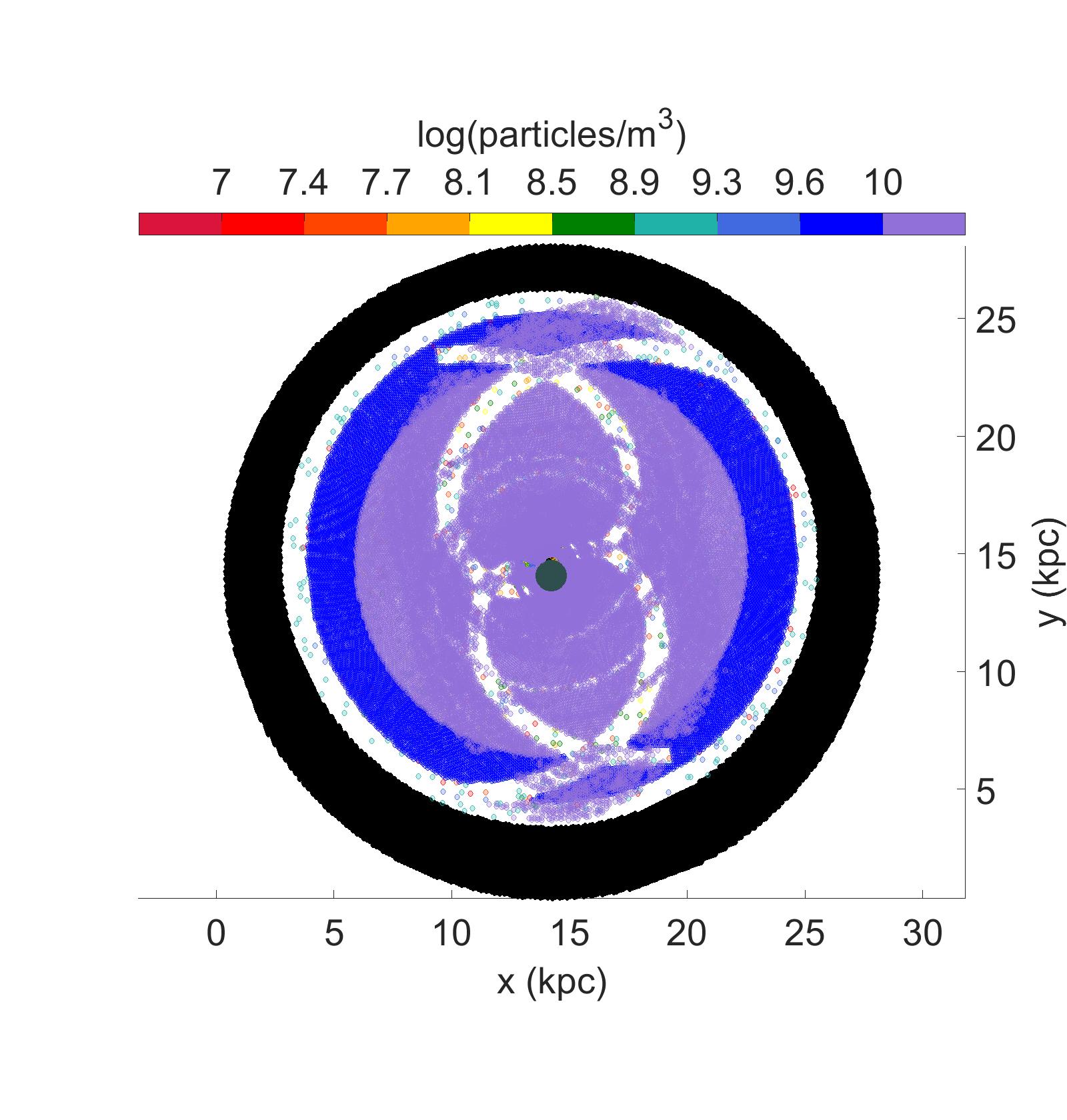}
        }\\ 
        \subfigure[]{%
            \label{fig:0x500density30myrfacelthd}

            \includegraphics[width=0.4\textwidth]{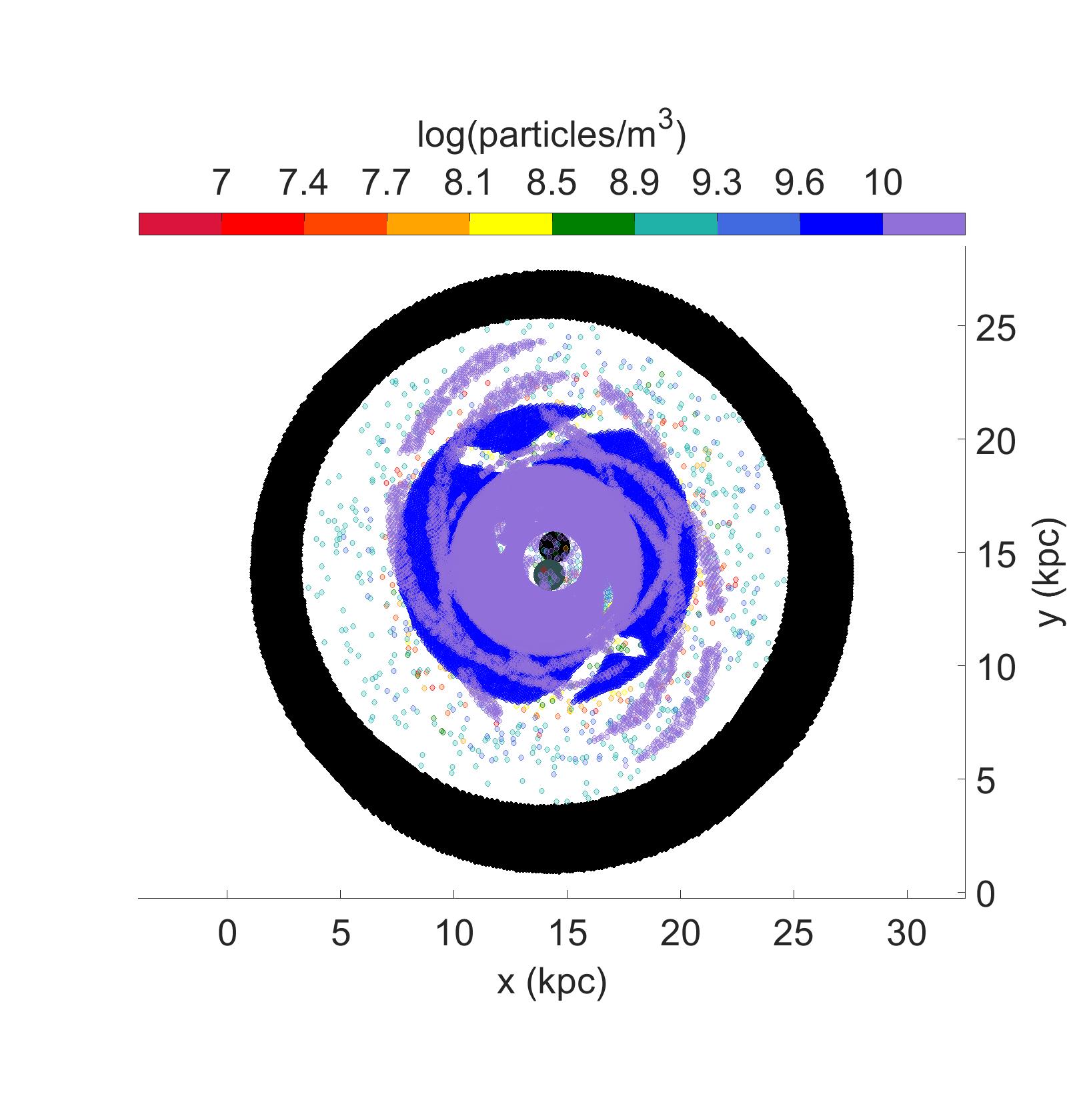}
        }%
    \end{center}

\end{figure}

\indent The post-shock temperatures immediately after the collision of the two discs are shown in panel a) of \cref{fig:0x500subfigurestemphtld} and in b) \cref{fig:0x500subfiguresdensitylthd} are the densities.  The \HI that collides with a spiral arm of \hh is heated to several million degrees whereas the denser \hh is only heated to an order of \SI{e5} Kelvin.  This disparity in temperature is due to the \HI experiencing a greater collision velocity in the centre of mass frame compared to a denser \hh cloud.

\begin{figure}

    \caption{\textbf{a)} 15 Myr cloud temperatures.  \textbf{b)} 30 Myr cloud temperatures.  \textbf{c)} 30 Myr CII.  See  \cref{gascooling} for more details.%
         }%
     \label{fig:0x500pctempedge}
     \begin{center}
        \subfigure[]{%
            \label{fig:0x500pcdensity15myrlthd}
            \includegraphics[width=0.4\textwidth]{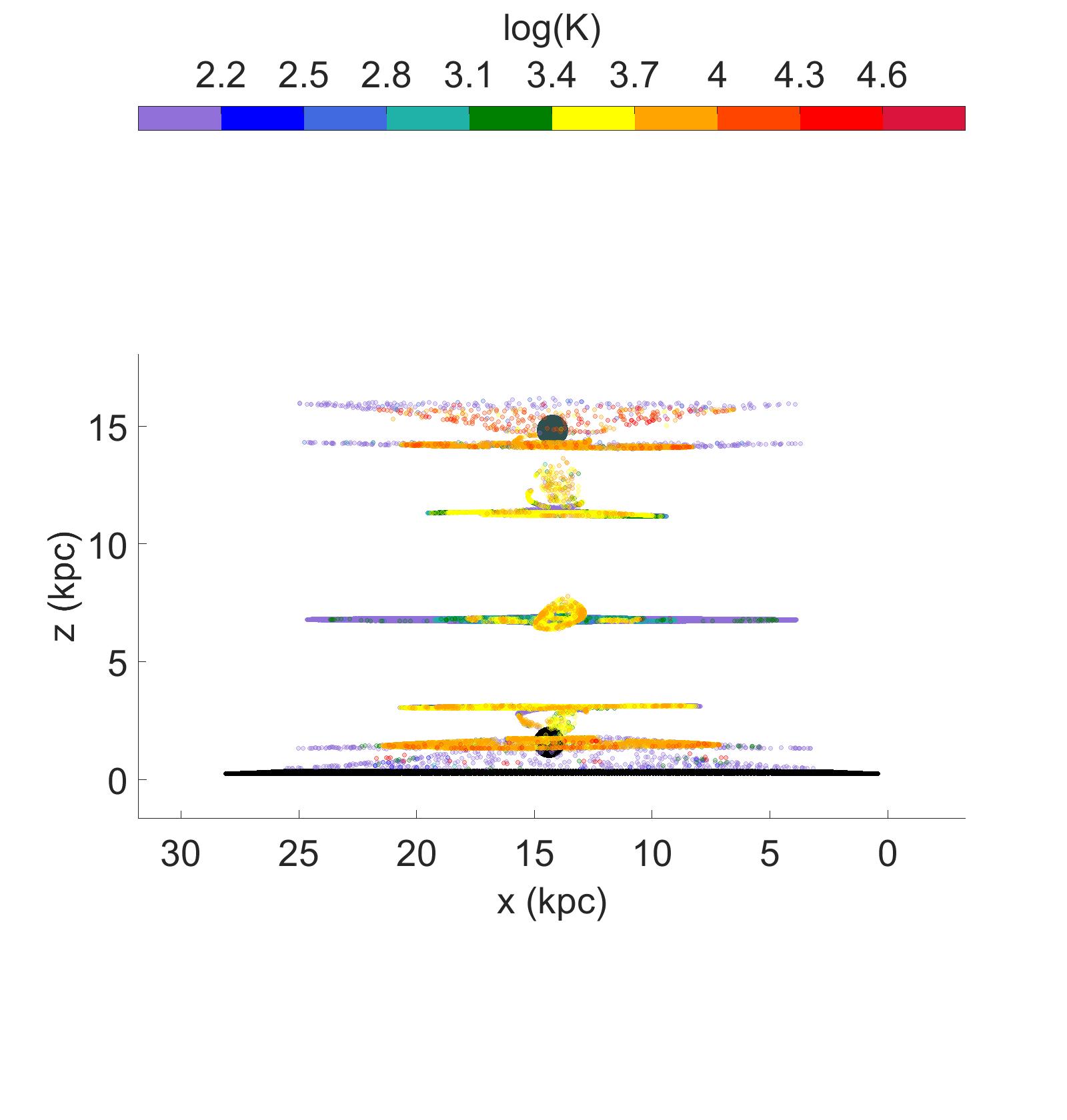}
            
        }\\ 
        
        \subfigure[]{%
            \label{fig:0x500pcdensity30myrlthd}
            \includegraphics[width=0.4\textwidth]{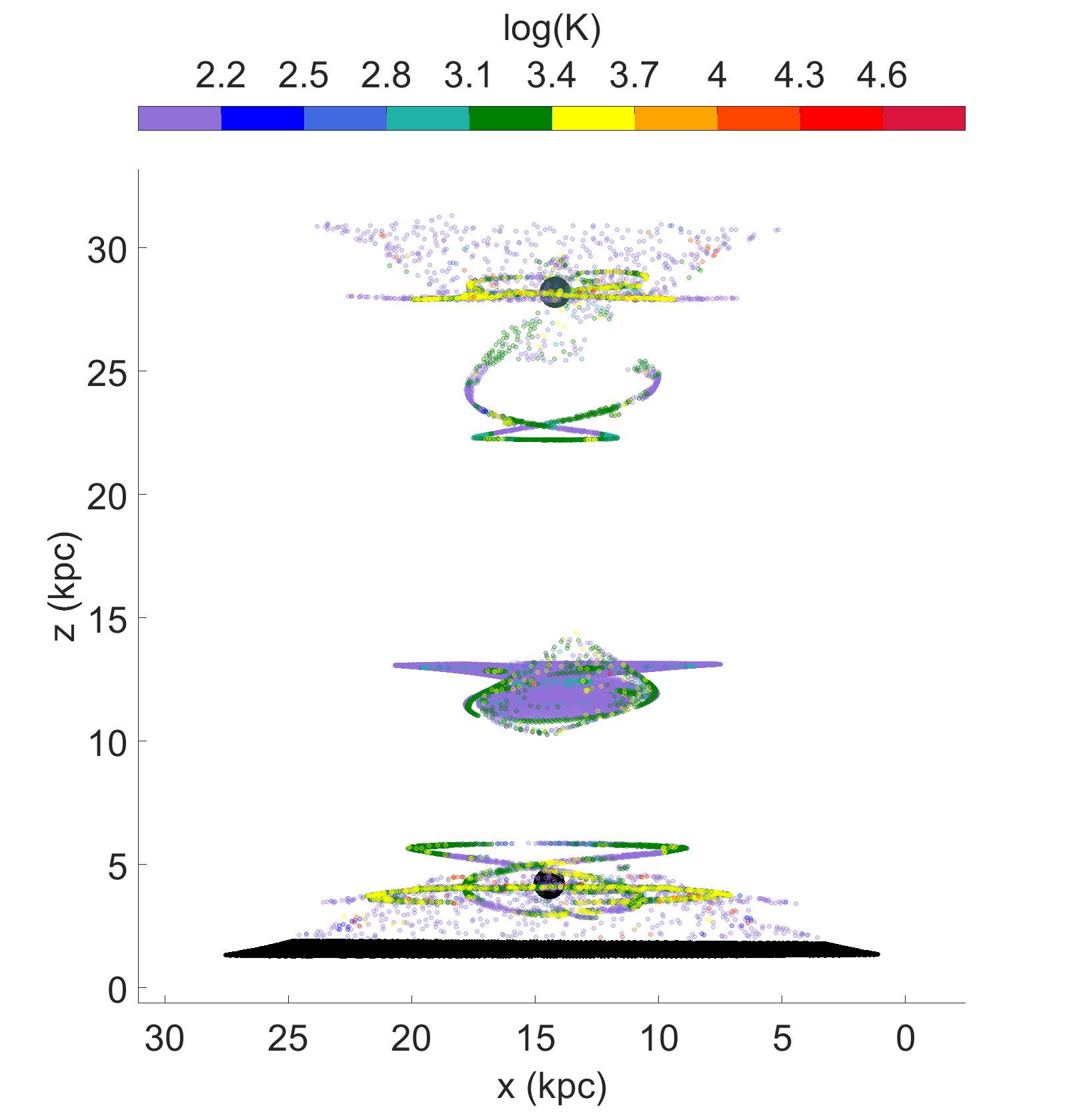}
        }\\ 
        
        \subfigure[]{%
           \label{fig:0x500pcdensity30myrCII}
           \includegraphics[width=0.4\textwidth]{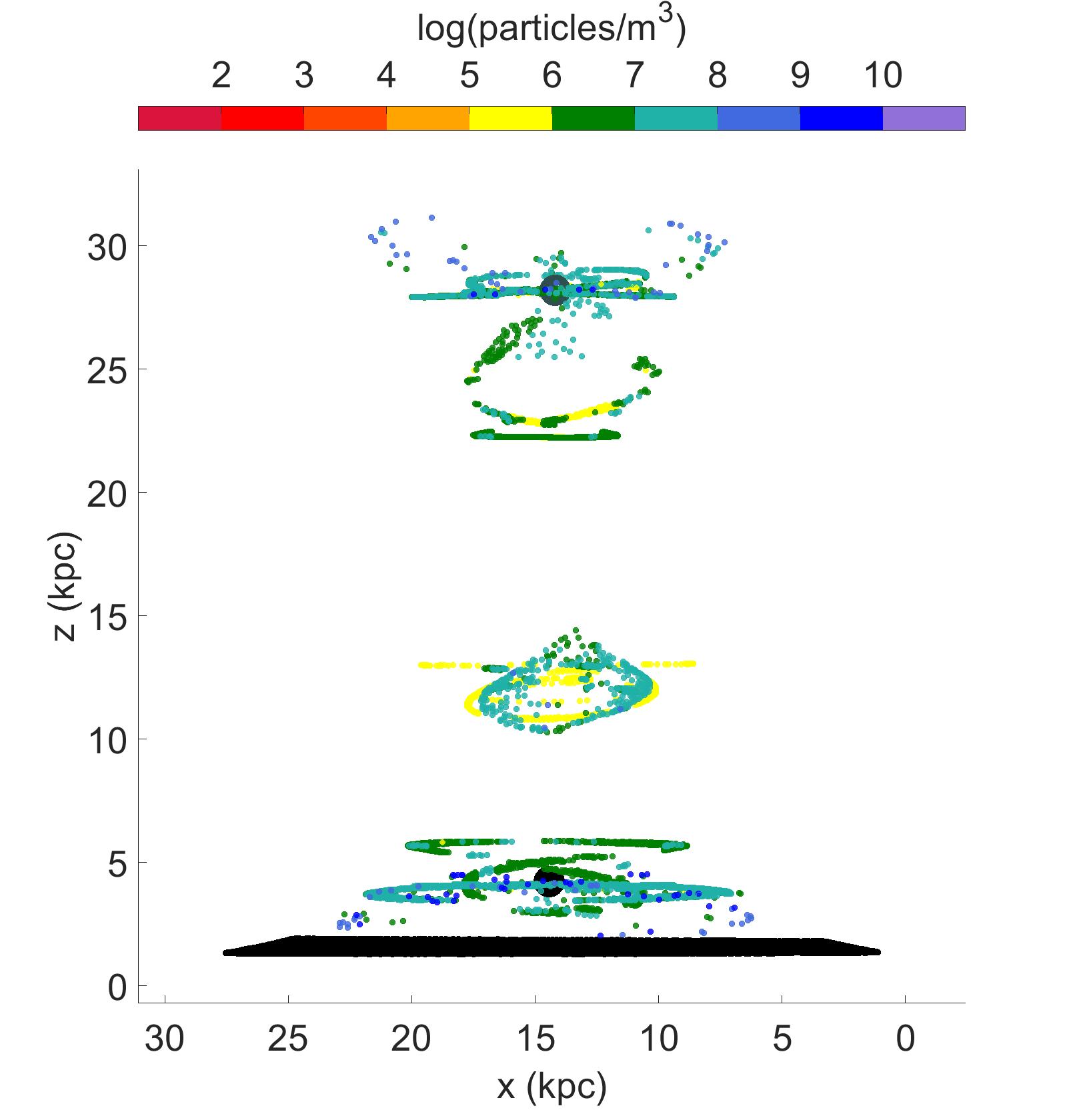}
        }%

    \end{center}

\end{figure}

\indent \Cref{fig:0x500subfigurestemphtld} b) and c) are face-on projections at 30 Myr views showing permitted line cooling and CII emitting clouds, respectively.  The densest gas cools to less than \SI{2e4} K on a timescale of \SI{e5} years where as gas of densities less than \SI{e6} m\textsuperscript{-3} remain at temperatures greater than \SI{3e6} K.  \Cref{fig:0x500pctempedge} a) shows where clouds currently emitting permitted line radiation are distributed.  These are entirely diffuse clouds with densities less than \SI{e4} m\textsuperscript{-3}.  \Cref{fig:0x500pctempedge} b) shows where clouds  expected to be currently emitting CII radiation are located.  It is important to note, this does not represent the coolest gas.  Clouds that have reached the temperature limit of 150 K before 30 Myr are not included in CII emission figures.  A good representation of where the coldest gas is located is in the figures showing densities above \SI{e7} m\textsuperscript{-3}, as gas above this threshold cools within a few million years.  

\indent \Cref{fig:0x500subfiguresdensitylthd} shows face-on projections with gas densities above \SI{1e7} \dunit at 0 Myr, 15 Myr and 30 Myr after the collision, respectively.  As in the single component run, angular momentum is canceled out and most of the abundant bridge \HI component remains in a disc that contracts radially.  Interestingly this contraction has not finished by 30 Myr, and could drive prolonged turbulence.  The densest gas is where the \hh arms of each galaxy are located.  Since the arms are a much higher density than the local \HI of the other disc they do not lose a significant fraction of their momentum and therefore do not collapse with the \HI disc.  Two separate spiral arms develop because the arms are a mixture of two different density components.  One of the components is 10 times and the other is 100 times the local density of \HI.  Each component is separated and displaced by a distance proportional to the density contrast.  The atomic gas that is 10 times the local \HI disc transfers a larger fraction of its momentum during the collision so it collapses significantly more than the \hh that is up to 100 times the local \HI density.  This separation of ISM phases by their density could explain why we see X-ray emission displaced from the CO emission in the Taffy bridge.

\begin{figure}

\caption{\textbf{a)} 15 Myr, gas $> \SI{e7} m^{-3}$.  \textbf{b)} At 30 Myr, gas $> \SI{e7} m^{-3}$ is shown.  \textbf{c)} At 30 Myr, showing gas $> \SI{e7} m^{-3}$.%
     }%
\label{fig:0x500pcdensityedge}     
     \begin{center}
        \subfigure[]{%
           \label{fig:0x500density15myrhtld}
           \includegraphics[width=0.4\textwidth]{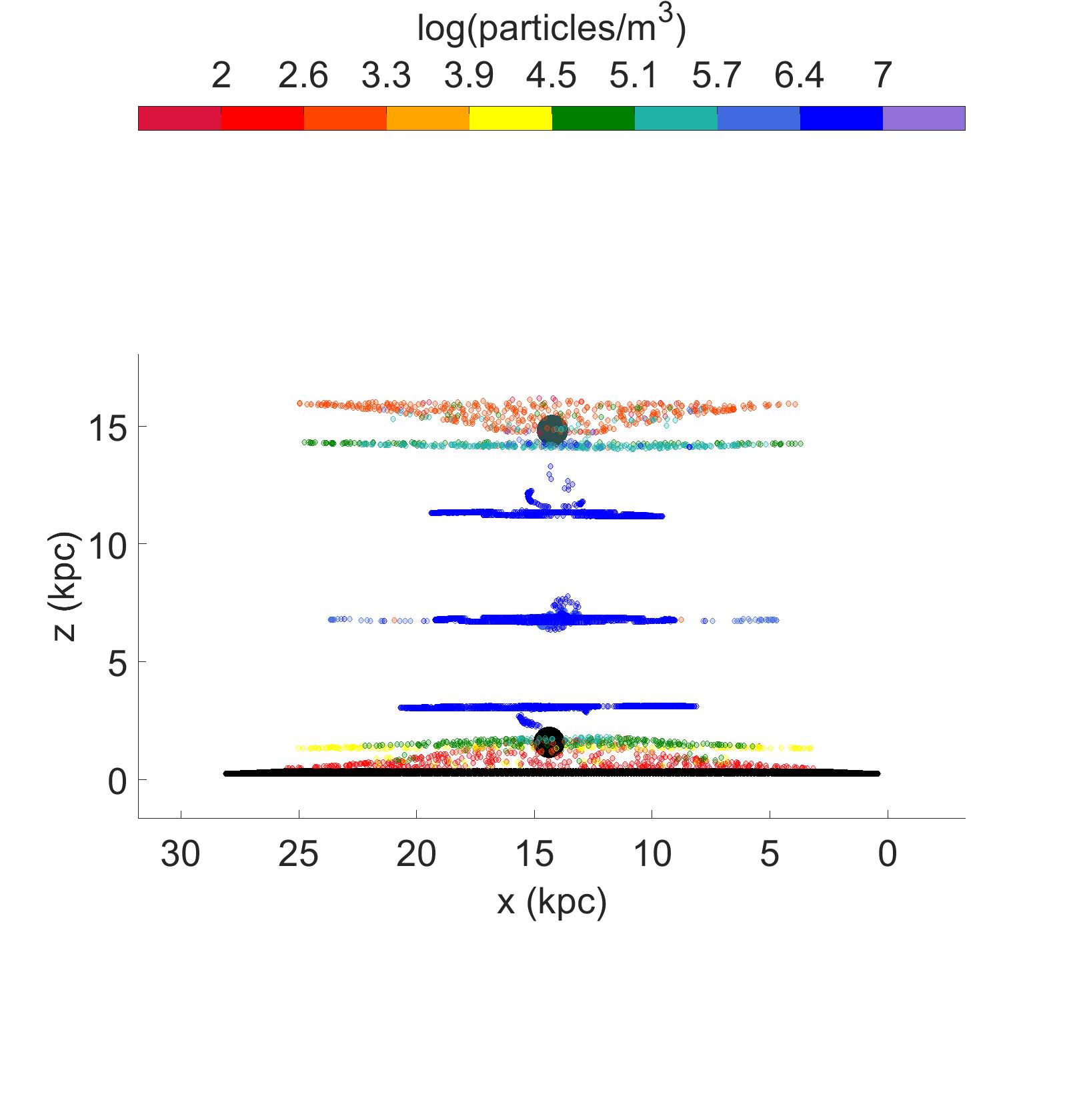}
        }\\ 
        \subfigure[]{%
           \label{fig:0x500density30myredgehtld}
           \includegraphics[width=0.4\textwidth]{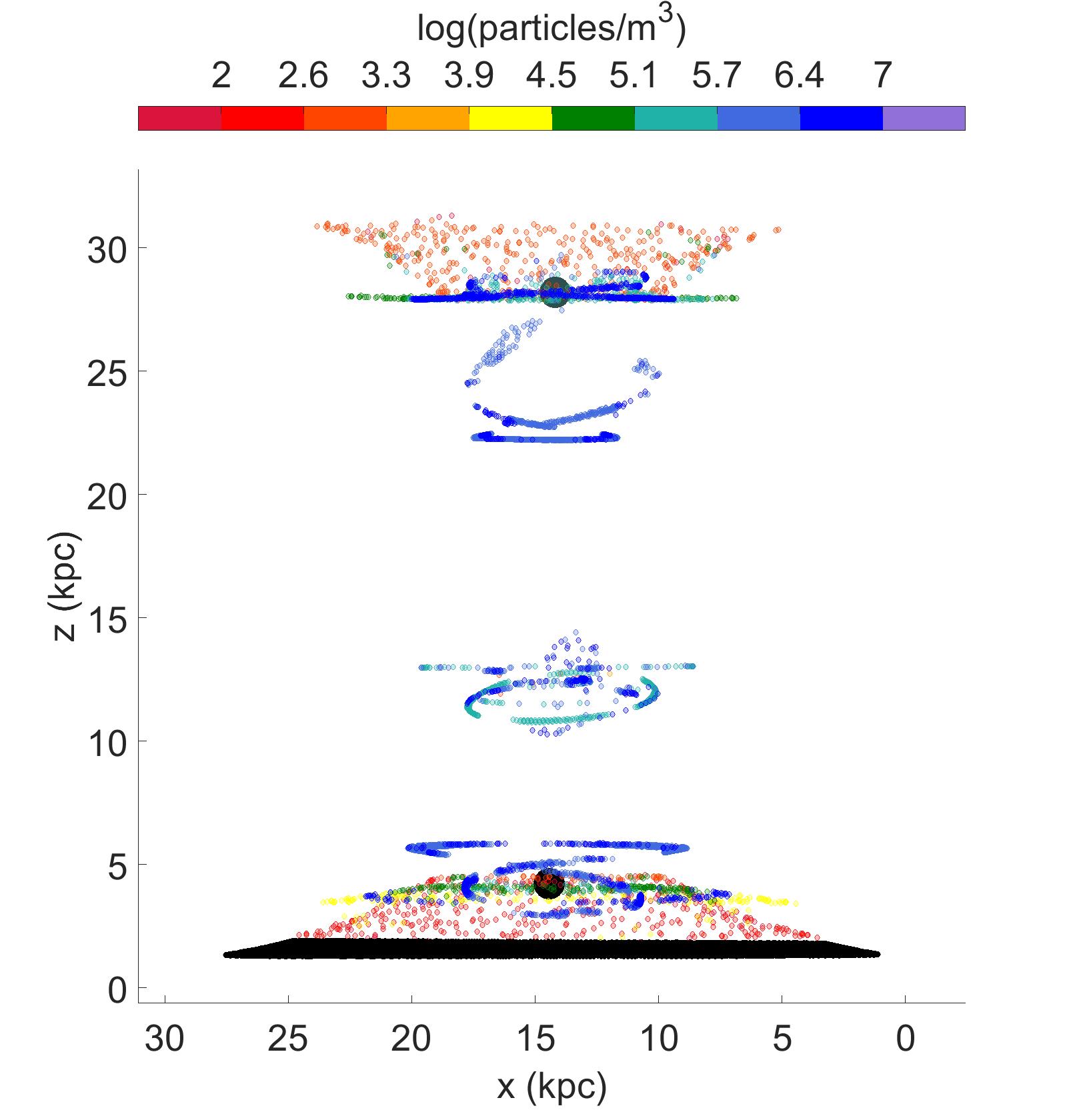}
        }\\ 
        \subfigure[]{%
            \label{fig:0x500density30myredgelthd}
            \includegraphics[width=0.4\textwidth]{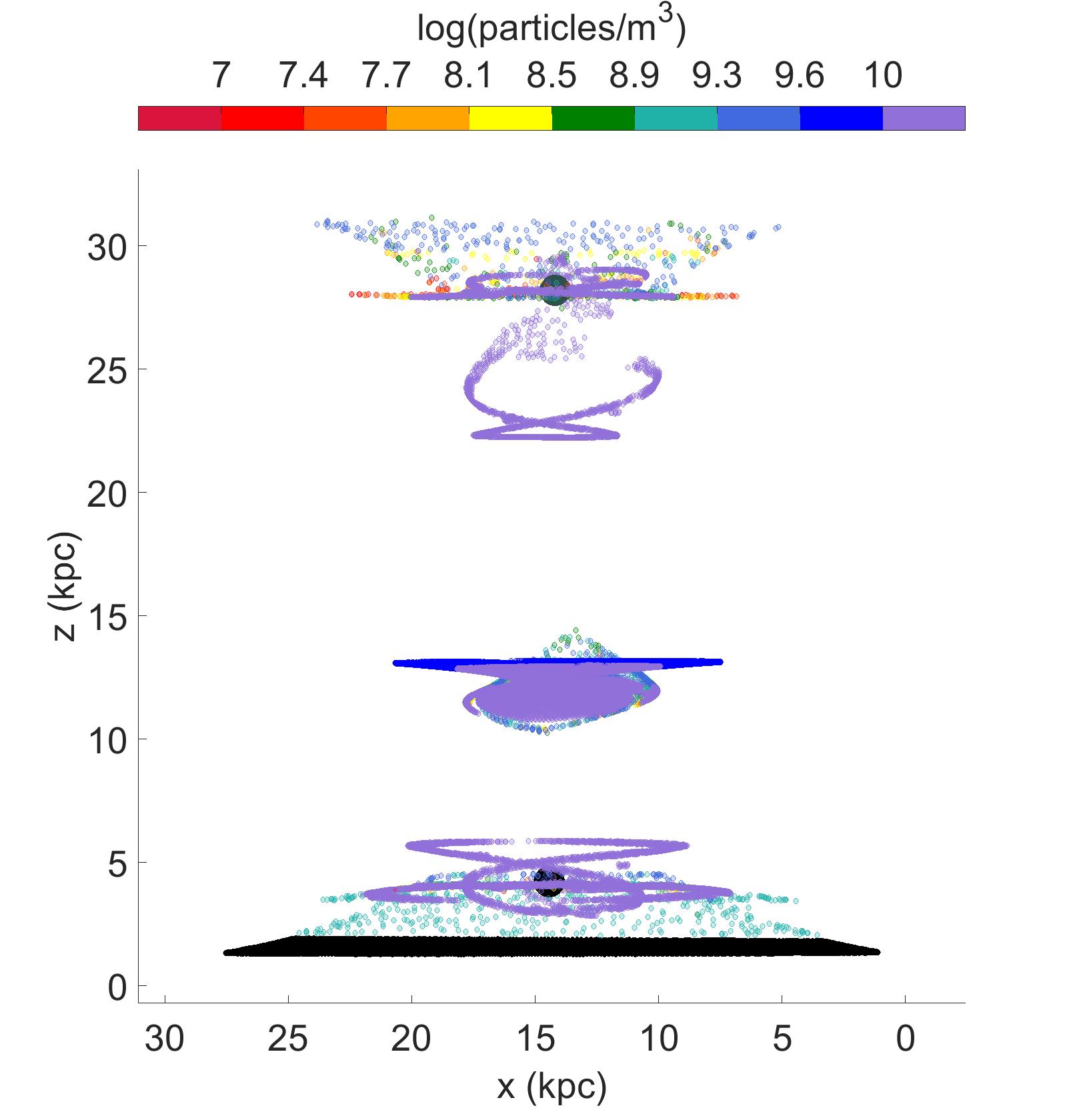}
        }%
    \end{center}
    
\end{figure}

\indent Edge-on projections at 30 Myr are shown in \cref{fig:0x500pctempedge,fig:0x500pcdensityedge}.  \Cref{fig:0x500pcdensityedge} shows the general increase in cloud density as contraction by cooling occurs from 0 to 15 Myr.  The hot diffuse hydrogen gas removed from one galaxy is mixed with cool dense gas of the other but does not follow the \HI-on-\HI material.  The dense H\tus{2} makes up twisted filaments of clouds enveloped inside a smooth hot diffuse medium and is seen spiraling away from the centres of each galaxy.  The arms wind up slightly over time as they are pulled toward the nearest galaxy.  This occurs in each run, but \cref{fig:0x500pcdensityedge} c) is a good illustration of the effect.  The dense, initially H\tus{2} clouds, experience a lesser change in momentum so they remain nearer their host galaxy than the \HI on \HI gas.  At 15 Myr the arms are only removed 2 kpc from their galaxy.  This is different from the \HI disc which is immediately left behind with nearly no velocity relative to the system's centre of mass. 

\begin{figure}
\caption{Edge-on view of a same sense run.  \textbf{a)} Temperature is shown. \textbf{b)} Densities $> \SI{e7} m^{-3}$ are shown.}
\label{fig:eqse0x500edge30myrdensitylthd}
    \subfigure[]{%
       \includegraphics[width=0.4\textwidth]{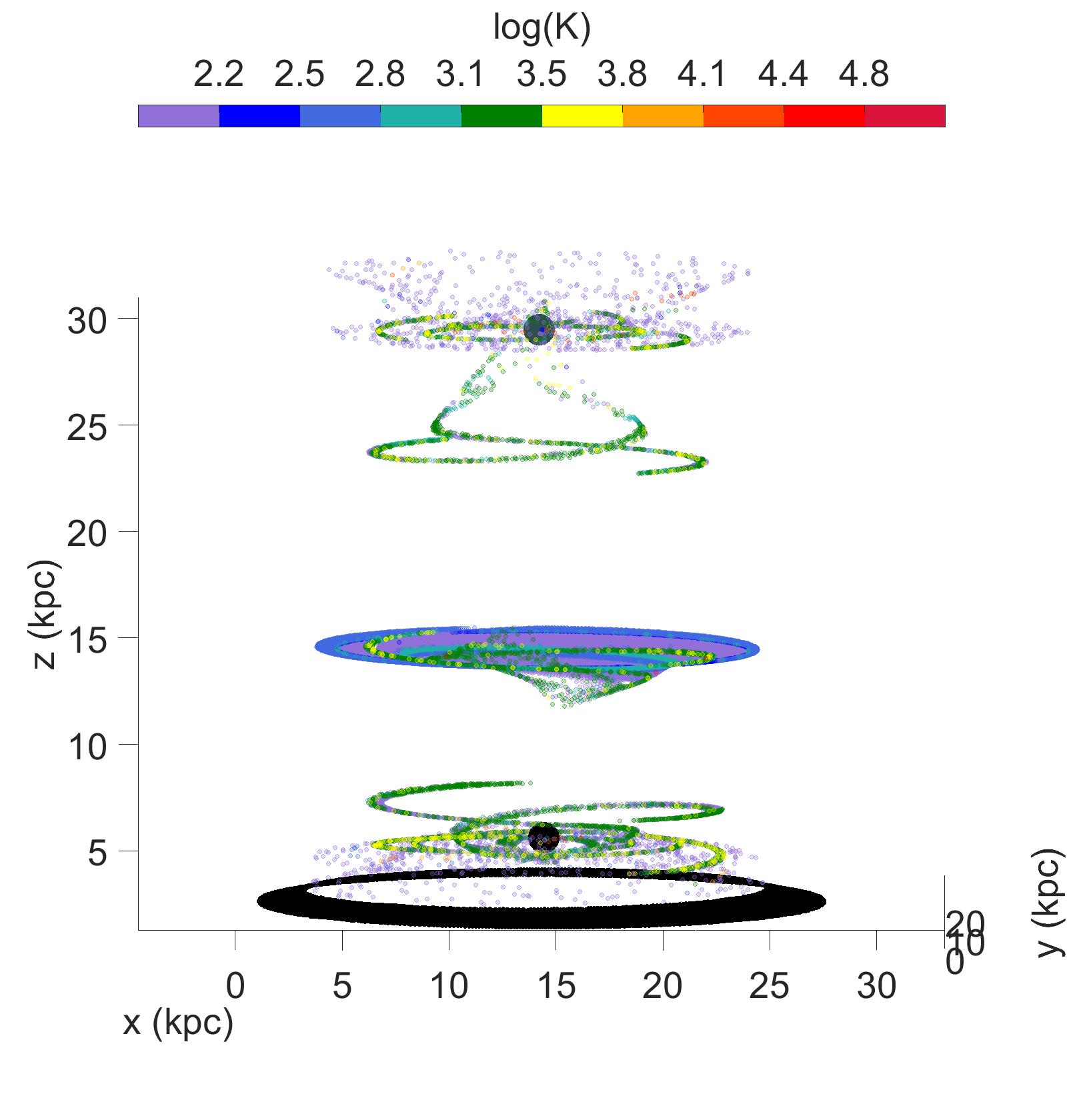}
    }\\ 
    \subfigure[]{%
       \includegraphics[width=0.4\textwidth]{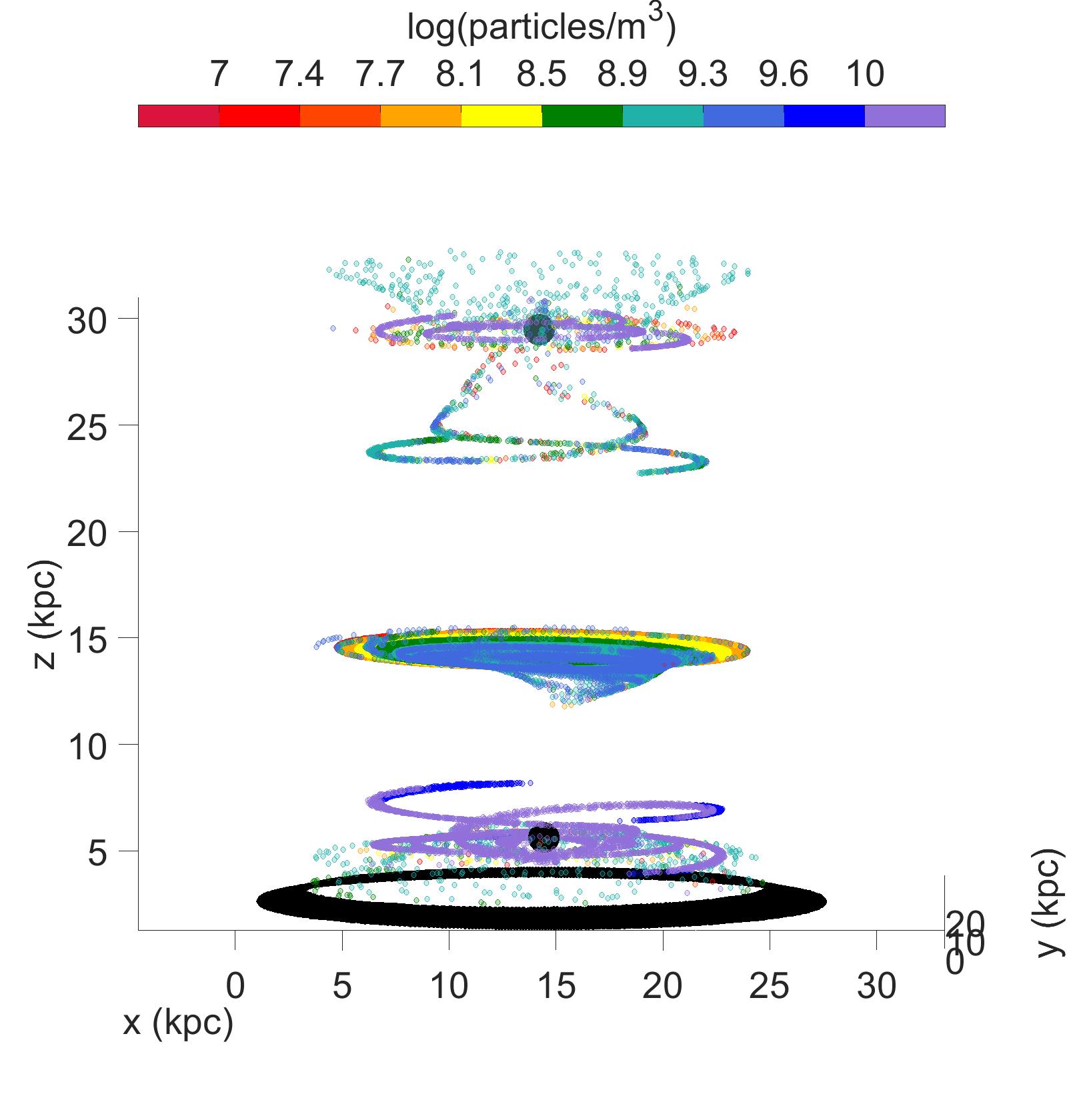}
    }
\end{figure}

\Cref{fig:eqse0x500edge30myrdensitylthd} shows the bridge resulting from the collision of two discs rotating in the same sense.  The \HI CBD is much more flat and extended than it is for the counter-rotating case, as seen previous in \cref{fig:0x500face30myrfigures,fig:eqsense0x500edge30myrlthd}.  For this run both the \HI CBD and spiral arms do not contract radially by any significant amount.  

\indent Disc-on-disc collisions with small offsets are the types of collisions expected to produce ring systems.  We find that in direct disc-on-disc collisions a substantial amount of ISM is completely removed from each disc.  The gas of G1 shown in black, best seen through the 500 pc offset runs, is not shocked and is only slightly perturbed from the gravitational interaction.  A ring wave would take several million years to expand into this outer ring of gas.  Once it does it could produce a significantly delayed epoch of star formation.

\subsection{Five component ISM models with 21 kpc offset}

\begin{figure}

\caption{At the time of collision: \textbf{a)} clouds with post shock temperatures below \SI{e6} K are shown and \textbf{b)} clouds with post shock temperatures above \SI{e6} K are shown.%
     }%
     
     \begin{center}

        \subfigure[]{%
           \label{fig:21kpcg1temp0myrlt}
           \includegraphics[width=0.4\textwidth]{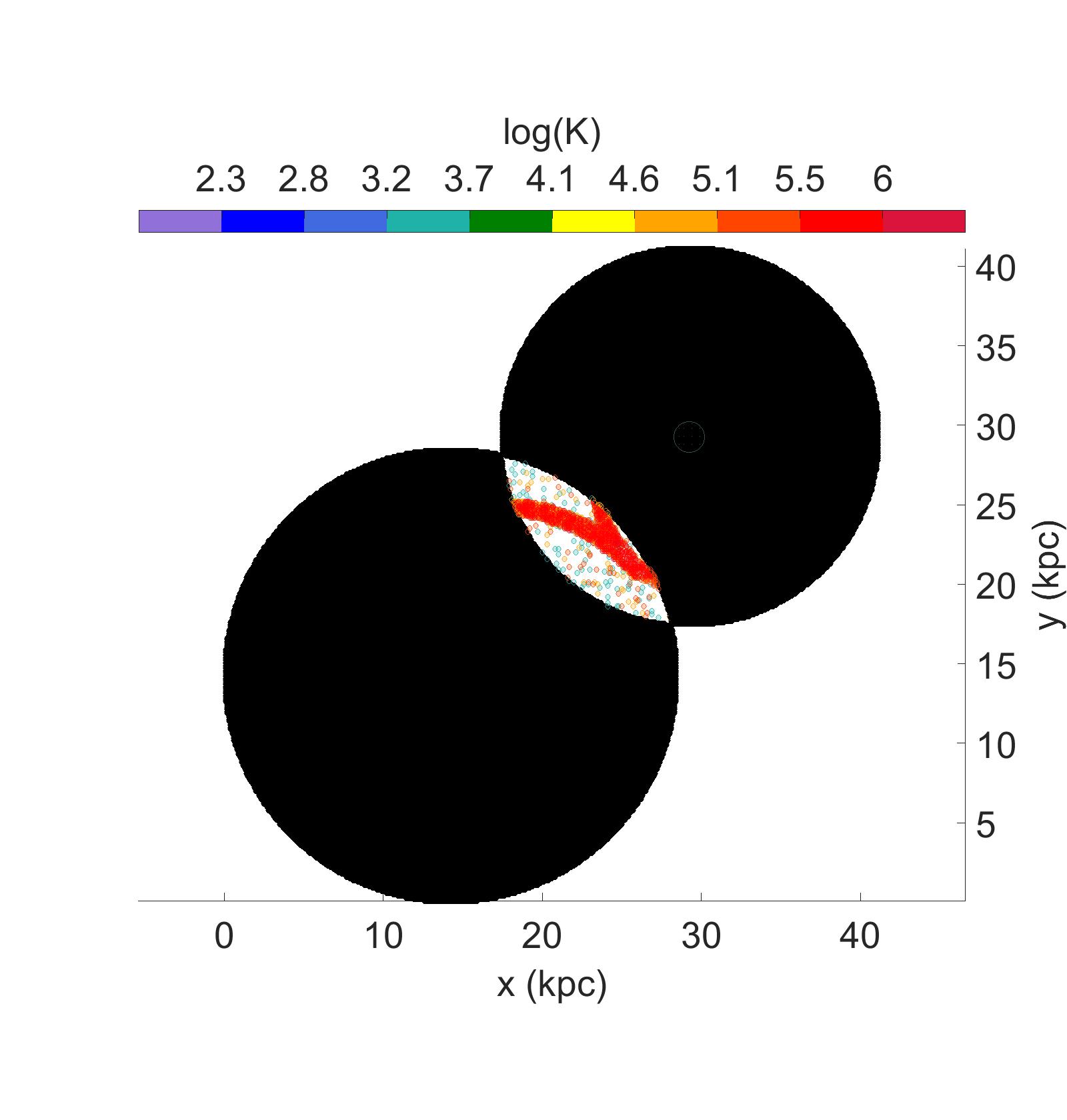}
        }\\ 
        \subfigure[]{%
            \label{fig:21kpctemp0myrHT}
            \includegraphics[width=0.4\textwidth]{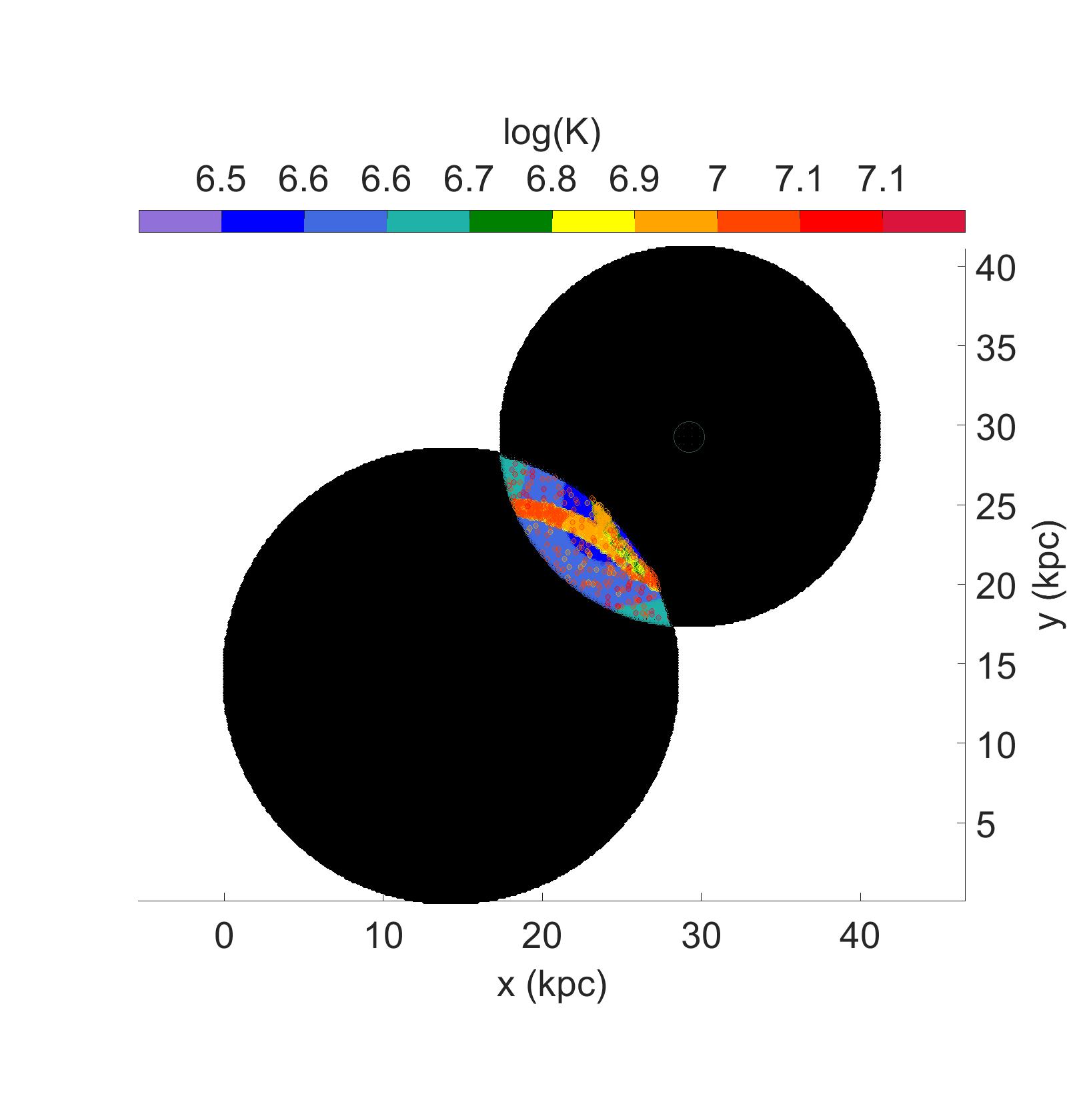}
        }%
    \end{center}
    
   \label{fig:21kpctemp0myrsubfigures}
\end{figure}

\begin{figure*}

\caption{At 30 Myr year after the collision.  Figures show clouds with \textbf{a)} densities $< \SI{e7} m^{-3}$ \textbf{b)} densities $> \SI{e7} m^{-3}$  \textbf{c)} temperature and \textbf{d)} CII cooling. See  \cref{gascooling} for more details.%
     }%
     
     \begin{center}

                \subfigure[]{%
          \label{fig:21kpcdensity30myrhtld}
          \includegraphics[width=0.4\textwidth]{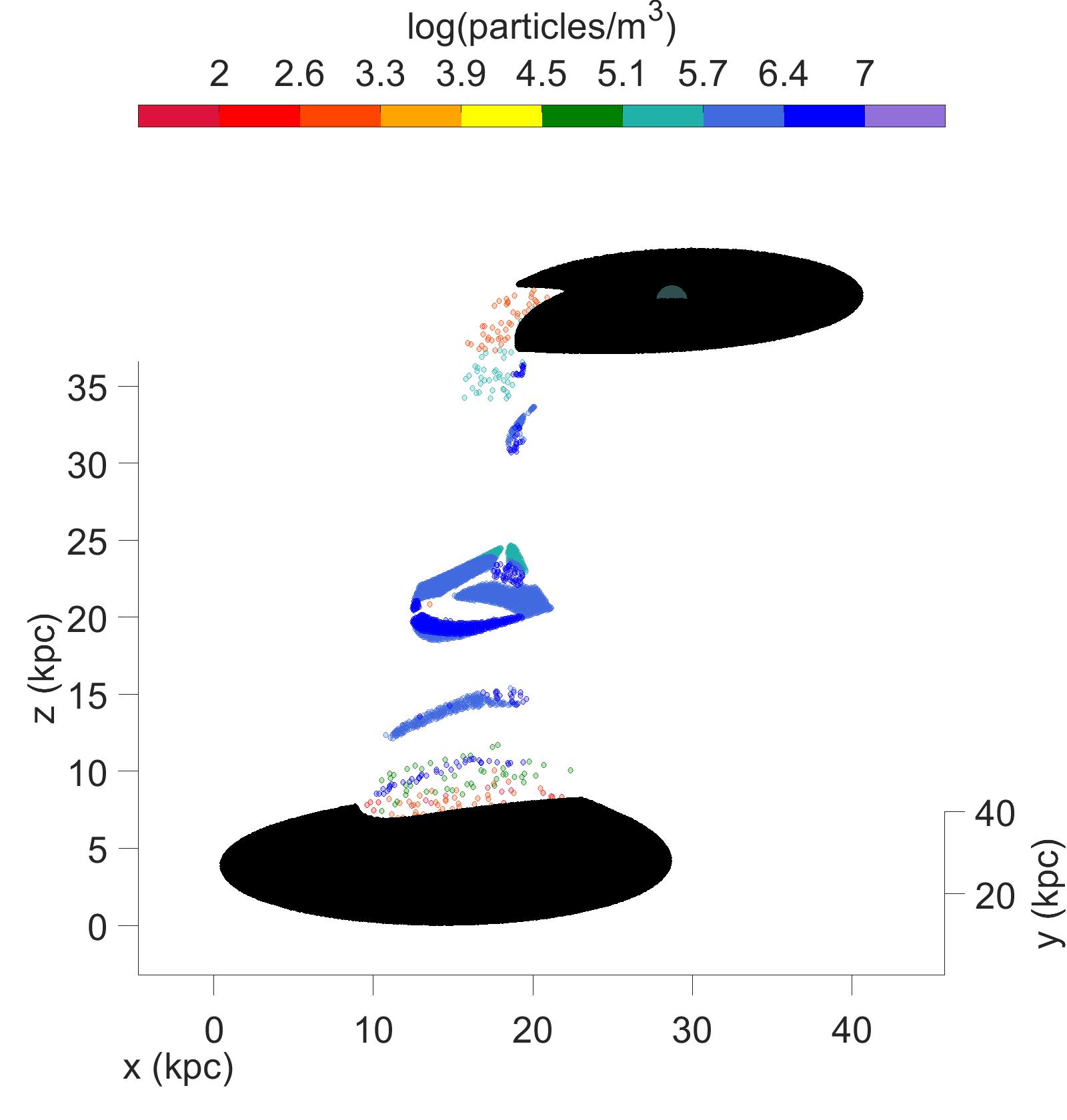}
        }
                \subfigure[]{%
          \label{fig:21kpcdensity30myrlthd}
          \includegraphics[width=0.4\textwidth]{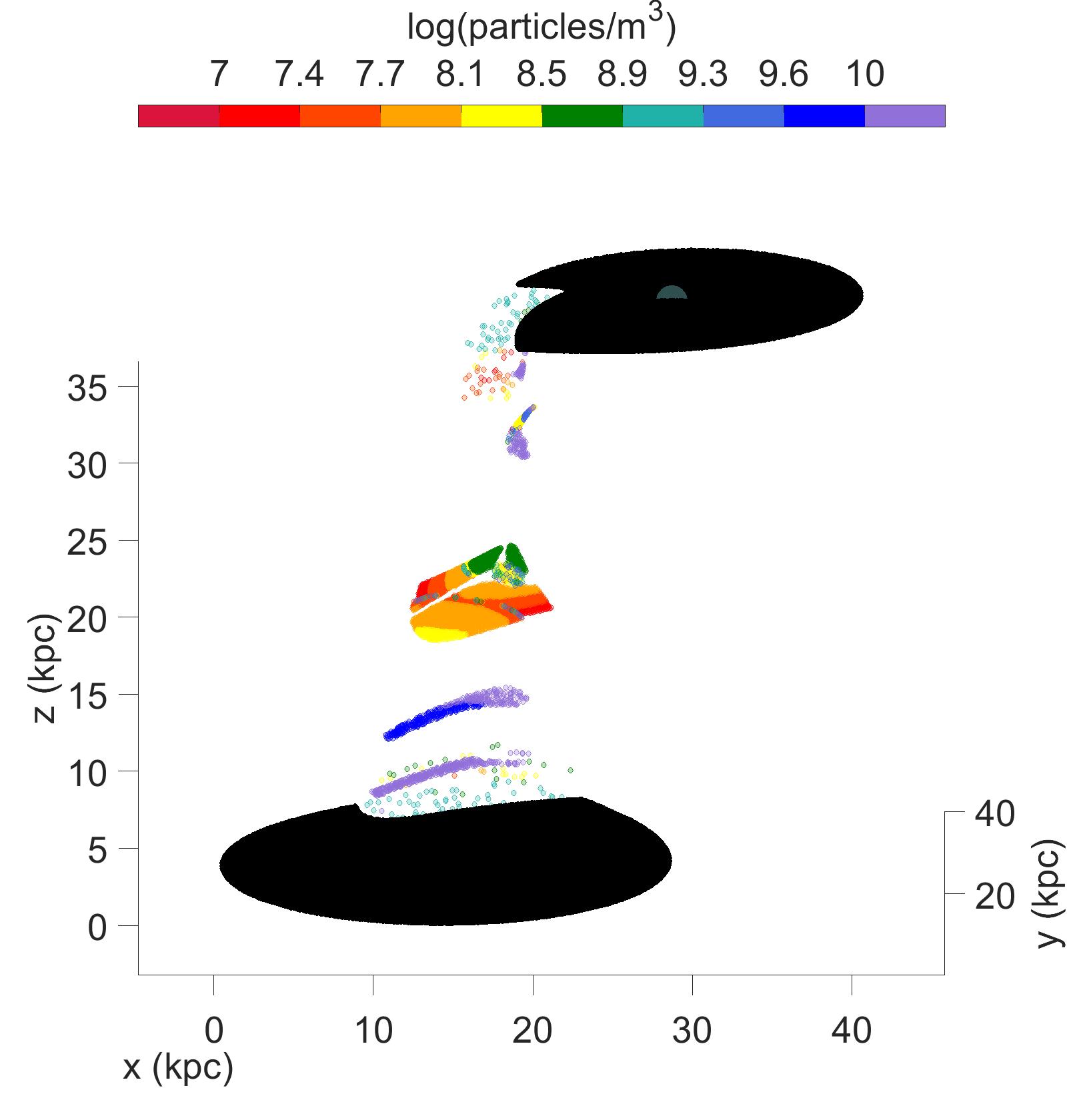}
        }\\ 
                \subfigure[]{%
          \label{fig:21kpcdensity30myrsingle}
          \includegraphics[width=0.4\textwidth]{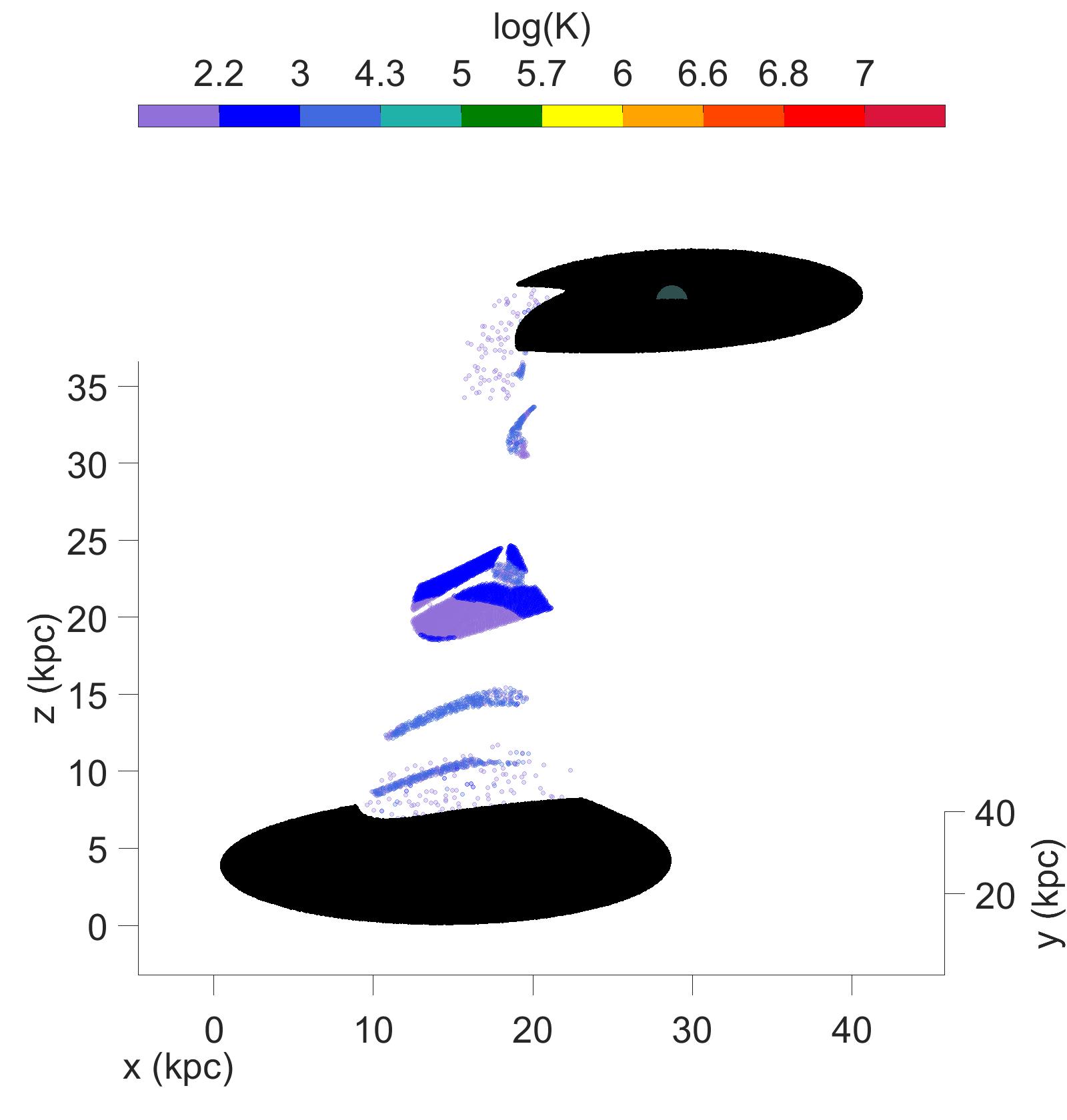}
        }
                \subfigure[]{%
          \label{fig:21kpcdensity30myrCII}
          \includegraphics[width=0.4\textwidth]{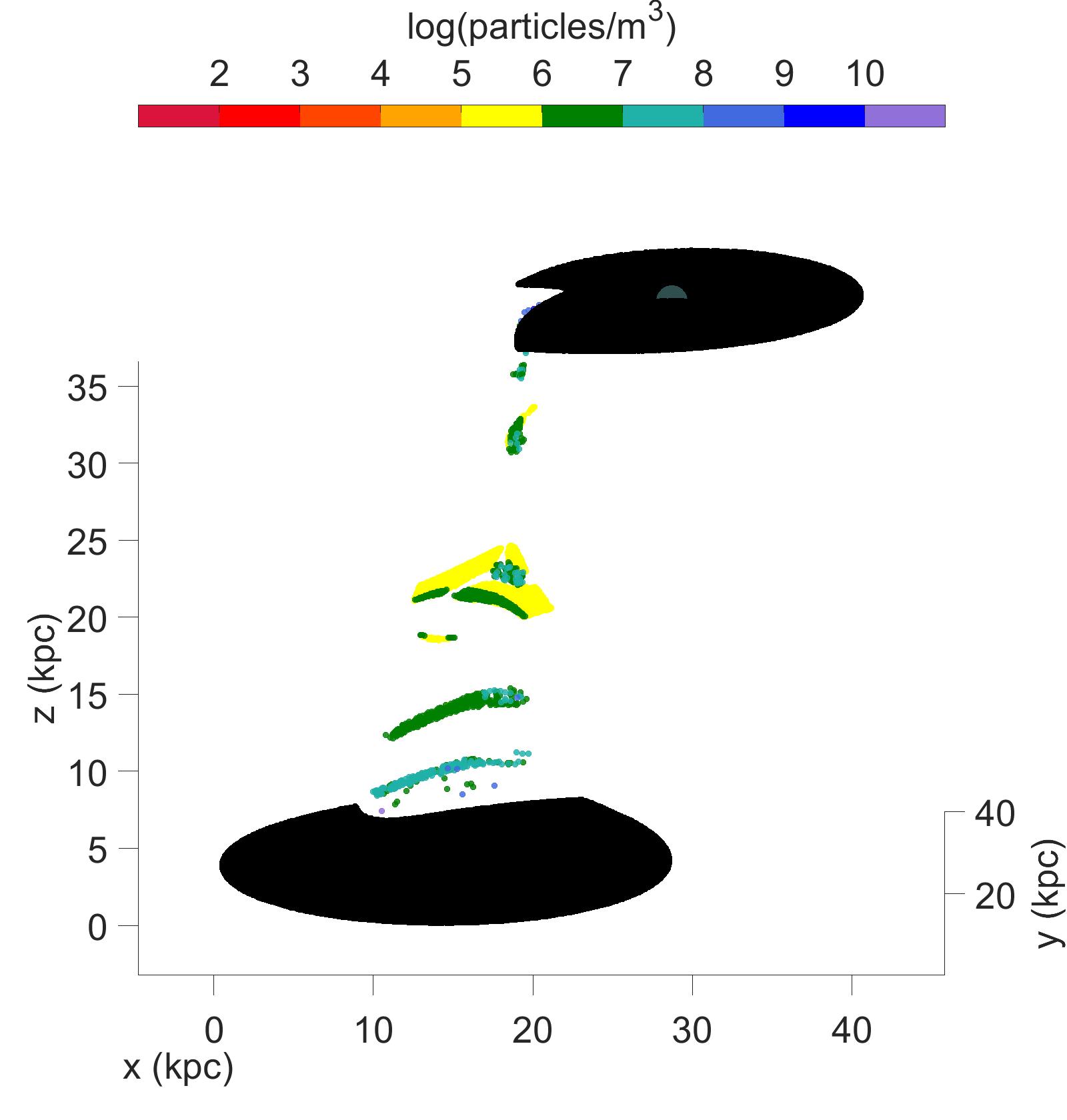}
        }\\ 
    \end{center}
    
  \label{fig:21kpctemp30myrsubfigures}
\end{figure*}

\begin{figure*}

\caption{At 30 Myr year after the collision.  Figures show clouds with \textbf{a)} densities $< \SI{e7} m^{-3}$ \textbf{b)} densities $> \SI{e7} m^{-3}$  \textbf{c)} temperature and \textbf{d)} CII cooling. See  \cref{gascooling} for more details.%
     }%
     
     \begin{center}

                \subfigure[]{%
          \label{fig:eqse21kpcdensity30myrhtld}
          \includegraphics[width=0.4\textwidth]{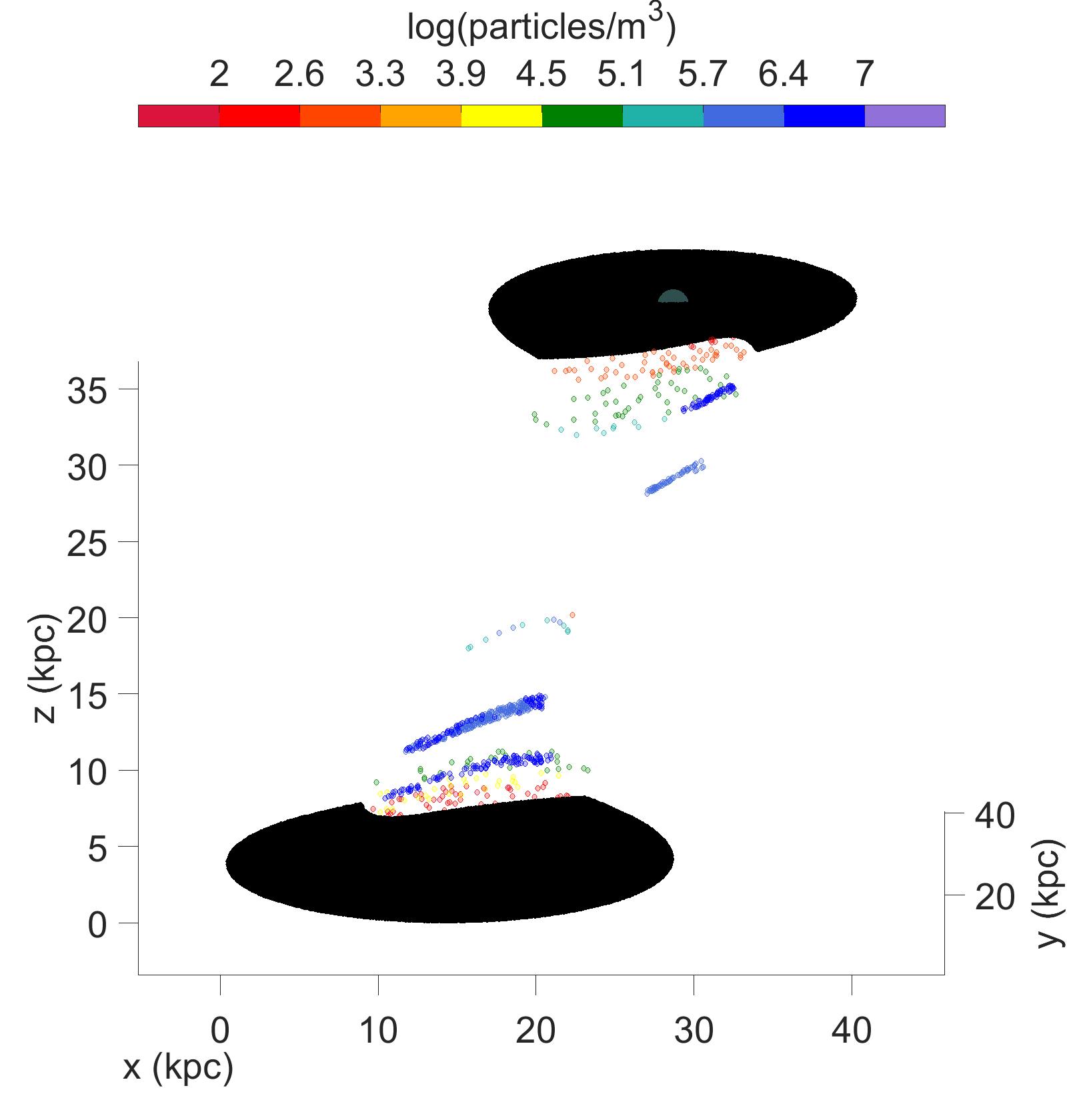}
        }
                \subfigure[]{%
          \label{fig:eqse21kpcdensity30myrlthd}
          \includegraphics[width=0.4\textwidth]{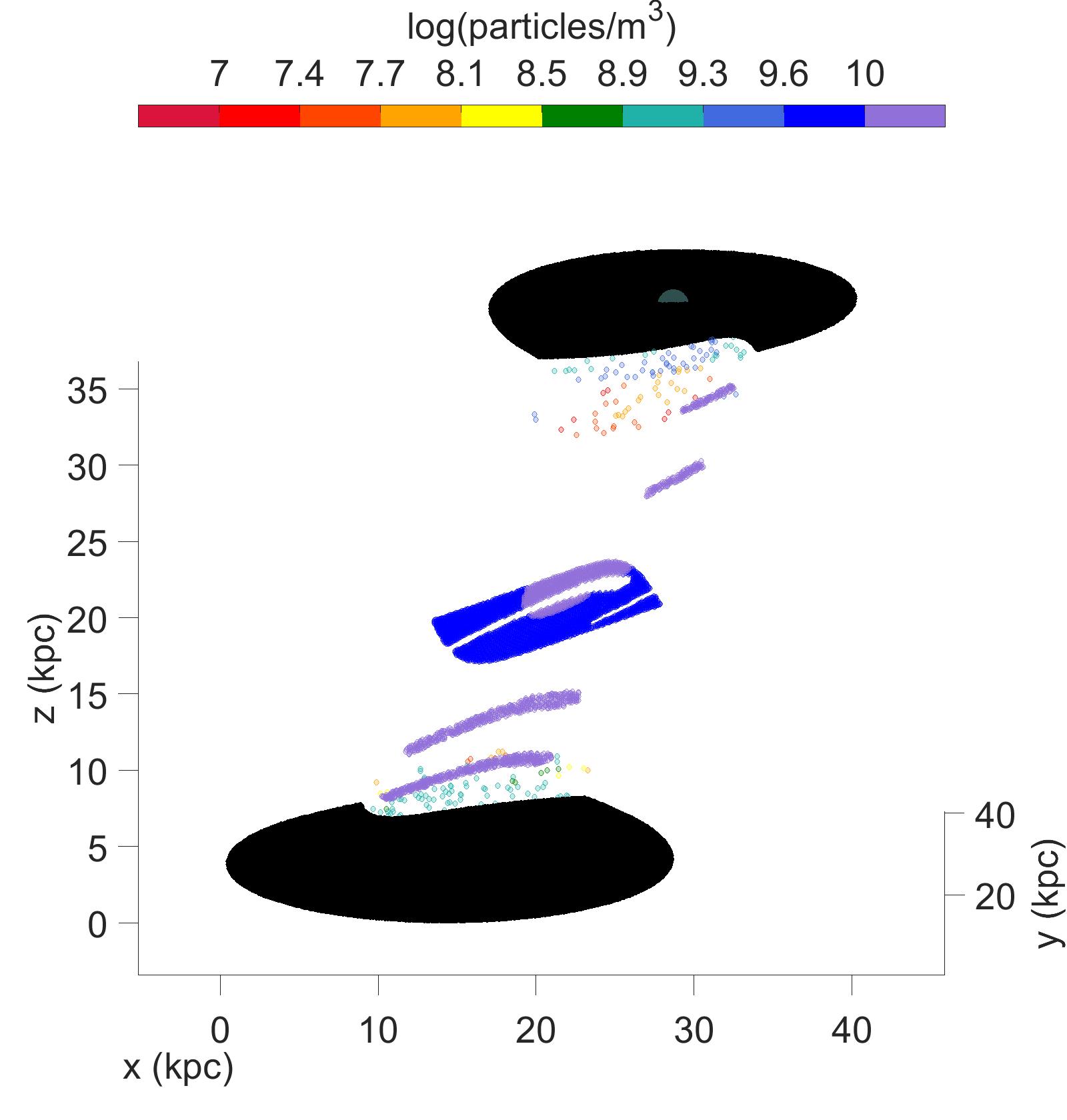}
        }\\ 
                \subfigure[]{%
          \label{fig:eqse21kpcdensity30myrsingle}
          \includegraphics[width=0.4\textwidth]{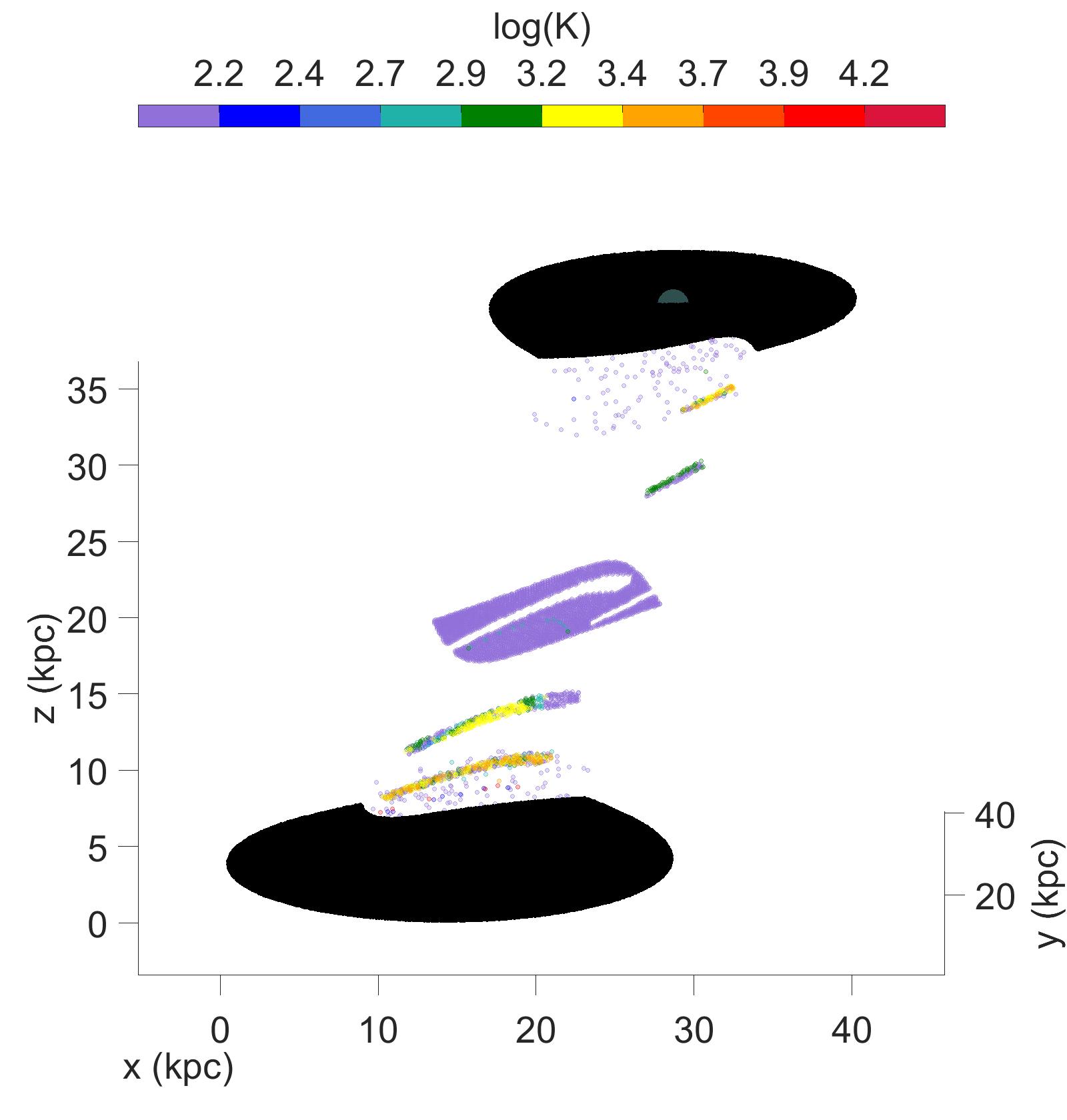}
        }
                \subfigure[]{%
          \label{fig:eqse21kpcdensity30myrCII}
          \includegraphics[width=0.4\textwidth]{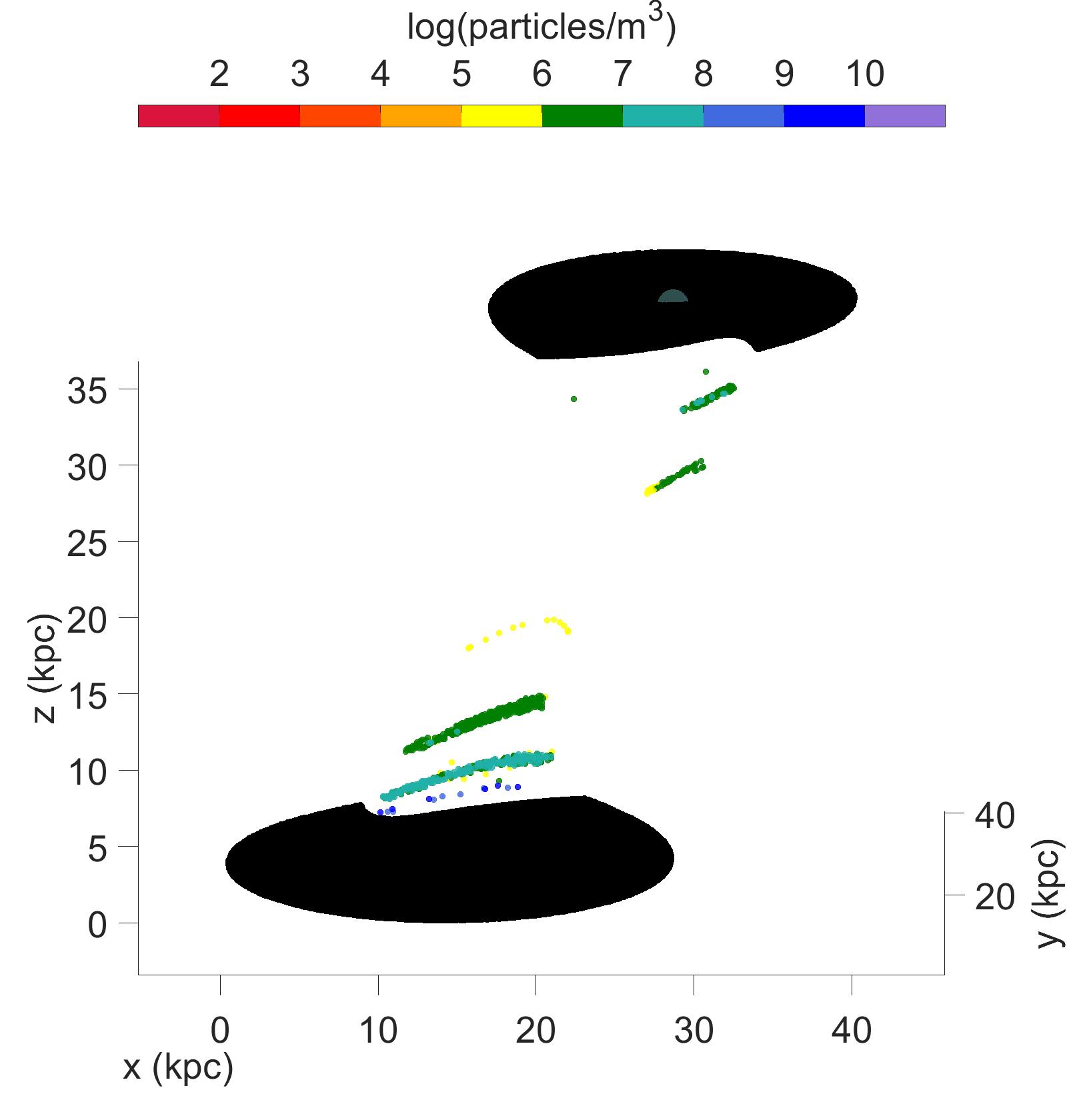}
        }\\ 
    \end{center}
    
  \label{fig:21kpcEQsensedensity}
\end{figure*}

\Cref{fig:21kpctemp0myrsubfigures,fig:21kpctemp30myrsubfigures} show collisions between counter-rotating discs and \cref{fig:21kpcEQsensedensity} shows the result of same sense rotation, with a 21 kpc offset in both cases.  Since these bridges look entirely different depending on viewing angle, multiple views of the same bridge are given to give a more complete picture.

\indent A completely different splash bridge is created when offsets of the two galaxies is increased so that neither galactic centre is in contact with the other's gas disc.  When the centres of counter-rotating discs collide their density profiles line up so that the azimuthal momentum of each gas particle left behind is minimal.  When the offset is increased to 21 kpc the mismatch of density profiles and rotation velocity leads to an asymmetrical bridge with a layered structure.  Interestingly these high offset collisions can still remove a significant amount of dense gas from one galaxy depending on the initial orientation of the two discs.  Gas seen in the clumps near each galaxy are generally dense material coupled with more diffuse gas of the other galaxy.  This seems to be true for every offset run in that the collisions are separated out layers corresponding to combinations of ISM phases involved.

\indent The temperatures at 30 Myr are very similar to what was seen in the 500 pc offset runs.  The counter rotation leads to cooler clouds at 30 Myr than the equal sense rotation.  This is likely caused by the equal sense rotation adding to the cloud collision velocities and therefore the post shock temperature.  With a large offset and counter rotating discs, the rotation will mostly move together and not contribute to the over all impact velocity.

\subsection{Taffy-Like Run}

\begin{figure*}

\caption{At 30 Myr after the collision with a 7 kpc offset.  \textbf{a)} Only gas $< \SI{e7} m^{-3}$ and \textbf{b)} Only gas $> \SI{e7} m^{-3}$.  \textbf{c)} Temperatures and  \textbf{d)} densities of CII emitting clouds.  See  \cref{gascooling} for more details.%
     }%
     
     \begin{center}

        \subfigure[]{%
          \label{fig:7kpc30myrhtld}
          \includegraphics[width=0.4\textwidth]{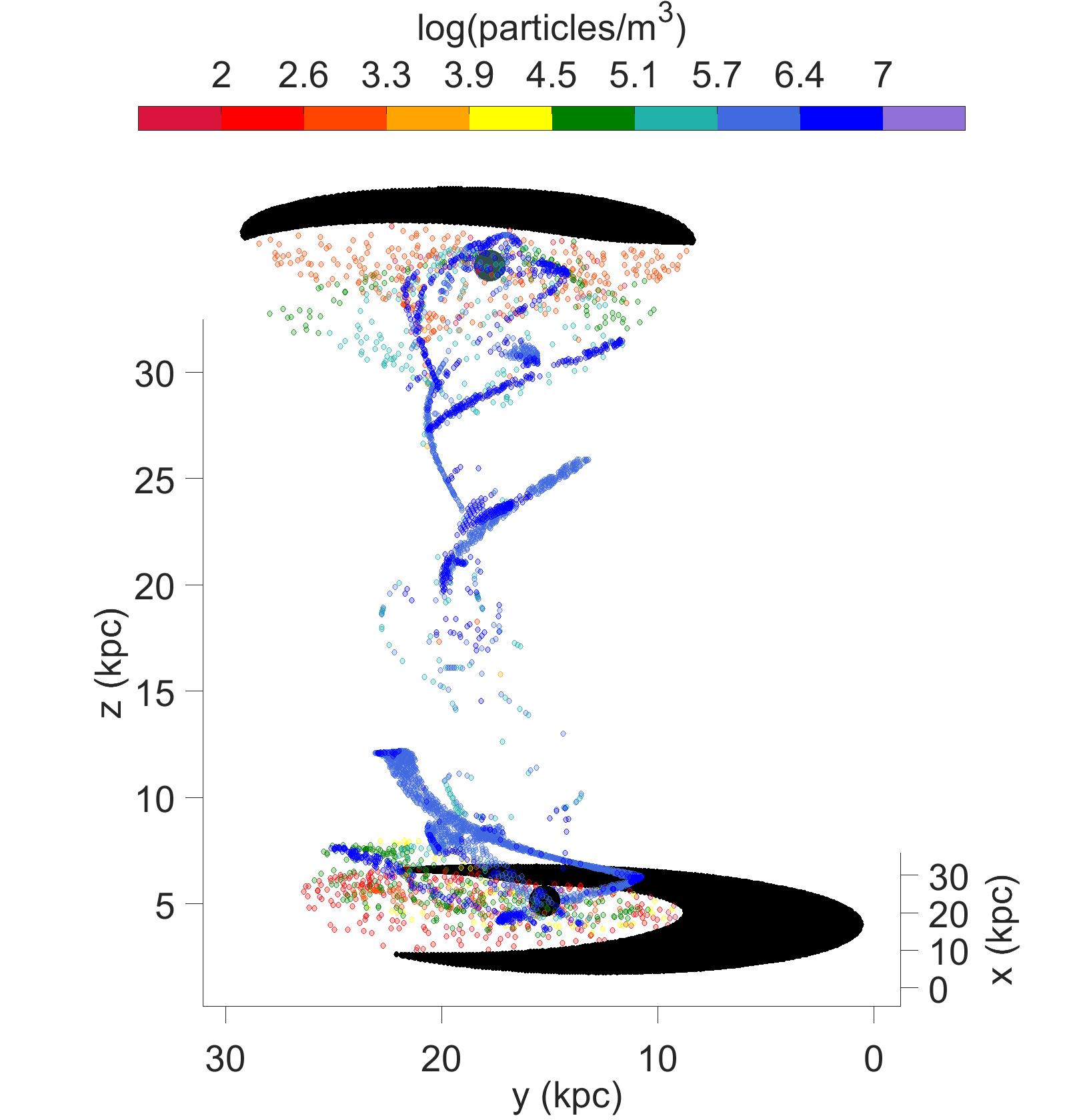}
        }
                \subfigure[]{%
          \label{fig:7kpc30myrlthd}
          \includegraphics[width=0.4\textwidth]{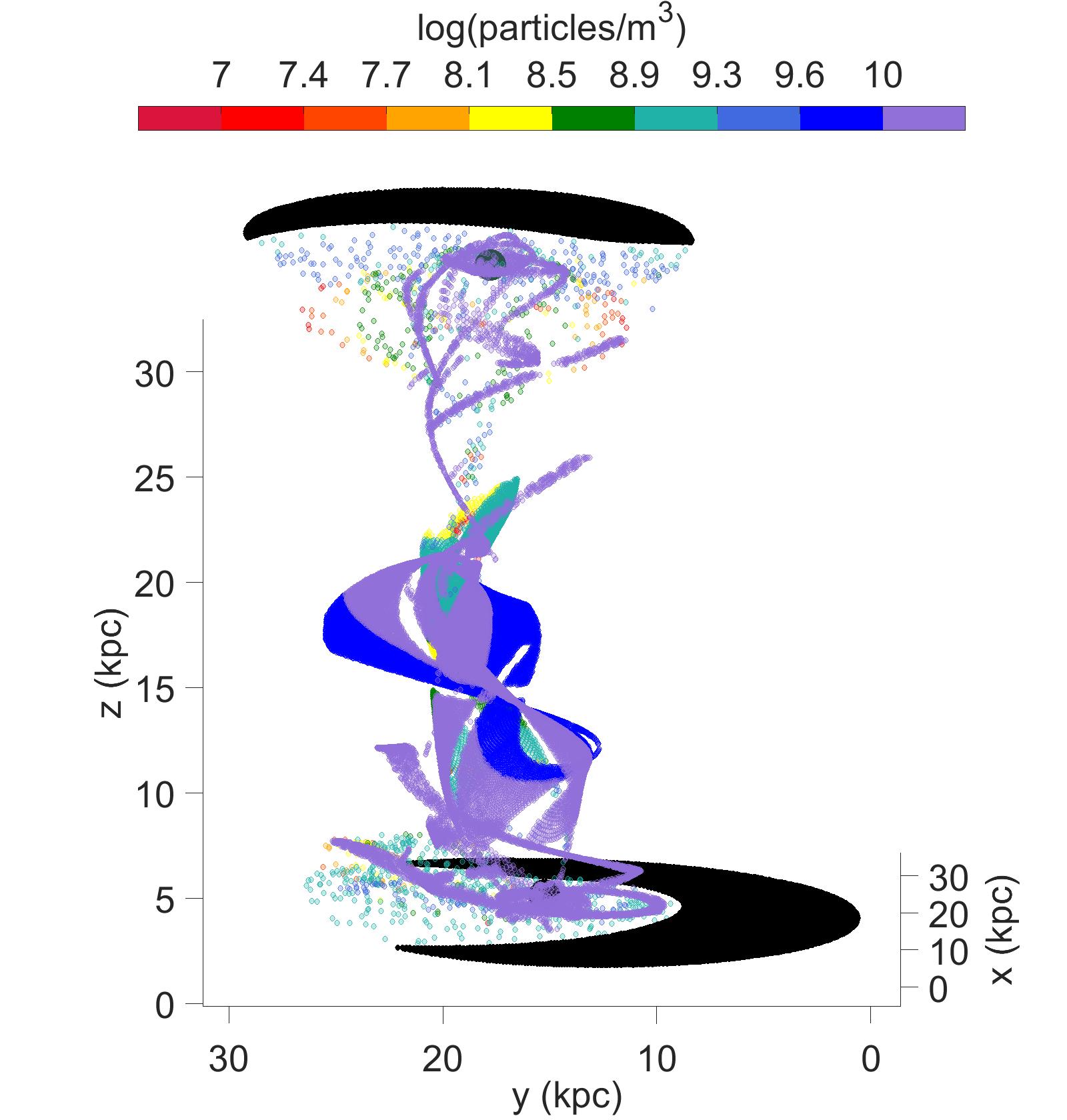}
        }\\ 
                \subfigure[]{%
          \label{fig:CIItemp7kpc30myrlthd}
          \includegraphics[width=0.4\textwidth]{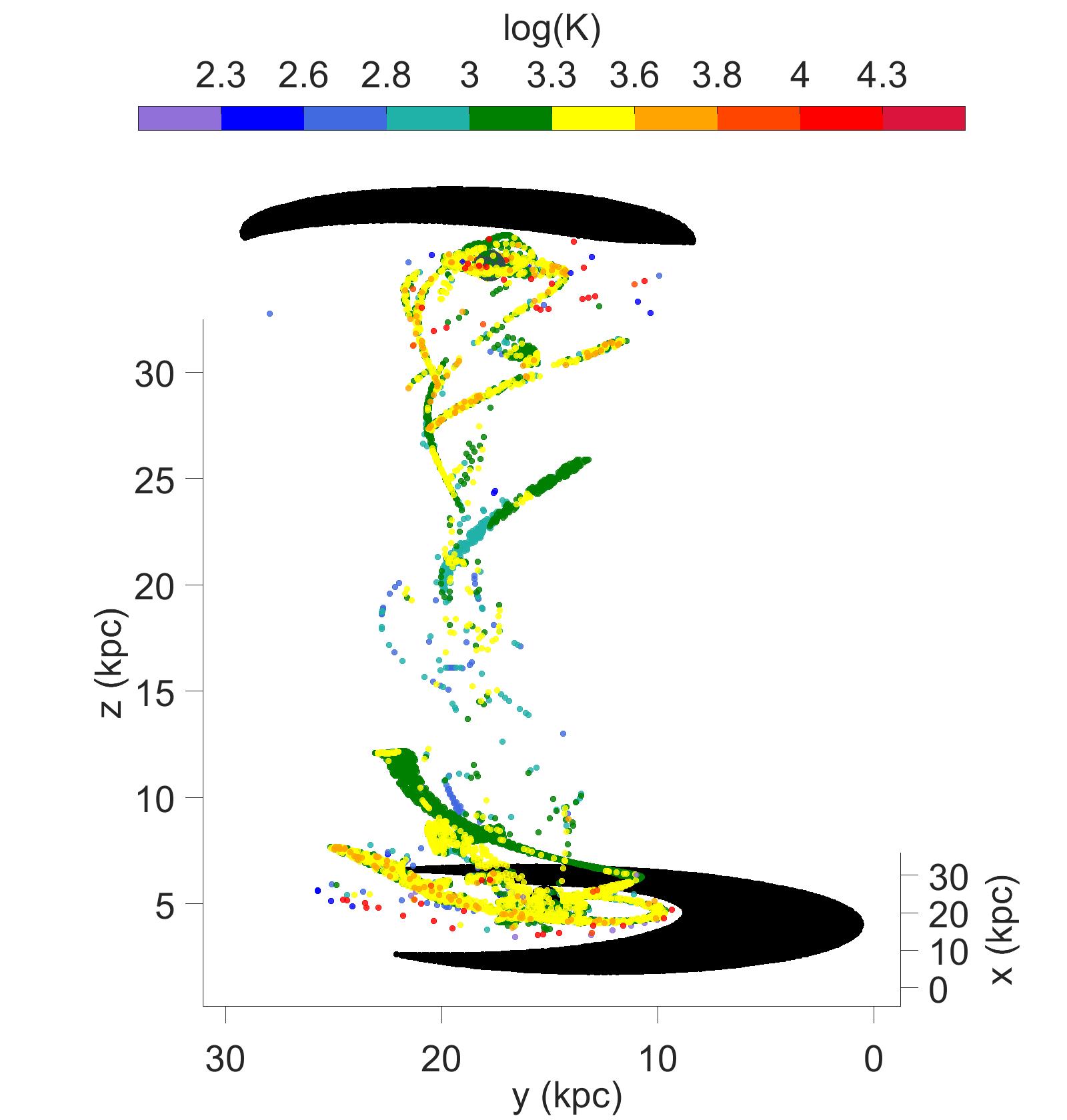}
        }
                \subfigure[]{%
          \label{fig:CIIdensity7kpc30myrlthd}
          \includegraphics[width=0.4\textwidth]{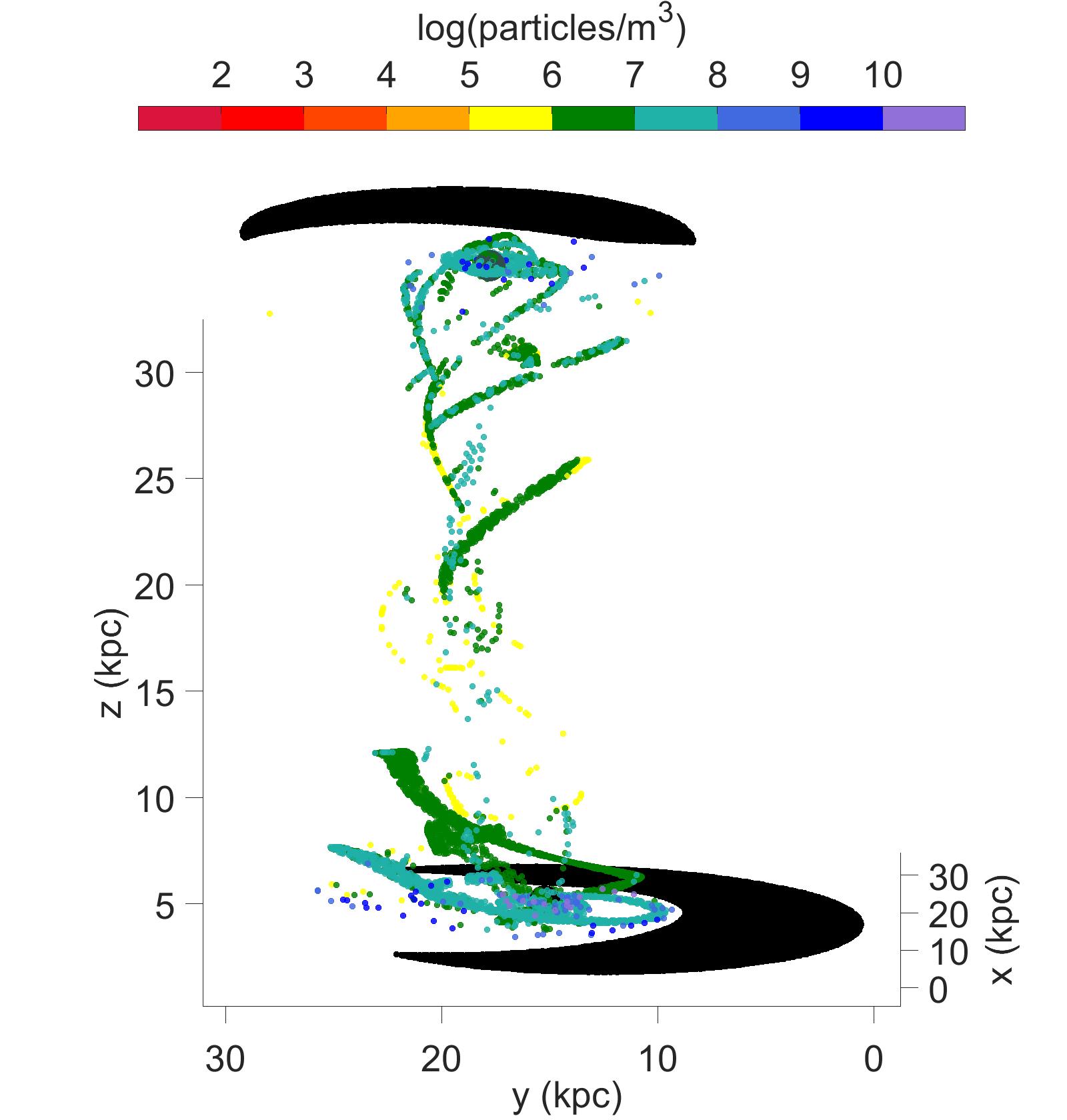}
        }\\ 

    \end{center}
    
  \label{fig:taffy7kpc30myrsubfigures}
\end{figure*}

The 7 kpc offset run shown in \cref{fig:taffy7kpc30myrsubfigures} is the best candidate for a collision like the Taffy galaxy system.  This runs shows very complicated gas structures in the bridge, with asymmetrical distributions.  These are like those observed and previously not well understood.  

\indent A gas mass of order \SI{e9} \msun of H\tus{2} is removed from the smaller UGC 12915-like galaxy.  Diffuse hot hydrogen gas spans most of the bridge and though hard to tell from the figures is twisted up in long filaments.  \Cref{fig:taffy7kpc30myrsubfigures} b) shows a large amount of the \HI gas residing in a large twisted sheet extending from G1.  This twisted sheet structure does not seem to extend much further passed the systems center of mass.  This is because the density in G2's initial disc is about 20$\%$ less dense.  This means none of the \HI from G2 is carrying enough momentum to be left near G2. Had G2 been more dense, this bridge would be skewed more towards G2 than G1.

\indent Though much of the gas has cooled to 150 K and is not shown in the temperature figures, there are a significant number of CII emitting clouds remaining at 30 Myr.  None of the gas remains at temperatures high enough to create the observed X-ray emission.  These complicated structures would likely lead to a great amount of turbulence which would reheat a fraction of this gas to higher temperatures.  There is also a significant amount of \hh removed from each galaxy which is left as long twisted filaments  The layering of the collided gas found in the 21 kpc offset run is also see here.  However, the smaller 7 kpc offset has lead to these layers being spread out more in the vertical direction.

\begin{figure*}

\caption{Fractional number of clouds out of total that have collided versus time, giving a sense of how cooling is develops.  In \textbf{blue} are clouds with density > \SI{2e9} \dunit, \textbf{red} are clouds T > \SI{e6} K, \textbf{orange} are clouds cooling through permitted H and He lines, \textbf{green} are clouds emitting CII, \textbf{black} are those that meet the criteria to adiabatically expand (see \cref{gascooling}), and \textbf{purple} are those that have reached 150 K and ceased all cooling.  Each frame is for a different collision, \textbf{a)} 500pc offset with counter rotation,  \textbf{b)} 21 kpc offset with counter rotation, \textbf{c)} 500 pc with equal rotation and \textbf{d)} 21 kpc with equal rotation.%
     }%
     \label{fig:cloudhistories}
     \begin{center}
                \subfigure[]{%
          \includegraphics[width=0.4\textwidth]{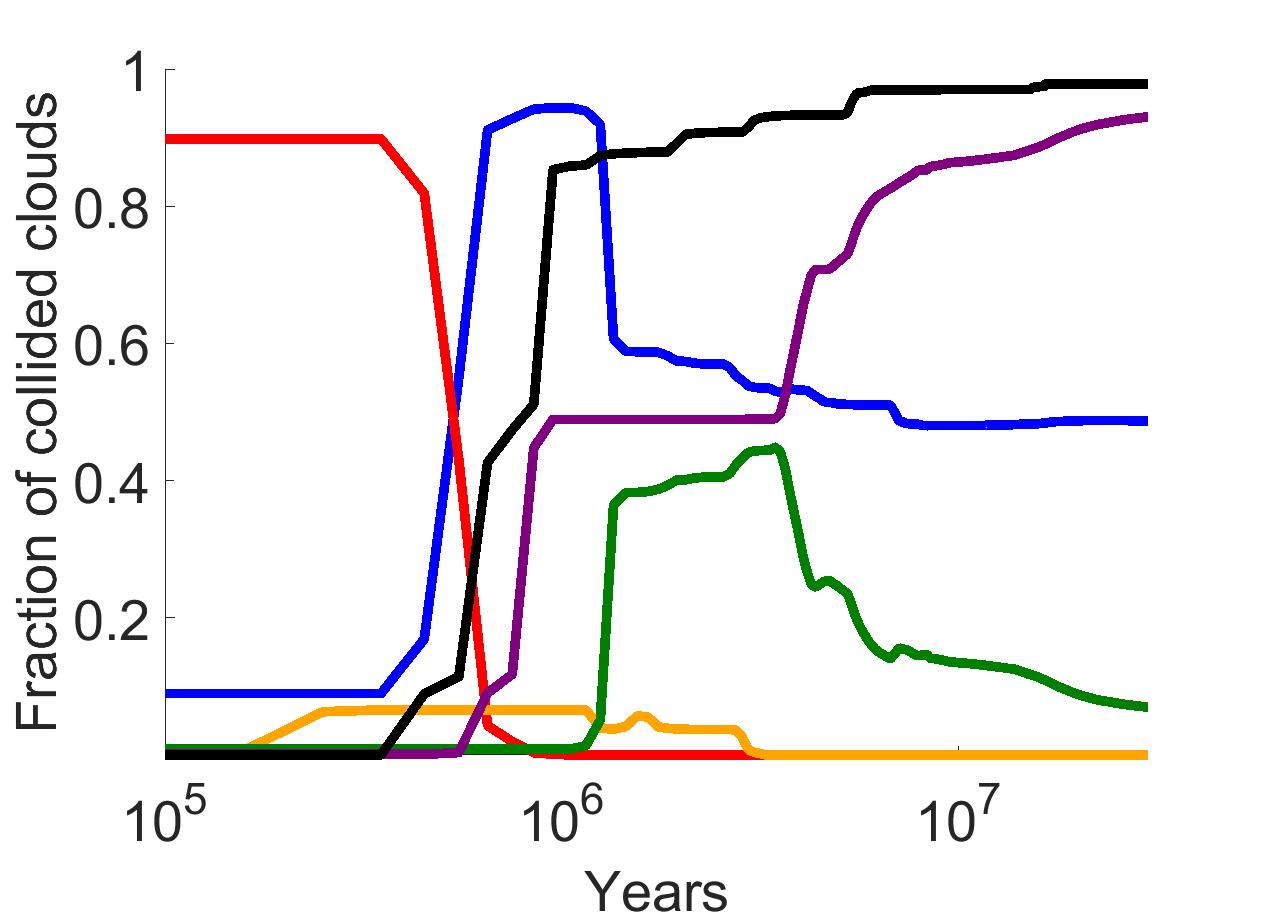}
        }
                \subfigure[]{%
          \includegraphics[width=0.4\textwidth]{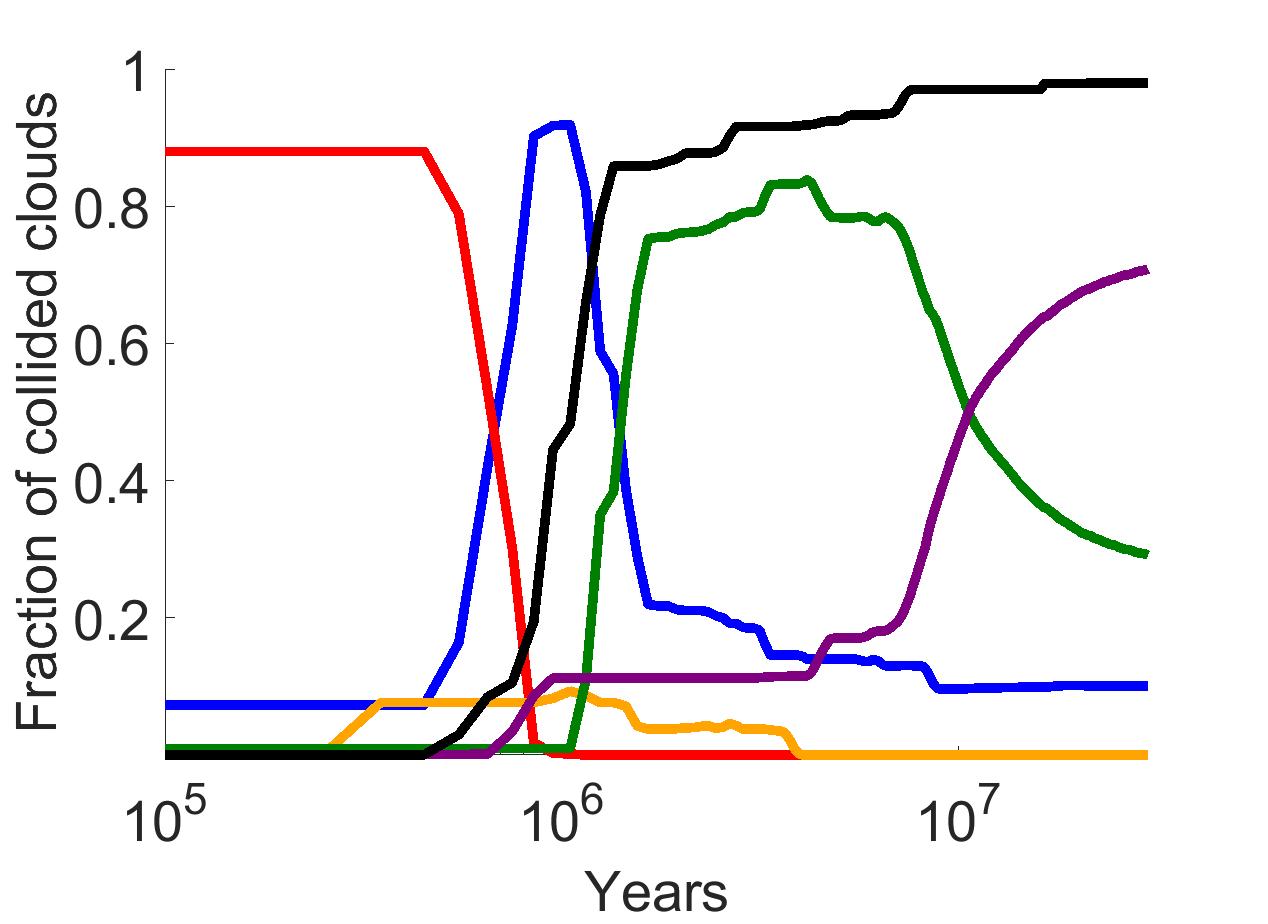}
        }\\ 
                \subfigure[]{%
          \includegraphics[width=0.4\textwidth]{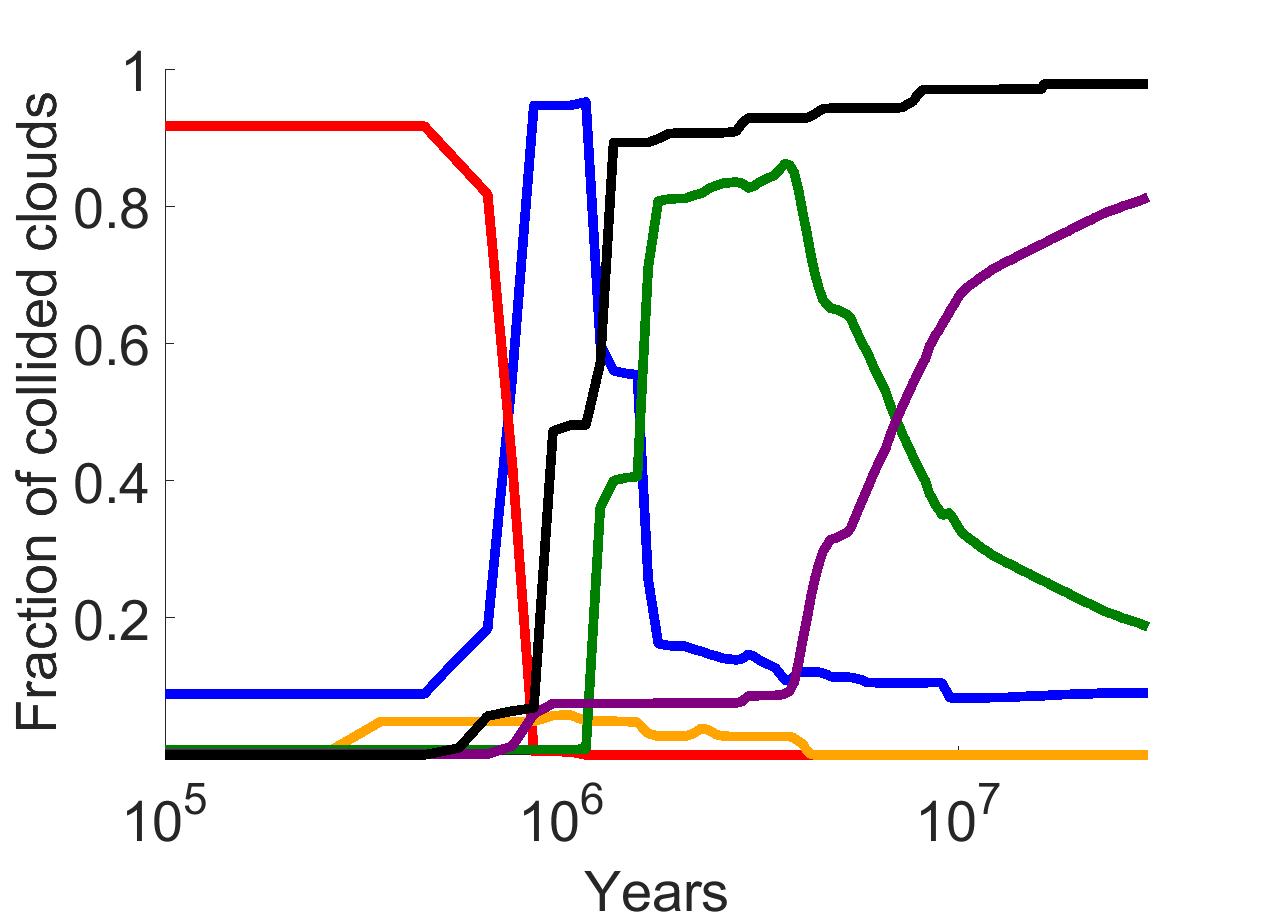}
        }
                \subfigure[]{%
          \includegraphics[width=0.4\textwidth]{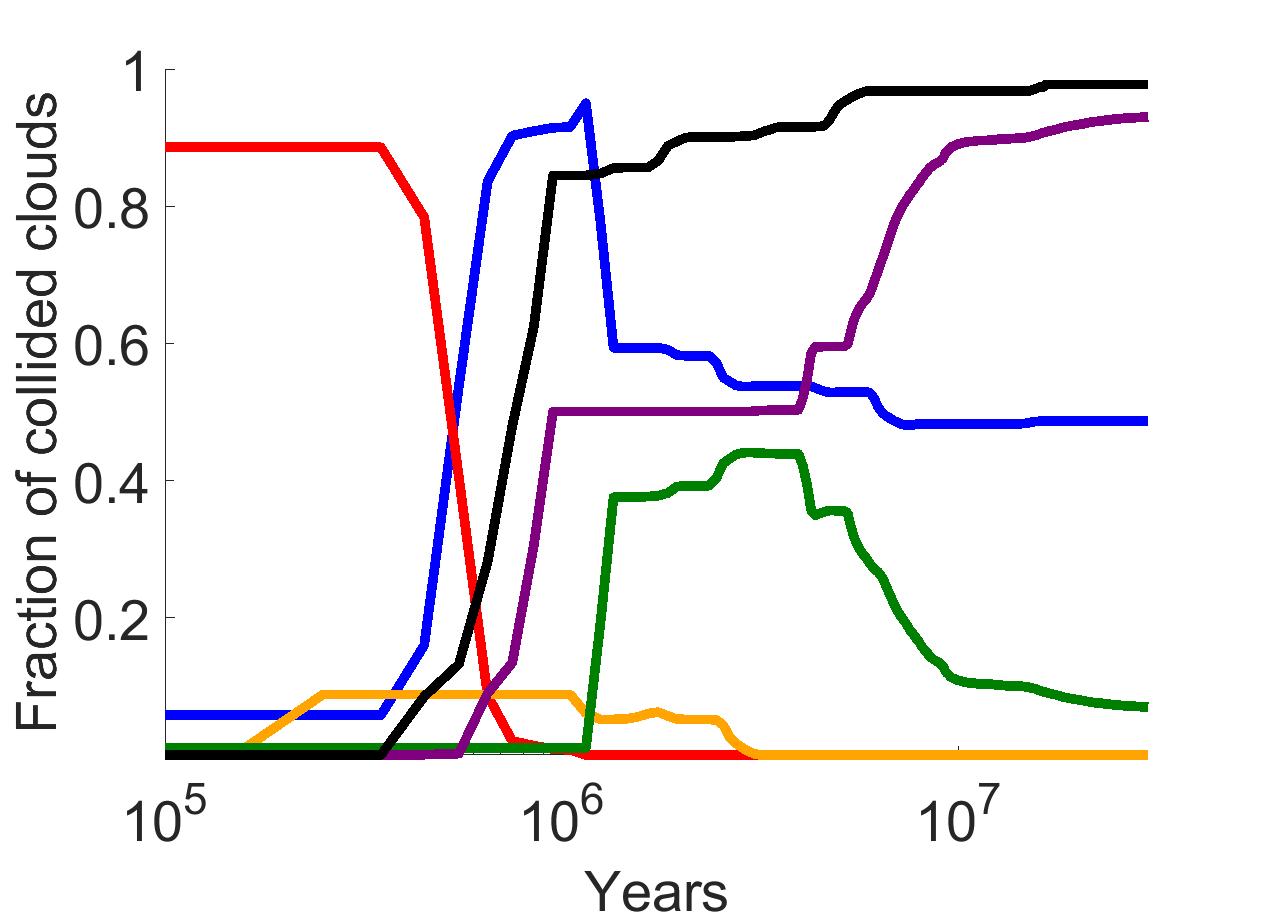}
        }\\ 
    \end{center}
    
\end{figure*}

\clearpage
\subsection{Cloud Populations}
\label{cloudpops}

\Cref{fig:cloudhistories} shows the fractional number of clouds that are within temperature or density ranges of interest over time for various runs.  One difference between the four runs, is that the highest fraction of clouds emitting CII at any particular time occurs in the 500 pc offset equal rotation collision and 21 kpc offset counter rotation collision.  These two runs have the lowest cloud collision velocities due to the rotation not contributing much.  This leads to lower overall post-shock temperatures.  The lower post shock temperatures mean there is less cloud contraction during permitted line cooling down to 20,000 K.  This allows CII cooling more room before it his a critical density of \SI{2e9} \dunit and therefore more CII cooling may occur.  The opposite is seen in the 500 pc offset with counter rotation and 21 kpc offset equal rotation runs.  Now due to the geometry of the two colliding discs their rotation has a maximal effect.  This leads to higher velocity cloud collisions and therefore higher post shock temperatures.  Since many more clouds will reach a density of \SI{2e9} \dunit during the line cooling phase, we see a suppressed number of clouds able to emit CII at any given time.  Something similar that occurs across all four runs is the steady growth of cloud adiabatic expansion.  By 30 Myr nearly every cloud meets the criteria of $nT > 10^8 m^{-3}K $ for adiabatic expansion.  The run with the largest fraction of high shock velocities is the 21 kpc equal-sense rotation collision.  This is because when the offset is small and the discs are counter rotating, the centres of the discs are not rotating at significant velocities leading to a gradient in shock strengths.

\subsection{Paths of \hh Formation}

\indent The Taffy galaxies have a very large amount of molecular material in their bridge as observed by the Spitzer Space Telescope \citep{peterson12}. One possible way to remove \hh from the discs is through the removal of the entire \hh cloud via ram pressure.  The likelihood of direct \hh cloud-cloud collisions is too small to explain a huge amount of molecular material.  However, as the model results have shown, collisions between gas clouds with densities within an order of magnitude of each other can result in the removal of both from their host galaxy.  This may, together with subsequent molecular formation, explain the massive amounts of \hh observed in the Taffy bridge as well as the anti-symmetric molecular region just below the disc of UGC 12915.  With such high energies involved in the shocks it is possible that pre-existing \hh would be completely disassociated. On the contrary, the dense cloud cores may be protected from the initial strong shocks which are weakened climbing the density gradient in the outer cloud envelope. The degree to which the outer \HI envelopes provide protection requires further study.  

\indent \citet{gui09} propose that the \hh does not need to survive the strong turbulent shocks involved in galaxy collisions to be observed several Myr after the collision.   They posit that with in a relatively short time \hh could form in the compressed fragmented \HI clouds left behind the collisional shocks.  Turbulence from the shocks is thought to drive the fragmentation and compression, which in turn can create an environment conducive to \hh formation.  This idea can be estimated by tracking the post-shock temperature and density of the clouds and comparing with the \hh formation time,

\begin{equation} t_{H_{2}} [yr] \simeq \SI{7e5} f_{dust} (\sfrac{ \num{1e5}  [K cm^{-3}]}{P_{th}})^{0.95}
\label{eq:h2formationtime}
\end{equation}

\citet{gui09} use \cref{eq:h2formationtime} to estimate the timescale for \hh formation in the Stephan' Quintet shock region.  $P_{th}$ is the thermal gas pressure and $f_{dust}$ is the dust mass fraction remaining in the shocked hydrogen.  The formation time of \hh scales roughly as the inverse of the gas density at \SI{e4} K.  By their estimation gas densities of \SI{e6} m\textsuperscript{-3} will be able to form \hh in 5 Myr.  The Taffy Galaxies collision age of 30 Myr will allow gas densities of \SI{e5} m\textsuperscript{-3} enough time to cool and form H\tus{2} (also see \citet{braine03}).  A density of \SI{e5} m\textsuperscript{-3} is over an order of magnitude less than most of the densities of the \HI disc in the collision remnants, so \hh is likely reformed even in regions where it does not survive the initial shocks.

\section{Discussion - Some Systems of Interest}

\indent The results above predict a large amount of the ISM in some collisions is splashed out, with complicated spatial distributions.  The unique morphologies produced in these models make for very promising comparisons to future observations of splash bridges.  The Taffy system was the original focus of our work, but as the models developed it became clear that they resemble a number of other systems.

\indent For example, several systems of galaxies in the Living HI Rogues Gallery contain \HI bridge features that might be explained through an inelastic disc collision \citep{hibbard01}.  These include: Arp 271, The Cartwheel Galaxy, ESO 353-G014/17 and ESO 138-IG029/30 (The Sacred Mushroom) which each contain a bridge of \HI with no strong signs of tidal tails, suggesting direct disc-disc collisions. 

\indent Optical images of Arp 194 show many clumps left in a bridge structure between the two galaxies which could be produced from an inelastic collision. Arp 271, for example, has a similar appearance, with filaments appearing to connect to parts of the vigorous spiral arms. The radio observations show extensive, but not well resolved HI in the system. In the models we see large clumps of dense material between the colliding discs left behind in filaments similar to what is seen in the Arp 194 system.  These large clumps are pulled out in small to moderate disc offsets and come from spiral arms of one galaxy colliding with the other galaxy's denser central \HI disc or spiral arms.

\indent The \HI disc that is formed by a low offset collisions is displaced far from either galaxy.  This could appear dark for a considerable time before star formation is initiated.  Thus, splash bridges may offer another explanation for the creation of VIRGO21HI and similar extragalactic clumps of dark hydrogen.

\section{Conclusions}

\indent Models with multiphase interstellar media, and varying distances between the centres of the discs at impact, show that many different splash bridge morphologies can be created.  In all models, regardless of offset or initial ISM distribution, every phase of ISM is displaced to a differing degrees based on its initial density.    The initially hot diffuse gas forms an independent structure from that of the warm diffuse, neutral, and cold \HI. ISM elements that are randomly mixed in the initial disc are entirely separated from nearby neighbors if they are of differing density.  The magnitude of the displacement varies from a few hundred parsecs to several kpc. 

\indent In small disc-disc offset cases all of the \HI gas from the smaller galaxy is removed by an equivalent amount of \HI from the other.  This leaves a flat central bridge disc (CBD) containing a large fraction of the \HI gas mass of both galaxies.  The \HI will be shocked to a few million degrees and will cool within a few million years.  Hydrogen of densities less than \SI{e6} \dunit, and shock heated to temperatures of \SI{e6} K will not entirely cool by 30 Myr. Twisted filaments of dense H\tus{2} develop across much of the distance between the two separating galaxies.  Relative rotations of these systems determines the magnitude of winding up of each feature.  If the rotations at small offsets are counter to one another the gas will lose much of its angular momentum.  This results in less rotation in the collided gas, and subsequently allows for more radial collapse of the CBD.  If the galaxies are rotating in a similar sense, much of the clouds angular momentum will be preserved and structures will swirl as they evolve and will not contract radially.  

\indent As the offset between the two gas discs grows beyond 10\% of the galactic radius we begin to lose the flat \HI CBD.  This is due to the density profiles of each galaxy no longer aligning with one another.  This leaves the dense \HI with more post-collision vertical momentum, giving the impacted gas a large spread of vertical distance from its host galaxy.  Spiral arms containing H\tus{2} clouds are removed from their galaxy by varying degrees and retain most of their spiral structure, leading to a long filaments throughout the bridge.  The affect of rotation at large offsets is inverted from what is seen at small offsets.  Counter rotating discs create a more twisted structure spanning from galaxy to galaxy.  Equal sense rotations still show some twisting, but the \HI is spread out much more into flat sheets. The initially hot diffuse gas no longer spans the entire \HI bridge and is not continuous between the galaxies.  The general picture is that in moving from small to large offsets we see a shrinking CBD, and more layered twisted sheets of gas.  

\indent The splashed interstellar gas is relevant to a range of ring galaxies and peculiar galaxies that show large amounts of dark HI. As the database of \HI observations of these interacting galaxies increases, more model comparisons will be possible.  Inelastic gas disc collisions can provide further insight into systems that only sideswipe each other.  Since the morphologies of the bridges are very sensitive to impact offsets, the observed morphology may be able to distinguish a central collision from a glancing one. With a significant amount of gas removed in these collisions, there are opportunities for the creation of dark galaxies.

\indent  Each feature of these bridges is a direct result of conservation of momentum between gas elements of different column density in the different phases of ISM.  Complicated structures, such as twisted sheets of \HI and flat gas discs at the bridge centre, are similar to those shown above and seen in observations. The treatment of inelastic gas collisions likely plays a significant role in shaping the morphology of directly colliding galaxy systems.  

\indent The modeling code described above is simple enough to be run on a personal computer but it has ability to replicate complicated structures from two multi-phase gas discs. In future work we intend to use it to: study tilted disc-disc collisions, make more detailed comparisons to observations of Taffy-like galaxies, to model likely locations of star-forming regions, and possibly to include shock distribution functions for modeling collisions between clouds with clumpy internal structure.

\indent We would like to thank Philip N. Appleton for conversations on our models and the Taffy Galaxies.

\bibliographystyle{mnras}
\bibliography{main}

\begin{thebibliography}{}
\makeatletter
\relax
\def\mn@urlcharsother{\let\do\@makeother \do\$\do\&\do\#\do\^\do\_\do\%\do\~}
\def\mn@doi{\begingroup\mn@urlcharsother \@ifnextchar [ {\mn@doi@}
  {\mn@doi@[]}}
\def\mn@doi@[#1]#2{\def\@tempa{#1}\ifx\@tempa\@empty \href
  {http://dx.doi.org/#2} {doi:#2}\else \href {http://dx.doi.org/#2} {#1}\fi
  \endgroup}
\def\mn@eprint#1#2{\mn@eprint@#1:#2::\@nil}
\def\mn@eprint@arXiv#1{\href {http://arxiv.org/abs/#1} {{\tt arXiv:#1}}}
\def\mn@eprint@dblp#1{\href {http://dblp.uni-trier.de/rec/bibtex/#1.xml}
  {dblp:#1}}
\def\mn@eprint@#1:#2:#3:#4\@nil{\def\@tempa {#1}\def\@tempb {#2}\def\@tempc
  {#3}\ifx \@tempc \@empty \let \@tempc \@tempb \let \@tempb \@tempa \fi \ifx
  \@tempb \@empty \def\@tempb {arXiv}\fi \@ifundefined
  {mn@eprint@\@tempb}{\@tempb:\@tempc}{\expandafter \expandafter \csname
  mn@eprint@\@tempb\endcsname \expandafter{\@tempc}}}

\bibitem[\protect\citeauthoryear{{Appleton} et~al.,}{{Appleton}
  et~al.}{2015}]{appleton15}
{Appleton} P.~N.,  et~al., 2015, \mn@doi [\apj] {10.1088/0004-637X/812/2/118},
  \href {https://ui.adsabs.harvard.edu/#abs/2015ApJ...812..118A} {812, 118}

\bibitem[\protect\citeauthoryear{{Braine} et~al.,}{{Braine}
  et~al.}{2003}]{braine03}
{Braine} J.,  et~al., 2003, in SF2A-2003: Semaine de l'Astrophysique Francaise.
  p.~231

\bibitem[\protect\citeauthoryear{{Condon}, {Helou}, {Sanders}  \&
  {Soifer}}{{Condon} et~al.}{1993}]{condon93}
{Condon} J.~J.,  {Helou} G.,  {Sanders} D.~B.,   {Soifer} B.~T.,  1993, \mn@doi
  [\aj] {10.1086/116549}, \href
  {https://ui.adsabs.harvard.edu/#abs/1993AJ....105.1730C} {105, 1730}

\bibitem[\protect\citeauthoryear{{Davies}, {Davies}  \& {Disney}}{{Davies}
  et~al.}{2008}]{davies08}
{Davies} J.~I.,  {Davies} J.~I.,   {Disney} M.~J.,  2008, in Dark Galaxies and
  Lost Baryons. pp 7--16, \mn@doi{10.1017/S1743921307013786}

\bibitem[\protect\citeauthoryear{Draine}{Draine}{2011}]{drainephysics}
Draine B.~T.,  2011, The Physics of the Interstellar and Intergalactic Medium.
Princeton University Press, 41 Williams Street, Princeton University Press, 6
  Oxford Street, Woodstock, Oxfordshire OX20 1TW

\bibitem[\protect\citeauthoryear{{Gao}, {Zhu}  \& {Seaquist}}{{Gao}
  et~al.}{2003}]{gao03}
{Gao} Y.,  {Zhu} M.,   {Seaquist} E.~R.,  2003, \mn@doi [\aj] {10.1086/378611},
  \href {https://ui.adsabs.harvard.edu/#abs/2003AJ....126.2171G} {126, 2171}

\bibitem[\protect\citeauthoryear{{Guillard}, {Boulanger}, {Pineau Des
  For{\^e}ts}  \& {Appleton}}{{Guillard} et~al.}{2009}]{gui09}
{Guillard} P.,  {Boulanger} F.,  {Pineau Des For{\^e}ts} G.,   {Appleton}
  P.~N.,  2009, \mn@doi [\aap] {10.1051/0004-6361/200811263}, \href
  {https://ui.adsabs.harvard.edu/#abs/2009A&A...502..515G} {502, 515}

\bibitem[\protect\citeauthoryear{{Hernquist}}{{Hernquist}}{1990}]{hernquist1990}
{Hernquist} L.,  1990, \mn@doi [\apj] {10.1086/168845}, \href
  {https://ui.adsabs.harvard.edu/\#abs/1990ApJ...356..359H} {356, 359}

\bibitem[\protect\citeauthoryear{{Hibbard}, {van Gorkom}, {Rupen}  \&
  {Schiminovich}}{{Hibbard} et~al.}{2001}]{hibbard01}
{Hibbard} J.~E.,  {van Gorkom} J.~H.,  {Rupen} M.~P.,   {Schiminovich} D.,
  2001, in {Hibbard} J.~E.,  {Rupen} M.,   {van Gorkom} J.~H.,  eds,
  Astronomical Society of the Pacific Conference Series Vol. 240, Gas and
  Galaxy Evolution. p.~657 (\mn@eprint {arXiv} {astro-ph/0110667})

\bibitem[\protect\citeauthoryear{{Nicastro} et~al.,}{{Nicastro}
  et~al.}{2018}]{Nicastro2018}
{Nicastro} F.,  et~al., 2018, \mn@doi [\nat] {10.1038/s41586-018-0204-1}, \href
  {https://ui.adsabs.harvard.edu/\#abs/2018Natur.558..406N} {558, 406}

\bibitem[\protect\citeauthoryear{{Peterson} et~al.,}{{Peterson}
  et~al.}{2012}]{peterson12}
{Peterson} B.~W.,  et~al., 2012, \mn@doi [\apj] {10.1088/0004-637X/751/1/11},
  \href {https://ui.adsabs.harvard.edu/#abs/2012ApJ...751...11P} {751, 11}

\bibitem[\protect\citeauthoryear{{Schombert}, {Wallin}  \&
  {Struck-Marcell}}{{Schombert} et~al.}{1990}]{schombert90}
{Schombert} J.~M.,  {Wallin} J.~F.,   {Struck-Marcell} C.,  1990, \mn@doi [\aj]
  {10.1086/115346}, \href
  {https://ui.adsabs.harvard.edu/#abs/1990AJ.....99..497S} {99, 497}

\bibitem[\protect\citeauthoryear{{Smith} \& {Struck}}{{Smith} \&
  {Struck}}{2001}]{ss01}
{Smith} B.~J.,  {Struck} C.,  2001, \mn@doi [\aj] {10.1086/318766}, \href
  {https://ui.adsabs.harvard.edu/#abs/2001AJ....121..710S} {121, 710}

\bibitem[\protect\citeauthoryear{{Smith}, {Struck}  \& {Pogge}}{{Smith}
  et~al.}{1997}]{smithstruck97}
{Smith} B.~J.,  {Struck} C.,   {Pogge} R.~W.,  1997, \mn@doi [\apj]
  {10.1086/304286}, \href
  {https://ui.adsabs.harvard.edu/#abs/1997ApJ...483..754S} {483, 754}

\bibitem[\protect\citeauthoryear{{Smith}, {Flynn}, {Candlish}, {Fellhauer}  \&
  {Gibson}}{{Smith} et~al.}{2015}]{smith15}
{Smith} R.,  {Flynn} C.,  {Candlish} G.~N.,  {Fellhauer} M.,   {Gibson} B.~K.,
  2015, \mn@doi [\mnras] {10.1093/mnras/stv228}, \href
  {https://ui.adsabs.harvard.edu/#abs/2015MNRAS.448.2934S} {448, 2934}

\bibitem[\protect\citeauthoryear{{Struck}}{{Struck}}{1997}]{struck97simulations}
{Struck} C.,  1997, \mn@doi [The Astrophysical Journal Supplement Series]
  {10.1086/313055}, \href
  {https://ui.adsabs.harvard.edu/#abs/1997ApJS..113..269S} {113, 269}

\bibitem[\protect\citeauthoryear{{Struck}}{{Struck}}{1999}]{struck99gc}
{Struck} C.,  1999, \mn@doi [\physrep] {10.1016/S0370-1573(99)00030-7}, \href
  {https://ui.adsabs.harvard.edu/#abs/1999PhR...321....1S} {321, 1}

\bibitem[\protect\citeauthoryear{{Toomre} \& {Toomre}}{{Toomre} \&
  {Toomre}}{1972}]{toomre72}
{Toomre} A.,  {Toomre} J.,  1972, \mn@doi [\apj] {10.1086/151823}, \href
  {https://ui.adsabs.harvard.edu/#abs/1972ApJ...178..623T} {178, 623}

\bibitem[\protect\citeauthoryear{{Vollmer}, {Braine}  \& {Soida}}{{Vollmer}
  et~al.}{2012}]{vollmer12}
{Vollmer} B.,  {Braine} J.,   {Soida} M.,  2012, \mn@doi [\aap]
  {10.1051/0004-6361/201219668}, \href
  {https://ui.adsabs.harvard.edu/#abs/2012A&A...547A..39V} {547, A39}

\bibitem[\protect\citeauthoryear{Wang, Ferland, Lykins, Porter, van Hoof  \&
  Williams}{Wang et~al.}{2014}]{wang15}
Wang Y.,  Ferland G.~J.,  Lykins M.~L.,  Porter R.~L.,  van Hoof P. A.~M.,
  Williams R. J.~R.,  2014, \mn@doi [Monthly Notices of the Royal Astronomical
  Society] {10.1093/mnras/stu514}, 440, 3100

\makeatother
\end{thebibliography}
\endthebibliography


\appendix
\section{{Code Tests}}
\begin{figure*}

\caption{Testing the code with a 50 kpc offset run so that the galaxies do not interact. \textbf{a)} 0 Myr \textbf{b)} 10 Myr \textbf{c)} 20 Myr \textbf{d)} 30 Myr%
     }%
     \label{fig:vollmertests}
     \begin{center}
                \subfigure[]{%
          \includegraphics[width=0.4\textwidth]{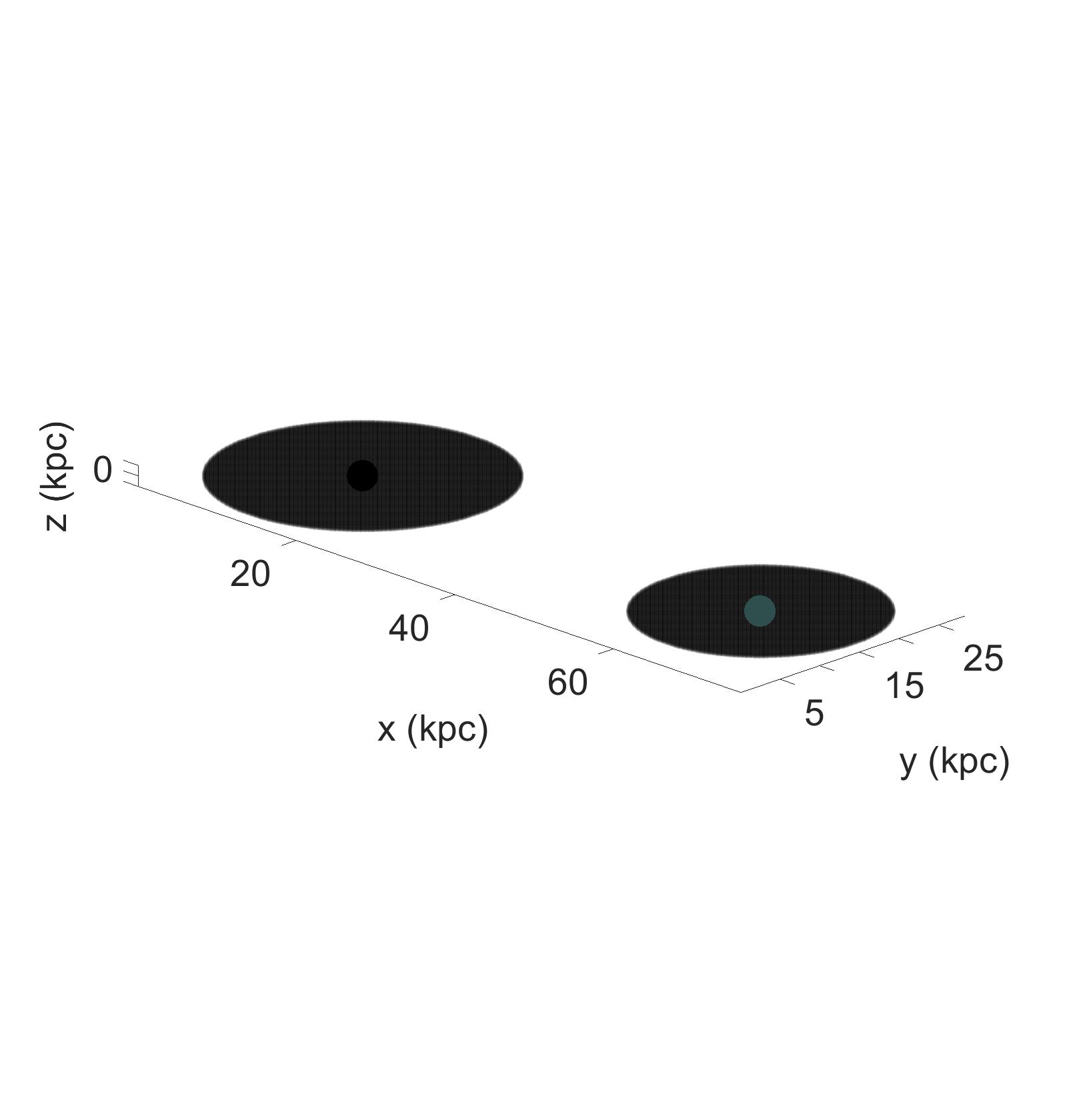}
        }
                \subfigure[]{%
          \includegraphics[width=0.4\textwidth]{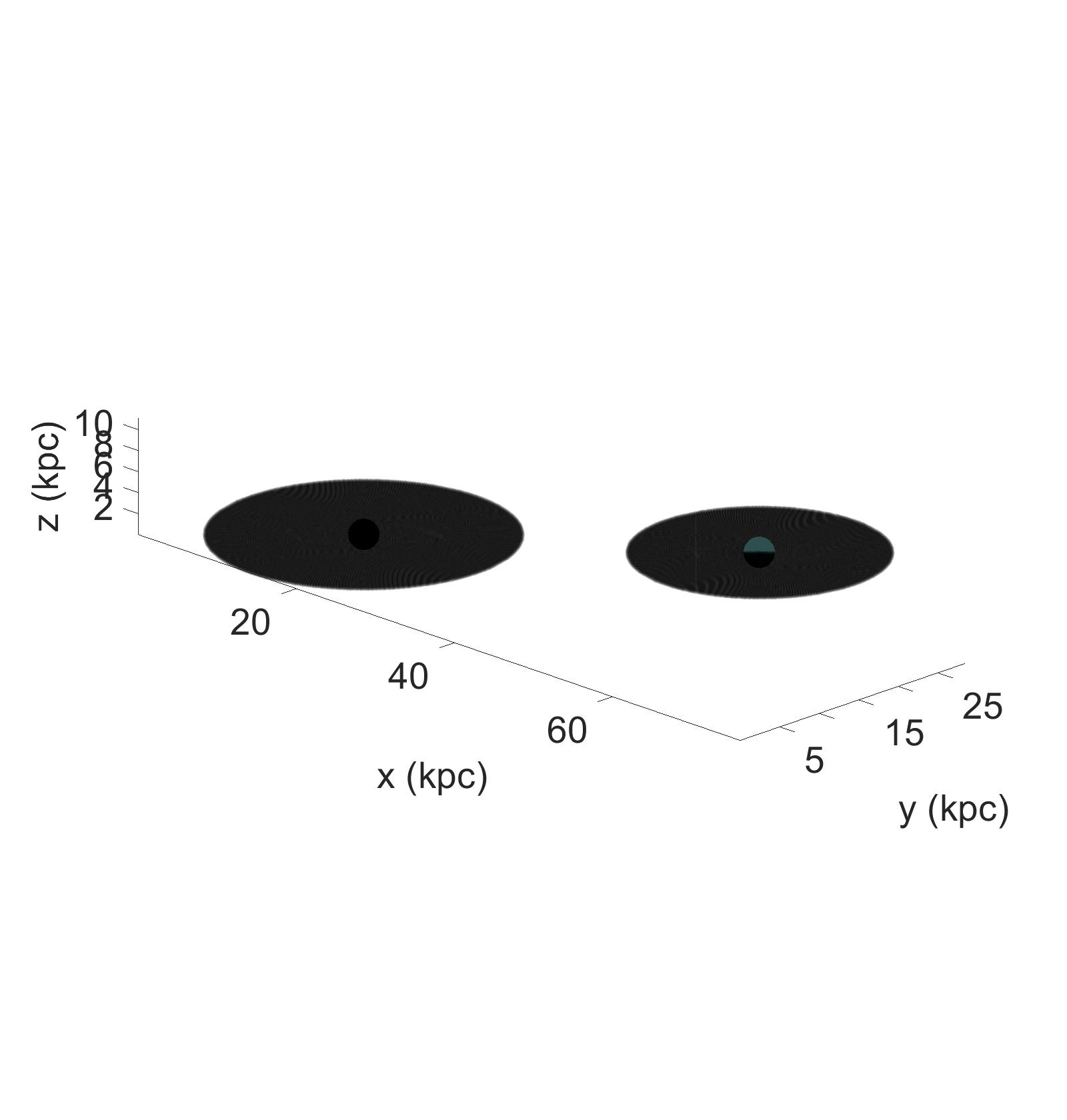}
        }\\ 
                \subfigure[]{%
          \includegraphics[width=0.4\textwidth]{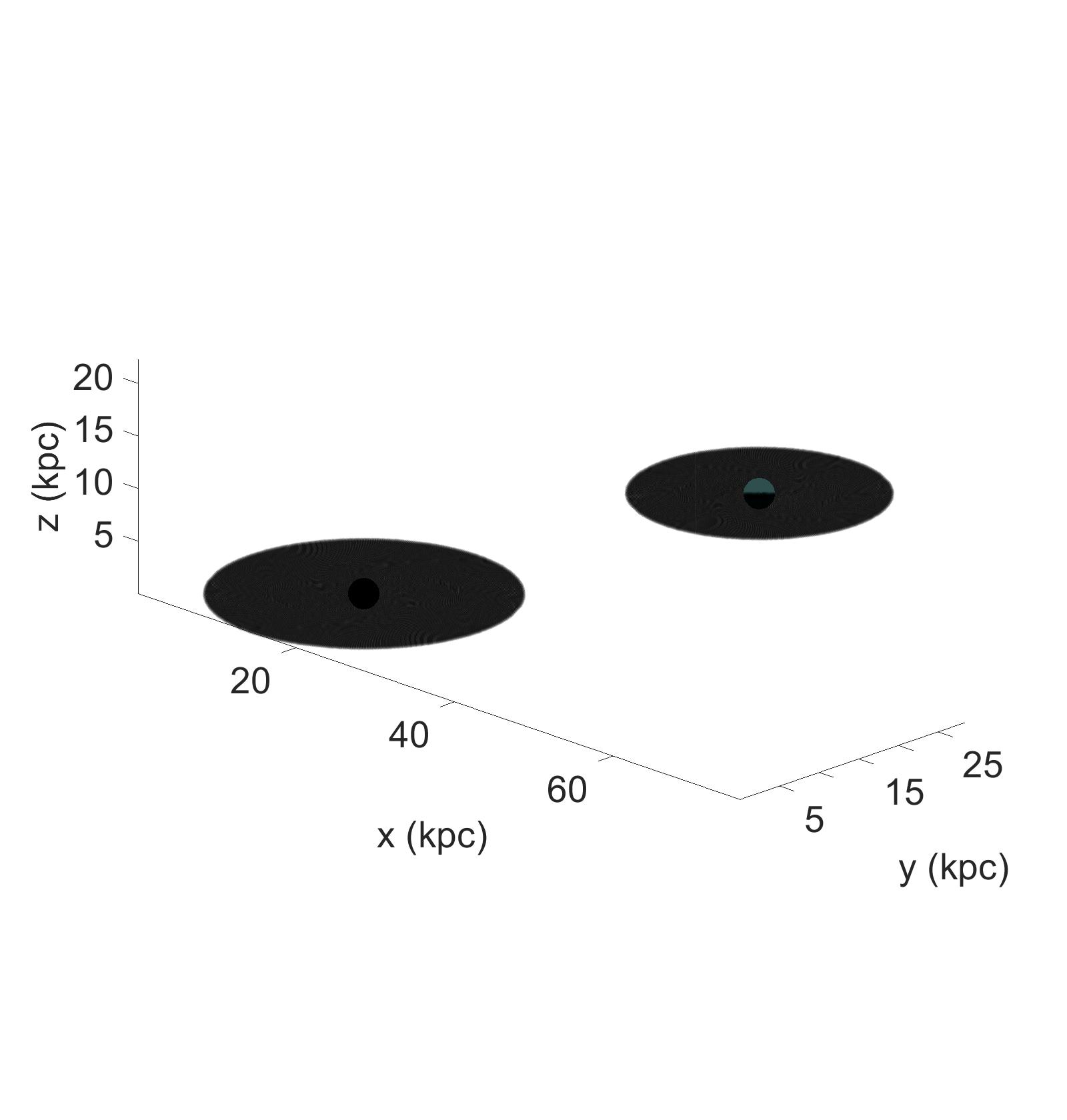}
        }
                \subfigure[]{%
          \includegraphics[width=0.4\textwidth]{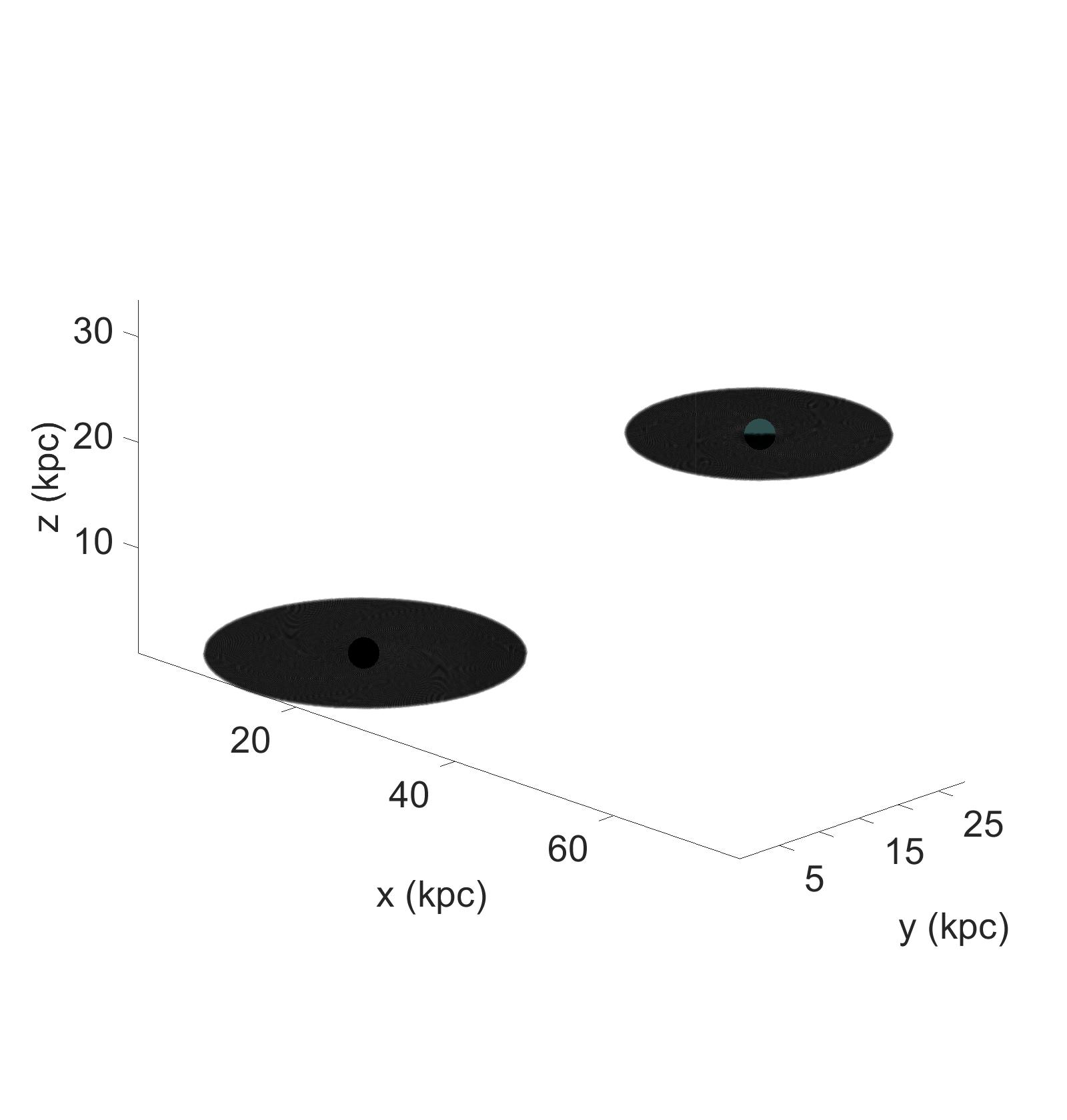}
        }\\ 
    \end{center}
    
\end{figure*}

\begin{figure*}

\caption{A collisionless 500 pc offset run, a) 30 Myr b) 100 Myr c) 200 Myr d) 300 Myr.  The z axis is re-scaled in the edge on frames.  This is to better see how the discs are evolving over a long period of time.}%
     
     \label{fig:vollmertests2}
     
     \begin{center}
                \subfigure[]{%
          \includegraphics[width=0.4\textwidth]{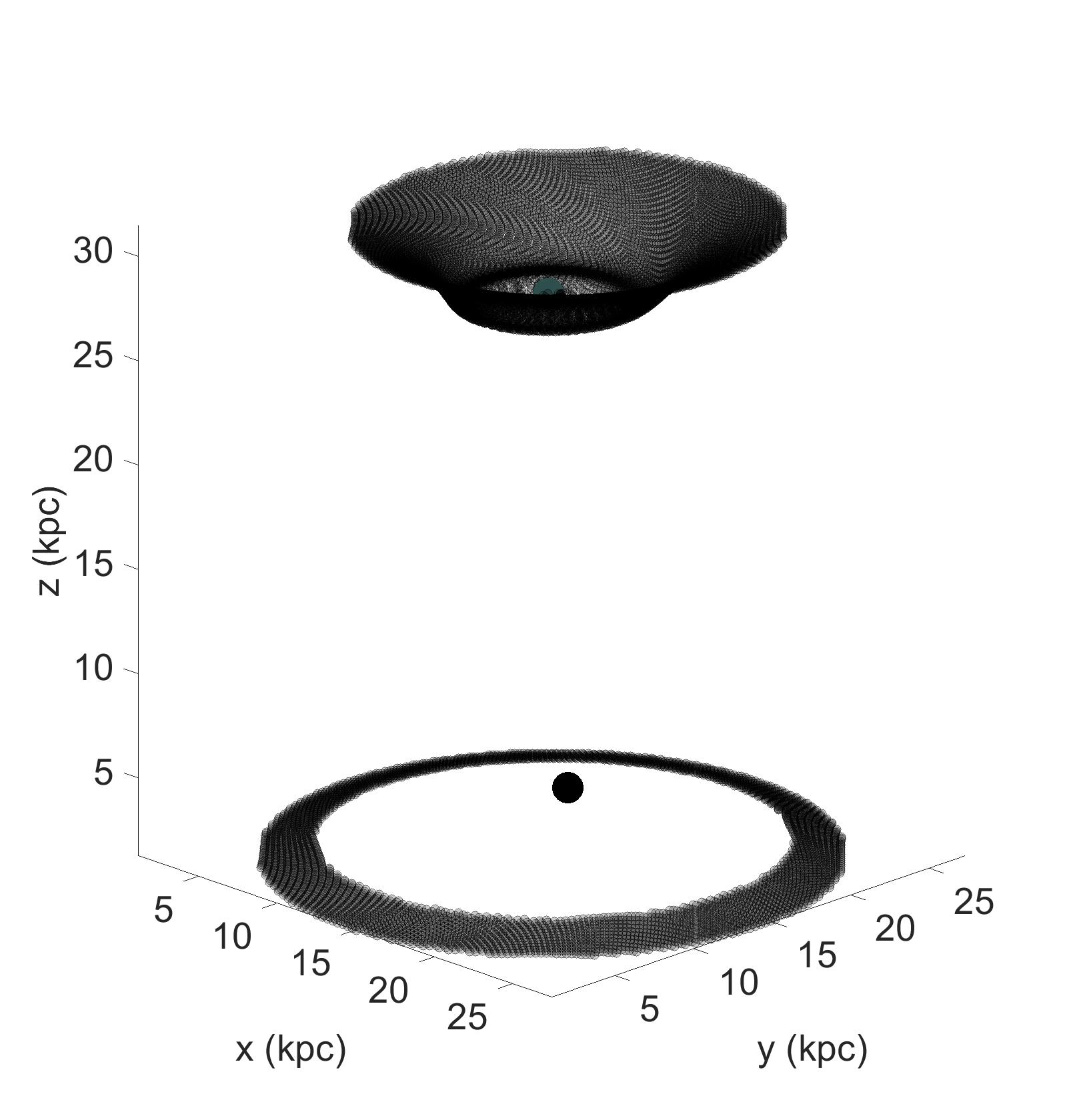}
        }
                \subfigure[]{%
          \includegraphics[width=0.4\textwidth]{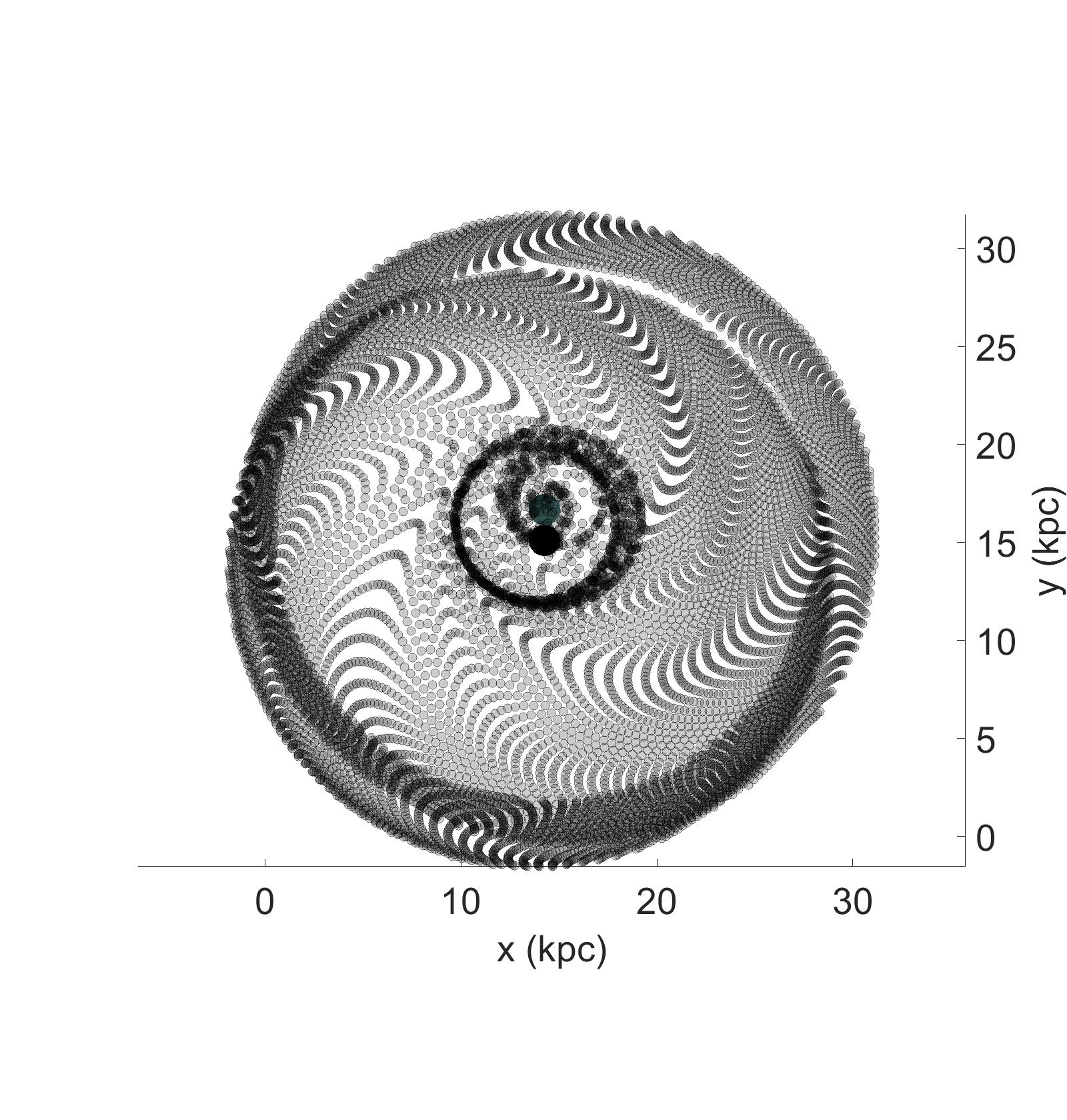}
        }\\ 
                \subfigure[]{%
          \includegraphics[width=0.4\textwidth]{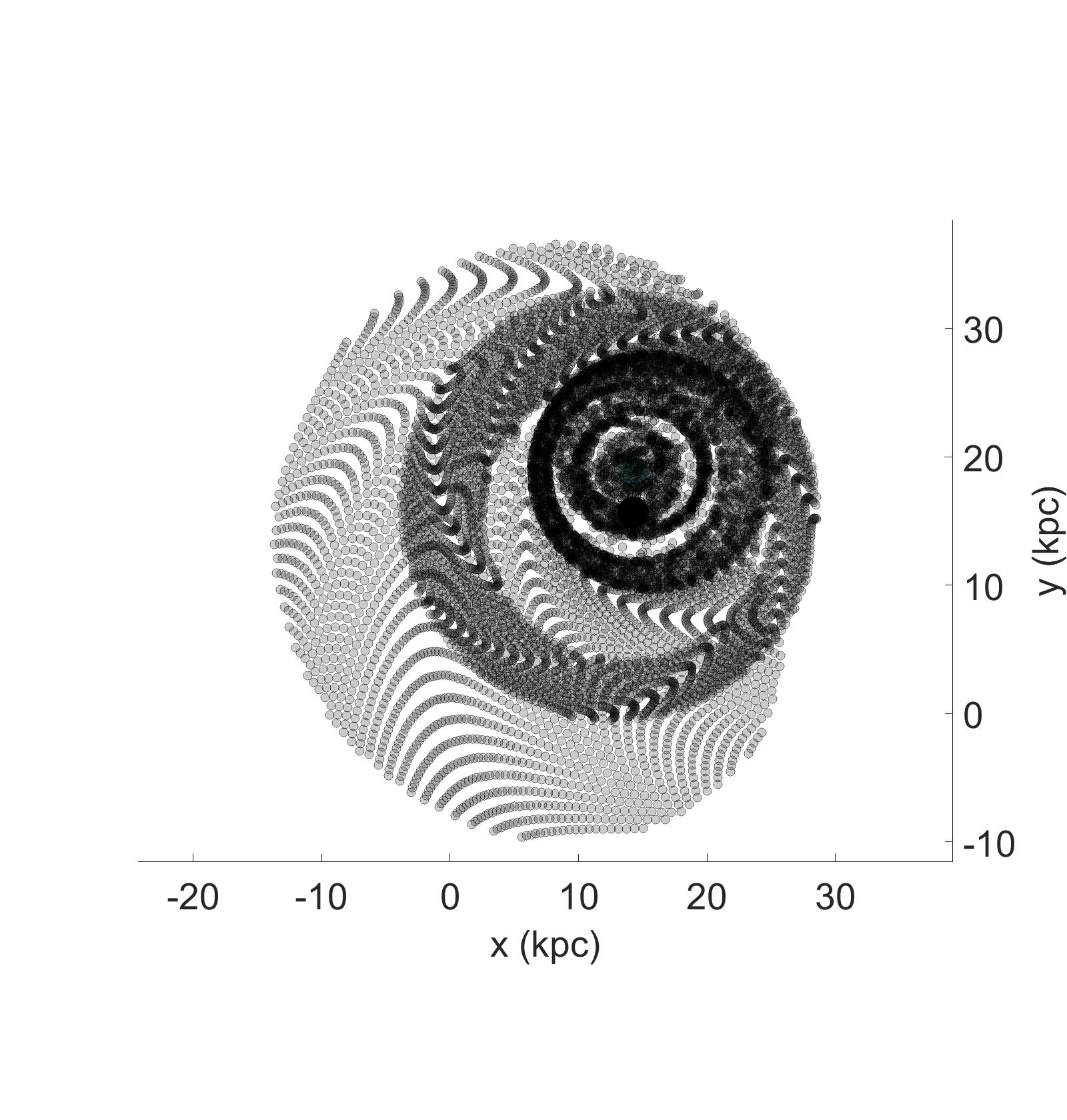}
        }
                \subfigure[]{%
          \includegraphics[width=0.4\textwidth]{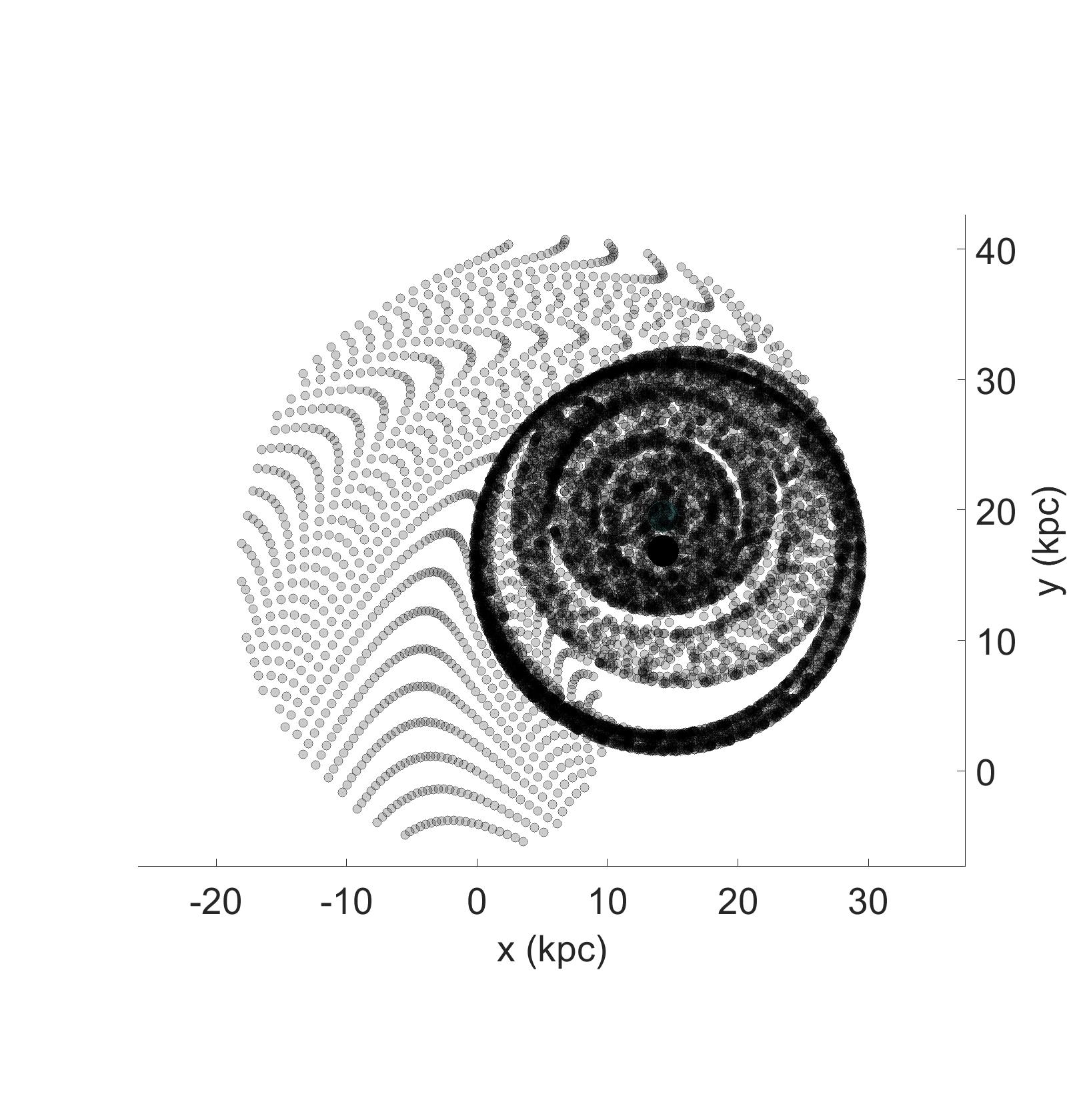}
        }\\ 
    \end{center}
    
\end{figure*}

\Cref{fig:vollmertests} is a model that was run for 30 Myr with a 50 kpc offset between the two galaxies to test orbital integrations.  This offset is large enough that neither disc interacts significantly with the other.  The orbits remain nearly circular in each disc as they separate from each other, as expected for this interaction.  The orbits of each galaxy's COM also move as expected.  The G1 COM, initially at rest, is minimally accelerated toward the departing G2 COM and the total separation between the two galaxies continues to grow.

\indent \Cref{fig:vollmertests2} is a kinematic test with gas cloud collisions turned off.  Here the test particles in grey that normally would have lost some momentum from the disc-disc collision are evolved unaffected.  The code normally merges all of the collided particles, but here there are no extra particles added to represent merged particles that would be left behind in G1.  This has no affect on the evolution of the G2 test particles.  This test was run for 300 Myr, as opposed to the 30 Myr for every other run, to provide enough time for gravitational effects between each galaxy to fully develop.  

In frame a) warping of the discs in the z-direction is clear. This is typical of the 'flapping' behavior seen in ring galaxy simulations due to the different forces felt by and different response times of the inner and outer discs. This effect can also be seen in some of the models above. In frame c) the disc of G2 can be seen beginning to create a small tail in the bottom left.  The discs themselves also slightly expand in the z direction.  In the inner part of the disc a ring wave has formed and can be clearly seen in the XY projection of frame b).  At 200 Myr there is a large amount of material now beyond the radius of the original disc.  More structure has developed in the region inside the ring wave.  The outermost ring seen at 100 Myr has grown by several kpc in diameter. The evolution of these ring waves is as expected from analytic and numerical models, confirming the general accuracy of the numerical integrations.

The second major part of the model code is computing the subgrid heating and cooling of gas elements. To check these calculations, the thermal histories of many gas elements were compared to analytical estimates of permitted line and CII cooling and are found to be in agreement. An example of some thermal histories is shown in \cref{fig:particletemperatures} of the main paper.

\bsp	
\label{lastpage}
\end{document}